\documentclass[preprint, aps, prd]{revtex4-1}

\def\pd{\partial}
\def\mc{\mathcal}

\usepackage[dvips]{graphicx}
\usepackage{amssymb}
\usepackage{amssymb,amsmath}
\usepackage{graphicx}
\usepackage{dsfont}
\usepackage{caption}
\usepackage{subcaption}
\usepackage{mathtools}
\usepackage{verbatim}
\usepackage{graphicx}
\usepackage{multirow}
\usepackage[outline]{contour}
\usepackage{xcolor,colortbl}
\makeatother
\begin{document}

\title{Holographic solutions from 5D $SO(2)\times ISO(3)$ $N=4$ gauged supergravity}

\author{Parinya Karndumri}\email[REVTeX Support:
]{parinya.ka@hotmail.com}
\affiliation{String Theory and Supergravity Group, Department of Physics, Faculty of Science,
Chulalongkorn University, 254 Phayathai Road, Pathumwan, Bangkok
10330, Thailand}

\date{\today}
\begin{abstract}
We study various types of holographic solutions from five-dimensional $N=4$ gauged supergravity coupled to three vector multiplets with $SO(2)\times ISO(3)$ gauge group. This gauged supergravity can be obtained from the maximal gauged supergravity in seven dimensions on a Riemann surface. For a negatively curved Riemann surface $H^2$, the resulting five-dimensional gauged supergravity admits a supersymmetric $N=4$ $AdS_5$ critical point. This $AdS_5$ vacuum is dual to an $N=2$ superconformal field theory (SCFT) arising from M5-branes wrapped on $H^2$. We study holographic RG flows between this SCFT and $N=2$ non-conformal phases by deformations involving relevant, marginal and irrelevant operators. Solutions describing conformal interfaces between these non-conformal phases and singular boundaries are also given. We finally study a number of supersymmetric $AdS_5$ black string and black hole solutions holographically dual to RG flows across dimensions from the $N=2$ SCFT to two-dimensional SCFTs and superconformal quantum mechanics in the IR. A number of solutions describing black strings and black holes in asymptotically domain wall space-time are also found. All of the solutions can be uplifted to M-theory by a consistent truncation on $H^2\times S^4$. 
\end{abstract}
\maketitle

\tableofcontents
\section{Introduction}
The AdS/CFT correspondence \cite{maldacena,Gubser_AdS_CFT,Witten_AdS_CFT} provides a very useful tool for investigating many aspects of strongly-coupled conformal field theories. The study of this holographic duality between a $(d+1)$-dimensional gravity theory and the dual field theory in $d$ dimensions has been effectively achieved via different types of solutions to gauged supergravities in various dimensions. These holographic solutions include domain walls with Minkowski and $AdS$ slices describing RG flows and conformal interfaces within the dual superconformal field theories (SCFTs). Another class of supergravity solutions interpolates between $AdS$ spaces of different dimensionalities. In the dual field theories, these solutions describe twisted compactifications of higher-dimensional conformal field theories on a compact manifold to other conformal field theories in lower dimensions.  
\\
\indent In this paper, we are interested in holographic solutions from half-maximal $N=4$ gauged supergravity in five dimensions with $SO(2)\times ISO(3)$ gauge group. This gauged supergravity is obtained by coupling the pure $N=4$ supergravity to three vector multiplets resulting in $SO(1,1)\times SO(5,3)$ global symmetry. By embedding the $SO(2)\times ISO(3)$ gauge group in $SO(5,3)$, we obtain the $N=4$ gauged supergravity with a supersymmetric $N=4$ $AdS_5$ vacuum at the origin of the scalar manifold. This gauged supergravity has been shown to arise from a compactification of the $SO(5)$ maximal gauged supergravity in seven dimensions on a Riemann surface with genus greater than one, $H^2$, in \cite{ISO3_5D_N4_gauntlett}. Using the consistent truncation of eleven-dimensional supergravity on $S^4$ to the $N=4$ seven-dimensional gauged supergravity \cite{M-theory_on_S4_1,M-theory_on_S4_2}, the $SO(2)\times ISO(3)$ gauged supergravity in five dimensions can be embedded in eleven dimensions via a consistent truncation on $H^2\times S^4$. Furthermore, with the formulation of exceptional field theory (EFT), it has also been shown that this is the only consistent truncation of M-theory on an $S^4$ fibration over a Riemann surface to $N=4$ gauged supergravity in five dimensions \cite{Malek_AdS5_N4_embed}. We will study holographic solutions from this gauged supergravity that describe various deformations of the $N=2$ SCFT dual to the aforementioned supersymmetric $AdS_5$ vacuum. 
\\
\indent We will first study holographic RG flows between the $N=4$ $AdS_5$ vacuum and non-conformal phases of the $N=2$ SCFT, see \cite{GPPZ} to \cite{5D_flowII} for holographic RG flow solutions in other five-dimensional gauged supergravities. Since there is no other supersymmetric $AdS_5$ vacua in this $SO(2)\times ISO(3)$ $N=4$ gauged supergravity, all supersymmetric RG flows essentially break conformal symmetry leading to solutions that interpolate between a conformal fixed point and non-conformal phases corresponding to domain wall geometries. These solutions describe deformations of the $N=2$ SCFT to non-conformal field theories or RG flows from non-conformal phases to a conformal fixed point in the IR depending on the types of deformations being relevant or irrelevant. In addition, we will also study supersymmetric Janus solutions in the form of $AdS_4$-sliced domain walls in constrast to the flat or Poincare domain walls in the case of RG flows. These solutions are dual to three-dimensional conformal interfaces, see \cite{Bak_Janus} to \cite{5D_N4_Janus} for Janus solutions in five-dimensional gauged supergravities.      
\\
\indent The final class of solutions considered in this paper is supersymmetric black strings and black holes in asymptotically $AdS_5$ space. These solutions interpolate between the $AdS_5$ vacuum and $AdS_3\times \Sigma$ and $AdS_2\times \mc{M}_3$ geometries in the IR. $\Sigma$ is a Riemann surface, and $\mc{M}_3$ is a 3-manifold with constant curvature. Holographically, the solutions describe RG flows across dimensions from $N=2$ SCFT in four dimensions to two-dimensional SCFTs and superconformal quantum mechanics in the IR. In order for these solutions to preserve some amount of supersymmetry, it is necessary to perform a topological twist by turning on certain gauge fields to cancel the spin connections on $\Sigma$ and $\mc{M}_3$ \cite{Witten_twist,MN_nogo}. Accordingly, the IR theories arise from twisted compactification of the $N=2$ SCFT on $\Sigma$ or $\mc{M}_3$. Similar solutions in other five-dimensional gauged supergravities can be found in \cite{black_string_Klemm1} to \cite{5D_N4_black_stringII}.
\\
\indent The paper is organized as follows. In section \ref{N4_SUGRA},
we review five-dimensional $N=4$ gauged supergravity coupled to three vector multiplets with $SO(2)\times ISO(3)$ gauge group. Holographic RG flow solutions will be considered in section \ref{RG_flows}. In section \ref{Janus_solutions}, we look for supersymmetric Janus solutions describing three-dimensional conformal interfaces within four-dimensional field theories. We also give a number of numerical Janus solutions. In sections \ref{string} and \ref{BH}, we find supersymmetric $AdS_5$ black strings and black holes with near horizon geometries $AdS_{3}\times \Sigma^2$ and $AdS_2\times \mc{M}_3$, respectively. We give some conclusions and comments in section \ref{conclusion}. In the appendix, we collect some relevant formulae for obtaining the eleven-dimensional metric. This is a useful tool to identify possible dual field theories and to determine whether a given IR singularity is physical or not. 
\section{Five-dimensional $N=4$ gauged supergravity with $SO(2)\times ISO(3)$ gauge group}\label{N4_SUGRA} 
In this section, we give a brief review of $N=4$ gauged supergravity constructed in \cite{N4_gauged_SUGRA,5D_N4_Dallagata}. We mainly focus on bosonic Lagrangian and supersymmetry tranformations of fermionic fields which are relevant for finding supersymmetric solutions. The complete construction can be found in \cite{N4_gauged_SUGRA,5D_N4_Dallagata} to which we refer for more detail.

\subsection{Five-dimensional $N=4$ gauged supergravity}
The $N=4$ supergravity multiplet consists of the graviton
$e^{\hat{\mu}}_\mu$, four gravitini $\psi_{\mu i}$, six vectors $(A^0_\mu,A_\mu^m)$, four spin-$\frac{1}{2}$ fields $\chi_i$ and one real scalar $\Sigma$, the dilaton. Space-time and tangent space indices are denoted respectively by $\mu,\nu,\ldots =0,1,2,3,4$ and
$\hat{\mu},\hat{\nu},\ldots=0,1,2,3,4$. The fundamental representation of $SO(5)_R\sim USp(4)_R$
R-symmetry is described by $m,n=1,\ldots, 5$ for $SO(5)_R$ and $i,j=1,2,3,4$ for $USp(4)_R$. A vector multiplet contains a vector field $A_\mu$, four gaugini $\lambda_i$ and five scalars $\phi^m$. For $N=4$ supergravity coupled to $n$ vector multiplets, we will use indices $a,b=1,\ldots, n$ to label these multiplets $(A^a_\mu,\lambda^{a}_i,\phi^{ma})$. 
\\
\indent In supergravity coupled to $n$ vector multiplets, there are $6+n$ vector fields denoted collectively by $A^{\mc{M}}_\mu=(A^0_\mu,A^m_\mu,A^a_\mu)$ and $5n+1$ scalars in the $SO(1,1)\times SO(5,n)/SO(5)\times SO(n)$ coset manifold. For later convenience, we have introduced a collective index $\mc{M}=(0,M)$ as in \cite{N4_gauged_SUGRA}. The $5n$ scalars parametrizing the $SO(5,n)/SO(5)\times SO(n)$ coset can be described by a coset representative $\mc{V}_M^{\phantom{M}A}$ transforming under the global $G=SO(5,n)$ and the local $H=SO(5)\times SO(n)$ by left and right multiplications, respectively. We use the global $SO(5,n)$ indices $M,N,\ldots=1,2,\ldots , 5+n$ while the local $H$ indices $A,B,\ldots$ can be split as $A=(m,a)$. The coset representative can then be written as
\begin{equation}
\mc{V}_M^{\phantom{M}A}=(\mc{V}_M^{\phantom{M}m},\mc{V}_M^{\phantom{M}a}).
\end{equation}
It is also useful to define a symmetric and $SO(5)\times SO(n)$ invariant matrix 
\begin{equation}
M_{MN}=\mc{V}_M^{\phantom{M}m}\mc{V}_N^{\phantom{M}m}+\mc{V}_M^{\phantom{M}a}\mc{V}_N^{\phantom{M}a}\, .
\end{equation}
All fermionic fields are symplectic Majorana spinors subject to the condition
\begin{equation}
\xi_i=\Omega_{ij}C(\bar{\xi}^j)^T 
\end{equation}
with $C$ and $\Omega_{ij}$ being the charge conjugation matrix and $USp(4)$ symplectic matrix, respectively.
\\
\indent As in other dimensions, gaugings of $N=4$ supergravity in five dimensions are efficiently obtained by using the embedding tensor formalism. In the present case, the corresponding embedding tensor has the components $\xi^{M}$, $\xi^{MN}=\xi^{[MN]}$ and $f_{MNP}=f_{[MNP]}$. These components determine the embedding of a gauge group $G_0$ in the global symmetry group $SO(1,1)\times SO(5,n)$. In this paper, we will consider only gaugings with $\xi^M=0$ which admit supersymmetric $AdS_5$ vacua as shown in \cite{AdS5_N4_Jan}. We will then set $\xi^{M}=0$ from now on. This also leads to considerable simplification in various expressions. In particular, the quadratic constraints on the embedding tensor simply reduce to
\begin{equation}
f_{R[MN}{f_{PQ]}}^R=0\qquad \textrm{and}\qquad {\xi_M}^Qf_{QNP}=0\, .\label{QC}
\end{equation}
Furthermore, for $\xi^{M}=0$, the gauge group is embedded entirely in $SO(5,n)$ with the corresponding gauge generators in $SO(5,n)$ fundamental representation given by
\begin{equation}
{(X_M)_N}^P=-{f_M}^{QR}{(t_{QR})_N}^P={f_{MN}}^P\quad \textrm{and}\quad {(X_0)_N}^P=-\xi^{QR}{(t_{QR})_N}^P={\xi_N}^P\, .
\end{equation}
We have chosen $SO(5,n)$ generators of the form ${(t_{MN})_P}^Q=\delta^Q_{[M}\eta_{N]P}$ with $\eta_{MN}=\textrm{diag}(-1,-1,-1,-1,-1,1,1,\ldots,1)$ being the $SO(5,n)$ invariant tensor. The gauge covariant derivative reads
\begin{equation}
D_\mu=\nabla_\mu+A_\mu^{M}X_M+A^0_\mu X_0=\nabla_\mu+A^{\mc{M}}X_{\mc{M}}
\end{equation}
with $\nabla_\mu$ being a space-time covariant derivative including $SO(5)\times SO(n)$ composite connection.  
\\
\indent The bosonic Lagrangian of a general gauged $N=4$ supergravity can be written as
\begin{eqnarray}
e^{-1}\mc{L}&=&\frac{1}{2}R-\frac{3}{2}\Sigma^{-2}D_\mu \Sigma D^\mu \Sigma +\frac{1}{16} D_\mu M_{MN}D^\mu
M^{MN}-V\nonumber \\
& &-\frac{1}{4}\Sigma^2M_{MN}\mc{H}^M_{\mu\nu}\mc{H}^{N\mu\nu}-\frac{1}{4}\Sigma^{-4}\mc{H}^0_{\mu\nu}\mc{H}^{0\mu\nu}+e^{-1}\mc{L}_{\textrm{top}}
\end{eqnarray}
where $e$ is the vielbein determinant.  
\\
\indent The covariant gauge field strength tensors read
\begin{equation}
\mc{H}^{\mc{M}}_{\mu\nu}=2\pd_{[\mu}A^{\mc{M}}_{\nu]}+{X_{\mc{N}\mc{P}}}^{\mc{M}}A^{\mc{N}}_\mu A^{\mc{P}}_\nu+Z^{\mc{M}\mc{N}}B_{\mu\nu\mc{N}}\label{covariant_field_strength}
\end{equation}
with 
\begin{equation}
Z^{MN}=\frac{1}{2}\xi^{MN}\qquad \textrm{and} \qquad Z^{0M}=-Z^{M0}=\frac{1}{2}\xi^M=0\, .
\end{equation}
\indent In the embedding tensor formalism, the two-form fields $B_{\mu\nu \mc{M}}$ are introduced off-shell. These fields do not have kinetic terms and couple to vector fields via the topological term $\mc{L}_{\textrm{top}}$. It is useful to note the first-order field equations for these two-form fields 
\begin{equation}
Z^{\mc{M}\mc{N}}\left[\frac{1}{6\sqrt{2}}\epsilon_{\mu\nu\rho\lambda\sigma}\mc{H}^{(3)\rho\lambda\sigma}_{\mc{N}}-\mc{M}_{\mc{N}\mc{P}}
\mc{H}^{\mc{P}}_{\mu\nu}\right]=0\label{2-form_field_eq}
\end{equation}
with $\mc{M}_{00}=\Sigma^{-4}$, $\mc{M}_{0M}=0$ and $\mc{M}_{MN}=\Sigma^2M_{MN}$. This gives a duality relation between vectors and two-form fields. The field strength $\mc{H}^{(3)}_{\mc{M}}$ is defined by
\begin{equation}
Z^{\mc{M}\mc{N}}\mc{H}^{(3)}_{\mu\nu\rho\mc{N}}=Z^{\mc{M}\mc{N}}\left[3D_{[\mu}B_{\nu\rho]\mc{N}}
+6d_{\mc{NPQ}}A^{\mc{P}}_{[\mu}\left(\pd_\nu A^{\mc{Q}}_{\rho]}+\frac{1}{3}{X_{\mc{RS}}}^{\mc{Q}}A^{\mc{R}}_\nu A^{\mc{S}}_{\rho]}\right)\right]\label{H3_def}
\end{equation}
for $d_{0MN}=d_{MN0}=d_{M0N}=\eta_{MN}$ and 
\begin{equation}
{X_{MN}}^P={f_{MN}}^P,\qquad {X_{M0}}^0=0,\qquad {X_{0M}}^N={\xi_M}^N\, . 
\end{equation}
\indent The scalar potential is given by
\begin{eqnarray}
V&=&-\frac{1}{4}\left[f_{MNP}f_{QRS}\Sigma^{-2}\left(\frac{1}{12}M^{MQ}M^{NR}M^{PS}-\frac{1}{4}M^{MQ}\eta^{NR}\eta^{PS}\right.\right.\nonumber \\
& &\left.+\frac{1}{6}\eta^{MQ}\eta^{NR}\eta^{PS}\right) +\frac{1}{4}\xi_{MN}\xi_{PQ}\Sigma^4(M^{MP}M^{NQ}-\eta^{MP}\eta^{NQ})\nonumber \\
& &\left.
+\frac{\sqrt{2}}{3}f_{MNP}\xi_{QR}\Sigma M^{MNPQR}\right]
\end{eqnarray}
where $M^{MN}$ is the inverse of $M_{MN}$, and $M^{MNPQR}$ is obtained from
\begin{equation}
M_{MNPQR}=\epsilon_{mnpqr}\mc{V}_{M}^{\phantom{M}m}\mc{V}_{N}^{\phantom{M}n}
\mc{V}_{P}^{\phantom{M}p}\mc{V}_{Q}^{\phantom{M}q}\mc{V}_{R}^{\phantom{M}r}
\end{equation}
by raising the indices with $\eta^{MN}$. 
\\
\indent As mentioned above, $\mc{L}_{\textrm{top}}$ is the topological term describing the kinetic terms for two-form fields and the coupling between two-form and gauge fields. Since all solutions given in this paper have vanishing two-form fields, we will not give the explicit form of $\mc{L}_{\textrm{top}}$ here. This can be found in \cite{N4_gauged_SUGRA}.
\\
\indent Supersymmetry transformations of fermionic fields are given by
\begin{eqnarray}
\delta\psi_{\mu i} &=&D_\mu \epsilon_i+\frac{i}{\sqrt{6}}\Omega_{ij}A^{jk}_1\gamma_\mu\epsilon_k\nonumber \\
& &-\frac{i}{6}\left(\Omega_{ij}\Sigma{\mc{V}_M}^{jk}\mc{H}^M_{\nu\rho}-\frac{\sqrt{2}}{4}\delta^k_i\Sigma^{-2}\mc{H}^0_{\nu\rho}\right)({\gamma_\mu}^{\nu\rho}-4\delta^\nu_\mu\gamma^\rho)\epsilon_k,\\
\delta \chi_i &=&-\frac{\sqrt{3}}{2}i\Sigma^{-1} D_\mu
\Sigma\gamma^\mu \epsilon_i+\sqrt{2}\Omega_{ij}A_2^{kj}\epsilon_k\nonumber \\
& &-\frac{1}{2\sqrt{3}}\left(\Sigma \Omega_{ij}{\mc{V}_M}^{jk}\mc{H}^M_{\mu\nu}+\frac{1}{\sqrt{2}}\Sigma^{-2}\delta^k_i\mc{H}^0_{\mu\nu}\right)\gamma^{\mu\nu}\epsilon_k,\\
\delta \lambda^a_i&=&i\Omega^{jk}({\mc{V}_M}^aD_\mu
{\mc{V}_{ij}}^M)\gamma^\mu\epsilon_k+\sqrt{2}\Omega_{ij}A_{2}^{akj}\epsilon_k-\frac{1}{4}\Sigma{\mc{V}_M}^a\mc{H}^M_{\mu\nu}\gamma^{\mu\nu}\epsilon_i
\end{eqnarray}
in which the fermion shift matrices are defined by 
\begin{eqnarray}
A_1^{ij}&=&-\frac{1}{\sqrt{6}}\left(\sqrt{2}\Sigma^2\Omega_{kl}{\mc{V}_M}^{ik}{\mc{V}_N}^{jl}\xi^{MN}+\frac{4}{3}\Sigma^{-1}{\mc{V}^{ik}}_M{\mc{V}^{jl}}_N{\mc{V}^P}_{kl}{f^{MN}}_P\right),\nonumber
\\
A_2^{ij}&=&\frac{1}{\sqrt{6}}\left(\sqrt{2}\Sigma^2\Omega_{kl}{\mc{V}_M}^{ik}{\mc{V}_N}^{jl}\xi^{MN}-\frac{2}{3}\Sigma^{-1}{\mc{V}^{ik}}_M{\mc{V}^{jl}}_N{\mc{V}^P}_{kl}{f^{MN}}_P\right),\nonumber
\\
A_2^{aij}&=&-\frac{1}{2}\left(\Sigma^2{{\mc{V}_M}^{ij}\mc{V}_N}^a\xi^{MN}-\sqrt{2}\Sigma^{-1}\Omega_{kl}{\mc{V}_M}^a{\mc{V}_N}^{ik}{\mc{V}_P}^{jl}f^{MNP}\right).
\end{eqnarray}
\indent $\mc{V}_M^{\phantom{M}ij}$ is defined in terms of ${\mc{V}_M}^m$ and $SO(5)$ gamma matrices ${\Gamma_{mi}}^j$ as
\begin{equation}
{\mc{V}_M}^{ij}=\frac{1}{2}{\mc{V}_M}^{m}\Gamma^{ij}_m
\end{equation}
with $\Gamma^{ij}_m=\Omega^{ik}{\Gamma_{mk}}^j$. Similarly, the inverse ${\mc{V}_{ij}}^M$ can be written as
\begin{equation}
{\mc{V}_{ij}}^M=\frac{1}{2}{\mc{V}_m}^M(\Gamma^{ij}_m)^*=\frac{1}{2}{\mc{V}_m}^M\Gamma_{m}^{kl}\Omega_{ki}\Omega_{lj}\,
.
\end{equation}
We will use the following representation of $SO(5)$ gamma matrices
\begin{eqnarray}
\Gamma_1&=&-\sigma_2\otimes \sigma_2,\qquad \Gamma_2=\mathbb{I}_2\otimes \sigma_1,\qquad \Gamma_3=\mathbb{I}_2\otimes \sigma_3,\nonumber\\
\Gamma_4&=&\sigma_1\otimes \sigma_2,\qquad \Gamma_5=\sigma_3\otimes \sigma_2
\end{eqnarray}
with $\sigma_i$, $i=1,2,3$, being the Pauli matrices.
\\
\indent The covariant derivative on $\epsilon_i$ is given by
\begin{equation}
D_\mu \epsilon_i=\pd_\mu \epsilon_i+\frac{1}{4}\omega_\mu^{ab}\gamma_{ab}\epsilon_i+{Q_{\mu i}}^j\epsilon_j
\end{equation}
with the composite connection defined by
\begin{equation}
{Q_{\mu i}}^j={\mc{V}_{ik}}^M\pd_\mu {\mc{V}_M}^{kj}-A^0_\mu\xi^{MN}\mc{V}_{Mik}{\mc{V}_N}^{kj}-A^M_\mu{\mc{V}_{ik}}^N\mc{V}^{kjP}f_{MNP}\, .
\end{equation}

\subsection{$N=4$ gauged supergravity with $SO(2)\times ISO(3)$ gauge group}
In this paper, we will consider $N=4$ gauged supergravity coupled to $n=3$ vector multiplets with $SO(2)\times ISO(3)\sim SO(2)\times (SO(3)\ltimes \mathbb{R}^3)$ gauge group. This gauged supergravity arises from a consistent truncation of eleven-dimensional supergravity on $H^2\times S^4$ \cite{ISO3_5D_N4_gauntlett}. In the notation of \cite{Malek_AdS5_N4_embed}, the corresponding embedding tensor is given by
\begin{eqnarray}
\xi^{\hat{m}\hat{n}}&=&g_1\epsilon_{\hat{m}\hat{n}},\qquad \hat{m},\hat{n}=1,2,\nonumber \\  
f_{\tilde{m}\tilde{n}\tilde{p}}&=&g\epsilon_{\tilde{m}\tilde{n}\tilde{p}},\qquad \tilde{m},\tilde{n},\tilde{p}=3,4,5,\nonumber \\
f_{a+5,b+5,c+5}&=&-2g\epsilon_{abc},\qquad f_{a+2,b+5,c+5}=-g\epsilon_{abc}, \qquad a,b,c=1,2,3
\end{eqnarray} 
with the gauge coupling constants $g_1$ and $g$. We have split the indices $m,n=1,2,\ldots, 5$ as $m=(\hat{m},\tilde{m})$ with $\hat{m}=1,2$ and $\tilde{m}=3,4,5$. From the embedding tensor, we find that the $SO(2)$ factor is generated by $\xi_{12}$ while the compact $SO(3)\subset ISO(3)$ is diagonally embedded in $SO(2)\times SO(3)\times SO(3)\subset SO(5,3)$ with gauge generators $X_{\tilde{m}}=(X_3,X_4,X_5)$. The three-dimensional translation group $\mathbb{R}^3$ is generated by $X_{\tilde{m}}-X_{a+5}=(X_3-X_6,X_4-X_7,X_5-X_8)$. At this point, it is useful to note that the truncation constructed in \cite{ISO3_5D_N4_gauntlett} is valid for any Riemann surfaces $\Sigma=S^2, H^2, \mathbb{R}^2$. We have taken a particular case of $\Sigma=H^2$ corresponding to $l=-1$ in \cite{ISO3_5D_N4_gauntlett} since this truncation leads to a supersymmetric $AdS_5$ vacuum.  
\\
\indent To give an explicit parametrization of the scalar coset $SO(5,3)/SO(5)\times SO(3)$, we take the $SO(5,3)$ non-compact generators to be
\begin{equation}
Y_{ma}=t_{m,a+5},\qquad m=1,2,\ldots, 5,\qquad a=1,2,3\, .
\end{equation}
Accordingly, the coset representative can be written as
\begin{equation}
\mc{V}=e^{\phi^{ma}Y_{ma}}\, .
\end{equation}
As shown in \cite{ISO3_5D_N4_gauntlett}, at the origin of $SO(5,3)/SO(5)\times SO(3)$ with $\phi^{ma}=0$, the $SO(2)\times ISO(3)$ gauged supergravity admits a supersymmetric $AdS_5$ vacuum with 
\begin{equation}
\Sigma=-\left(\frac{g}{\sqrt{2}g_1}\right)^{\frac{1}{3}},\qquad V_0=-3\left(\frac{g^2g_1}{2}\right)^{\frac{2}{3}},\qquad L=\left(-\frac{4\sqrt{2}}{g^2g_1}\right)^{\frac{1}{3}}\, .
\end{equation}
The $AdS_5$ radius $L$ is related to the cosmological constant $V_0$ via
\begin{equation}
L=\sqrt{-\frac{6}{V_0}}\, .
\end{equation}
By choosing $g=-\sqrt{2}g_1$ or equivalently scaling $\Sigma$ to $\Sigma=1$, we find
\begin{equation}
V_0=-\frac{3}{2}g^2\qquad \textrm{and}\qquad L=\frac{2}{g}
\end{equation}
in which we have chosen $g>0$. The $AdS_5$ vacuum preserves $N=4$ supersymmetry and $SO(2)\times SO(3)\subset SO(2)\times ISO(3)$ symmetry. This vacuum can be identified as an $AdS_5\times H^2\times S^4$ solution of eleven-dimensional supergravity. 
\\
\indent It is also useful to note all scalar masses at this $N=4$ vacuum given in \cite{ISO3_5D_N4_gauntlett}. These are shown in table \ref{table1}. We denote scalars by the representations under the residual symmetry $SO(2)\times SO(3)$ at the vacuum. The dilaton $\Sigma$ is the singlet $(\mathbf{1},\mathbf{1})_\Sigma$. The other representations are obtained by considering the embedding of $SO(2)\times SO(3)$ in $SO(5)\times SO(3)\subset SO(5,3)$. Under $SO(5)\times SO(3)$, the $15$ scalars transform as $(\mathbf{5},\mathbf{3})$. By branching $SO(5)\rightarrow SO(2)\times SO(3)$ with $\mathbf{5}\rightarrow (\mathbf{2},\mathbf{1})+(\mathbf{1},\mathbf{3})$, we find that 
\begin{equation}  
(\mathbf{5},\mathbf{3})\rightarrow (\mathbf{2},\mathbf{1},\mathbf{3})+(\mathbf{1},\mathbf{3},\mathbf{3})
\end{equation}
under $SO(2)\times SO(3)\times SO(3)$. Finally, by taking the diagonal subgroup of the two $SO(3)$ factors, we end up with
\begin{equation}  
(\mathbf{5},\mathbf{3})\rightarrow (\mathbf{2},\mathbf{3})+(\mathbf{1},\mathbf{1})+(\mathbf{1},\mathbf{3})+(\mathbf{1},\mathbf{5}).
\end{equation}
In the table, we have also given the dimensions of the operators dual to these scalars given by the relation $m^2L^2=\Delta(\Delta-4)$. The three massless scalars in $(\mathbf{1},\mathbf{3})$ are Goldstone bosons corresponding to the symmetry breaking $ISO(3)\rightarrow SO(3)$ at the vacuum.
\begin{table}[h]
\centering
\begin{tabular}{|c|c|c|}
  \hline
Scalars & $m^2L^2\phantom{\frac{1}{2}}$ & $\Delta$  \\ \hline
 $(\mathbf{1},\mathbf{1})_\Sigma$ &  $-4$ &  $2$  \\
 $(\mathbf{1},\mathbf{1})$ & $12$ &  $6$  \\
 $(\mathbf{1},\mathbf{3})$ &  $0$ &  $4$  \\
$(\mathbf{1},\mathbf{5})$ & $0$ &  $4$  \\
$(\mathbf{2},\mathbf{3})$ & $5$ &  $5$  \\
  \hline
\end{tabular}
\caption{Scalar masses at the $N=4$ supersymmetric $AdS_5$ vacuum with $SO(2)\times SO(3)$ symmetry and the
corresponding dimensions of the dual operators.}\label{table1}
\end{table}

Furthermore, it has also been pointed out in \cite{ISO3_5D_N4_gauntlett} that there are no other supersymmetric $AdS_5$ vacua.     
\section{Holographic RG flows}\label{RG_flows}
We begin with the simplest holographic solutions describing RG flows from the $N=2$ SCFT dual to the supersymmetric $AdS_5$ vacuum. To simplify the computation, we consider a truncation to $SO(2)_{\textrm{diag}}$ singlet sector. This $SO(2)_{\textrm{diag}}$ is a diagonal subgroup of $SO(2)\times SO(2)\subset SO(2)\times SO(3)\subset SO(2)\times ISO(3)$ generated by $\xi_{12}+X_3$. There are five singlets under $SO(2)_{\textrm{diag}}$ symmetry with the corresponding coset representative given by 
\begin{equation}
\mc{V}=e^{\phi_1(Y_{12}+Y_{23})}e^{\phi_2(Y_{13}-Y_{22})}e^{\phi_3(Y_{42}+Y_{53})}e^{\phi_4(Y_{43}-Y_{52})}e^{\phi_5Y_{31}}\, .\label{SO2d_coset}
\end{equation}
In this section, we are interested in holographic RG flow solutions with the metric ansatz given by
\begin{equation}
ds^2=e^{2A(r)}dx^2_{1,3}+dr^2\, .
\end{equation}
$dx^2_{1,3}$ is the flat metric on the Minkowski space in four dimensions with the warp factor $A$ depending only on the radial coordinate $r$. To preserve four-dimensional Poincare symmetry of $dx^2_{1,3}$, we take the non-vanishing scalars to depend only on $r$ and set all the other fields to zero.
\\
\indent It turns out that in order to consistently truncate out all the vector fields, we need to set $\phi_2=\phi_4=0$. The latter lead to non-vanishing Yang-Mills currents that become the sources for the gauge fields. With $\phi_2=\phi_4=0$, we find the scalar potential
\begin{eqnarray}
V&=&\frac{1}{8}g_1^2\cosh^2\phi_3\sinh^2\phi_1\Sigma^4(3+\cosh2\phi_1+2\cosh2\phi_3\sinh^2\phi_1)\nonumber \\
& &+\sqrt{2}g_1g\Sigma\left(\cosh^2\phi_1\cosh^2\phi_3\cosh\phi_5+2\sinh^2\phi_3\sinh\phi_5\right)\nonumber \\
& &+\frac{1}{256}g^2\Sigma^{-2}\left[42-48\cosh2\phi_1+6\cosh4\phi_1+4\cosh(4\phi_1-2\phi_3) \right.\nonumber \\
& &-24\cosh(2\phi_1-2\phi_3)+\cosh(4\phi_1-4\phi_3)-152\cosh2\phi_3-2\cosh4\phi_3\nonumber \\
& &-24\cosh(2\phi_1+2\phi_3)+\cosh(4\phi_1+4\phi_3)+4\cosh(4\phi_1+2\phi_3)\nonumber \\
& & +128\cosh^2\phi_1\sinh^22\phi_3\sinh2\phi_5+\left\{2(5+4\cosh2\phi_1+\cosh4\phi_1)\times \right.\nonumber \\
& &\times \cosh4\phi_3 +6\cosh4\phi_1 +8\cosh2\phi_3\sinh^2\phi_1(5+4\cosh2\phi_1) \nonumber \\
& &\left.\left. -6-4\cosh2\phi_1\right\} 4\cosh2\phi_5 \right].
\end{eqnarray}
The $A^{ij}_1$ tensor takes a diagonal form
\begin{equation}
A^{ij}_1=\textrm{diag}(\alpha,\beta,\alpha^*,\beta)\label{A1}
\end{equation}
with 
\begin{eqnarray}
\alpha&=&\frac{2}{\sqrt{2}}\Sigma^{-1}(\cosh\phi_3+i\sinh\phi_1-\cosh\phi_1\sinh\phi_3)\left[g_1\Sigma^3(\cosh\phi_3-i\sinh\phi_1\right.\nonumber \\
& &+\cosh\phi_1\sinh\phi_3)  -\sqrt{2}g\left\{\cosh\phi_3\cosh\phi_5+(\cosh\phi_5-2\sinh\phi_5)\times \right.\nonumber \\
& &\left.\left.\times (\cosh\phi_1\sinh\phi_3-i\sinh\phi_1)\right\}\right],\\
\beta&=&\frac{1}{16\sqrt{3}}\Sigma^{-1}\left[\sqrt{2}ge^{-\phi_5}\left\{11+4e^{2\phi_5}(\sinh\phi_3-\cosh\phi_1\cosh\phi_3)^2 \right.\right.\nonumber \\
& & \left.-6\cosh2\phi_1\cosh^2\phi_3-\cosh2\phi_3+4\cosh\phi_1\sinh2\phi_3\right\}\nonumber \\
& &\left.-2g_1\Sigma^3(3+\cosh2\phi_1+2\cosh2\phi_3\sinh^2\phi_1)\right].
\end{eqnarray}
The real eigenvalue $\beta$ gives rise to the superpotential $W$ in terms of which the scalar potential can be written as
\begin{equation}
V=\frac{3}{2}\Sigma^2\left(\frac{\pd W}{\pd \Sigma}\right)^2+\frac{9}{4}\textrm{sech}^2\phi_3\left(\frac{\pd W}{\pd \phi_1}\right)^2+\frac{9}{4}\left(\frac{\pd W}{\pd \phi_3}\right)^2+\frac{9}{2}\left(\frac{\pd W}{\pd \phi_5}\right)^2-6W^2
\end{equation}
with $W=\sqrt{\frac{2}{3}}\beta$. It should be noted that for $\phi_1=0$, we have $\alpha=-\beta$. In this case, the solutions preserve $N=4$ supersymmetry. In general, the solutions preserve only $N=2$ supersymmetry corresponding to the Killing spinors $\epsilon^2$ and $\epsilon^4$. 
\\
\indent Setting $\epsilon^1=\epsilon^3=0$ and imposing the projector
\begin{equation}
\gamma_{\hat{r}}\epsilon_2=\mp i\epsilon_4\qquad \textrm{and}\qquad \gamma_{\hat{r}}\epsilon_4=\pm i\epsilon_2,\label{gamma_r_proN2}
\end{equation}
we find the BPS equations from the conditions $\delta \psi^i_{\hat{\mu}}=0$ with $\hat{\mu}=0,1,2,3$, $\delta\chi^i=0$ and $\delta \lambda^i_a=0$ of the form
\begin{eqnarray}
\phi_1'&=&\mp\frac{3}{2}\textrm{sech}^2\phi_3\frac{\pd W}{\pd \phi_1},\qquad \phi'_3=\mp\frac{3}{2}\frac{\pd W}{\pd \phi_3},\nonumber \\
\phi'_5&=&\mp 3\frac{\pd W}{\pd \phi_5},\qquad \Sigma'=\mp\Sigma^{2}\frac{\pd W}{\pd \Sigma},\qquad A'=\pm W\, .
\end{eqnarray}
Throughout this paper, we use $'$ to denote $r$-derivatives. The condition $\delta\psi^i_r=0$ leads to the usual Killing spinors of the domain wall of the form
\begin{equation}
\epsilon^{2,4}=e^{\frac{A}{2}}\epsilon^{2,4}_0
\end{equation} 
with $\epsilon^{2,4}_0$ being constant spinors satisfying the projector \eqref{gamma_r_proN2}.

\subsection{Holographic RG flows with $SO(2)\times SO(3)$ symmetry}
We begin with a simple solution with $SO(2)\times SO(3)$ symmetry. In this case, we set 
\begin{equation}
\phi_1=0\qquad \textrm{and}\qquad \phi_5=\phi_3\, .
\end{equation}
With $\phi_1=0$, all the eigenvalues of $A_1^{ij}$ are degenerate up to an overall sign. The solutions then perserve $N=4$ supersymmetry corresponding to $\epsilon^i$ with $i=1,2,3,4$. However, the $\gamma_r$ projector, which in this case takes the form of
\begin{equation}
\gamma_r\epsilon_i=\mp{(\sigma_2\otimes \sigma_3)_i}^j\epsilon_j, \label{gamma_r_proN4}
\end{equation} 
will reduce the number of supercharges from $16$ to $8$. We also note that setting $\epsilon_1=\epsilon_3=0$ in this projector, we recover the projector given in \eqref{gamma_r_proN2}. Therefore, the solutions perserve $N=2$ Poincare supersymmetry in four dimensions. These solutions would describe holographic RG flows from the $N=2$ SCFT to non-conformal $N=2$ field theories.   
\\
\indent The explicit form of the relevant BPS equations are given by
\begin{eqnarray}
\phi'_3&=&g\Sigma^{-1}e^{-2\phi_3}\sinh\phi_3,\nonumber \\
\Sigma'&=&-\frac{1}{6}\left[ge^{-3\phi_3}(1-3e^{2\phi_3})-2\sqrt{2}g_1\Sigma^3\right],\nonumber \\
A'&=&-\frac{1}{6}\Sigma^{-1}\left[g(e^{-3\phi_3}-3e^{-\phi_3})+\sqrt{2}g_1\Sigma^3\right]. \label{SO3_RG_BPS}
\end{eqnarray}
We have chosen a specific choice of sign in \eqref{gamma_r_proN4} such that the UV $N=2$ SCFT appears in the limit $r\rightarrow \infty$. From table \ref{table1}, we know that the dilaton $\Sigma$ and the $SO(2)\times SO(3)$ singlet scalar given by $\phi_5=\phi_3$ are dual to operators of dimensions $\Delta=2$ and $\Delta=6$, respectively. This is also confirmed by linearizing the BPS equations given above which results in  
\begin{equation}
\phi_3\sim e^{gr}\sim e^{\frac{2r}{L}}\qquad \textrm{and}\qquad \Sigma\sim -\left(\frac{g}{\sqrt{2}g_1}\right)^{\frac{1}{3}}+e^{-\frac{2r}{L}}
\end{equation}
in which we have used the relations $g_1=-\frac{g}{\sqrt{2}}$ and $L=\frac{2}{g}$. This behavior implies that $\Sigma$ corresponds to a vacuum expectation value of an operator of dimension $\Delta=2$ while $\phi_3$ leads to a source or deformation by an irrelevant operator of dimension $\Delta=6$.
\\
\indent We first consider an RG flow solution with only a relevant deformation given by an operator of dimension $\Delta=2$ dual to $\Sigma$. In this case, the $N=2$ SCFT dual to the $AdS_5$ vacuum appears in the UV of the RG flow. We then set $\phi_3=0$ and obtain a simpler set of BPS equations
\begin{eqnarray}
\Sigma'=\frac{1}{3}(g+\sqrt{2}g_1\Sigma^3)\qquad \textrm{and}\qquad A'=\frac{1}{6}\Sigma^{-1}(2g-\sqrt{2}g_1\Sigma^3).
\end{eqnarray}  
With $g_1=-\frac{g}{\sqrt{2}}$, the solution can be readily obtained
\begin{eqnarray}
2g(r-r_0)&=&2\sqrt{3}\tan^{-1}\left(\frac{1+2\Sigma}{\sqrt{3}}\right)+\ln\left[\frac{1+\Sigma+\Sigma^2}{(\Sigma-1)^2}\right],\\
A&=&\ln \Sigma-\frac{1}{2}\ln(g-g\Sigma^3).
\end{eqnarray}
In this solution, we have neglected an additive integration constant for $A$ that can be absorbed in rescaling of $dx^2_{1,3}$ coordinates. The $r_0$ constant is also irrelevant and can be set to zero by shifting radial coordinate $r$.
\\
\indent As $r\rightarrow \infty$, we find
\begin{equation}
\Sigma\sim 1+e^{-gr}\qquad \textrm{and}\qquad A\sim \frac{g}{2}r
\end{equation}
as expected from the $AdS_5$ asymptotic geometry. The solution is singular at $\Sigma=0$. Near this singularity, we find 
\begin{equation}
\Sigma\sim \frac{2}{3}g(r-r_*)\qquad \textrm{and}\qquad A\sim \ln(r-r_*)
\end{equation}
with $r_*=r_0+\frac{\sqrt{3}\pi}{6g}$. We can determine whether the IR singularity is physically acceptable or not by using the criterion given in \cite{Gubser_Sing}. Near the singularity, as $r\rightarrow r_*$, the scalar potential is bounded from above 
\begin{equation}
V\sim -\frac{1}{8(r-r_*)}\rightarrow -\infty\, .
\end{equation}
Therefore, the singularity is physical, and the solution describes an RG flow from the $N=2$ SCFT to a non-conformal field theory in the IR. 
\\
\indent Since the gauged supergravity under consideration here is a consistent truncation of eleven-dimensional supergravity on $H^2\times S^4$, we can also determine whether the singularity in the uplifted solution is physical by the criterion of \cite{MN_nogo}. We will choose the $S^4$ cordinates $\mu^{\hat{a}}$, $\hat{a}=1,2,\ldots,5$, to be
\begin{eqnarray}
\mu^1&=&\cos\vartheta\cos\theta,\qquad \mu^2=\cos\vartheta\sin\theta,\qquad \mu^3=\sin\vartheta\sin\beta\cos\xi,\nonumber \\
\mu^4&=&\sin\vartheta \sin\beta\sin\xi,\qquad \mu^5=\sin\vartheta\cos\beta
\end{eqnarray}  
which satisfy $\mu^{\hat{a}}\mu^{\hat{a}}=1$. Using the relations given in the appendix, we find
\begin{eqnarray}
\hat{g}_{00}=\left(e^{-6\lambda}\cos^2\vartheta+e^{4\lambda}\sin^2\vartheta\right)^{\frac{1}{3}}e^{2A-4\varphi}\sim (r-r_*)^2\rightarrow 0
\end{eqnarray}
which also implies that the IR singularity is physical.
\\
\indent We then consider the solution with non-vanishing $\phi_3$. In this case, there is a source of an operator of dimension $\Delta=6$ near $AdS_5$ geometry. Therefore, the dual $N=2$ SCFT must be an IR fixed point. We now explicitly solve the BPS equations given in \eqref{SO3_RG_BPS}. By combining $\Sigma'$ and $\phi_3'$ equations, we can solve for $\Sigma$ as a function of $\phi_3$. The result is given by
\begin{equation}
\Sigma^3=-\frac{2ge^{\phi_3}(e^{2\phi_3}-1)}{\sqrt{2}g_1e^{4\phi_3}+2gC_0}
\end{equation}
with an integration constant $C_0$. In order to make the solution approach the $N=4$ $AdS_5$ vacuum, we need to choose the constant $C_0$ to be 
\begin{equation}
C_0=-\frac{g_1}{\sqrt{2}g}\, .
\end{equation}
The solution for $\Sigma$ then becomes
\begin{equation}
\Sigma^3=-\frac{\sqrt{2}ge^{\phi_3}}{g_1(1+e^{2\phi_3})}\, .\label{Sigma_sol1}
\end{equation}
Similarly, by combining $A'$ and $\phi'_3$ equations and using the solution for $\Sigma$, we can solve for $A$ as a function of $\phi_3$ 
\begin{equation}
A=\frac{1}{3}\phi_3+\frac{1}{3}\ln(1-e^{2\phi_3})+\frac{1}{6}\ln\left[\sqrt{2}g_1(e^{4\phi_3}-1)\right]
\end{equation}
in which we have again neglected an additive integration constant. Finally, by using \eqref{Sigma_sol1} in $\phi_3'$ equation and defining a new radial coordinate $\rho$ via $\frac{d\rho}{dr}=\Sigma^{-1}$, we find
\begin{equation}
g(\rho-\rho_0)=-\ln(1+e^{\phi_3})+\ln (1-e^{\phi_3})+2e^{\phi_3}
\end{equation}
with $\rho_0$ being another integration constant. $\rho_0$ can also be set to zero by shifting the coordinate $\rho$. This solution holographically describes an RG flow to an $N=2$ conformal fixed point in the IR. The RG flow is driven by a deformation involving a source term of a dimension 6-operator dual to $\phi_3$ in the presence of non-vanishing vacuum expectation value of a dimension-2 operator dual to $\Sigma$.  
\\
\indent In order to determine the possible UV field theory, we look at the behavior of the solution as $\rho\rightarrow \infty$. In this limit, the five-dimensional solution becomes
\begin{equation}
\phi_3\sim \ln \rho,\qquad \Sigma\sim \rho^{-\frac{1}{3}},\qquad A\sim \frac{5}{3}\ln \rho\, .\label{UV_asymp1}
\end{equation}
We now uplift the solution to eleven dimensions using the $S^4\times H^2$ truncation ansatz of \cite{M-theory_on_S4_2} and \cite{ISO3_5D_N4_gauntlett} with, in the notation of \cite{ISO3_5D_N4_gauntlett} and see also more detail in the appendix,
\begin{equation}
ds^2_{11}=\Delta^{\frac{1}{3}}ds^2_7+\frac{1}{m^2}\Delta^{-\frac{2}{3}}T^{-1}_{\hat{a}\hat{b}}D\mu^{\hat{a}}D\mu^{\hat{b}}
\end{equation}
for
\begin{equation}
ds_7^2=e^{-4\phi}\left(e^{2A}dx^2_{1,3}+dr^2\right)+e^{6\phi}ds^2(H_2)\, .
\end{equation}
In these equations, $m$ is the seven-dimensional gauge coupling constant. The warp factor $\Delta$ is defined in terms of the $SL(5)$ matrix $T_{\hat{a}\hat{b}}$ as
\begin{equation}
\Delta=T_{\hat{a}\hat{b}}\mu^{\hat{a}}\mu^{\hat{b}}
\end{equation}
with $\mu^{\hat{a}}$, $\hat{a}=1,2,\ldots ,5$, being coordinates on $S^4$. In the present case, $T_{\hat{a}\hat{b}}$ takes the form, see more detail in the appendix,
\begin{equation}  
T_{\hat{a}\hat{b}}=\textrm{diag}(e^{-6\lambda},e^{-6\lambda},e^{4\lambda},e^{4\lambda},e^{4\lambda}).
\end{equation}
The scalars $\phi$ and $\lambda$ are related to $\phi_3$ and $\Sigma$ via
\begin{equation}
\phi=\frac{1}{10}(3\phi_3-\ln \Sigma)\qquad \textrm{and}\qquad \lambda=-\frac{1}{10}(\phi_3+3\ln \Sigma).
\end{equation}
\indent With the asymptotic behavior, upon using \eqref{UV_asymp1},
\begin{equation}
\phi\sim \frac{1}{3}\ln \rho\qquad \textrm{and}\qquad \lambda\sim\lambda_0\sim \textrm{constant} 
\end{equation}
and the $S^4$ coordinates of the form
\begin{equation}
\mu^{\hat{a}}=(\cos\xi\cos\theta,\cos\xi\sin\theta,\sin\xi \hat{\mu}^\alpha),\qquad \alpha=1,2,3
\end{equation}
with $\hat{\mu}^\alpha$ being coordinates on $S^2$ satisfying $\hat{\mu}^\alpha\hat{\mu}^\alpha=1$, we find the eleven-dimensional metric near the UV limit of the form
\begin{eqnarray}
ds^2_{11}&\sim& \rho^{2}\left[dx^2_{1,3}+ds^2(H_2)\right]+\frac{d\rho^2}{\rho^2}+\left(e^{6\lambda_0}\sin^2\xi+ e^{-4\lambda_0}\cos^2\xi\right) d\xi^2\nonumber \\
& &+e^{6\lambda_0}\cos^2\xi (d\theta-\omega_H)^2+e^{-4\lambda_0}\sin^2\xi d\hat{\mu}^\alpha d\hat{\mu}^\alpha
\end{eqnarray}
with $\omega_H$ being the spin connection on $H^2$. We can see that, in the UV, the extra two dimensions along $H^2$ become large. The eleven-dimensional metric takes the form of a locally $AdS_6\times \widetilde{S}^4$ with $\widetilde{S}^4$ being a deformed $S^4$ such that only the $SO(2)\times SO(3)$ subgroup of the $SO(5)$ isometry is preserved. It is also useful to point out that the $SO(2)\times SO(3)$ symmetry of the solution corresponds to the isometry of the $S^1\times S^2$ parametrized by $\theta$ and $\hat{\mu}^\alpha$. The UV field theory is then expected to be the $N=(0,2)$ six-dimensional SCFT arising from the world-volume theory of M5-branes wrapped on a hyperbolic space $H^2$. This is similar to the solutions describing D6-branes wrapped on supersymmetric four-cycles in \cite{Calos_D6_M4}.  

\subsection{Holographic RG flows with $SO(2)\times SO(2)$ symmetry}
We now consider a slightly more complicated solution with a smaller residual symmetry $SO(2)\times SO(2)$. In this case, we still have $\phi_1=0$ but unlike in the previous case $\phi_5\neq \phi_3$. The solutions preserve $N=4$ supersymmetry as in the previous case due to vanishing $\phi_1$. The explicit form of the corresponding BPS equations is given by
\begin{eqnarray}
\phi'_3&=&\frac{1}{2}g\Sigma^{-1}e^{-2\phi_3-\phi_5}(e^{2\phi_5}-1),\label{phi3_eq1}\\
\phi'_5&=&-\frac{1}{2}g\Sigma^{-1}e^{-2\phi_3-\phi_5}(1-2e^{2\phi_3}+e^{2\phi_5}),\label{phi5_eq11}\\
\Sigma'&=&\frac{1}{6}\left[2\sqrt{2}g_1\Sigma^3+ge^{-2\phi_3-\phi_5}(e^{2\phi_5}+2e^{2\phi_3}-1)\right],\label{Sigma_eq11}\\
A'&=&\frac{1}{6}\Sigma^{-1}\left[ge^{-2\phi_3-\phi_5}(e^{2\phi_5}+2e^{2\phi_3}-1)-\sqrt{2}g_1\Sigma^3\right].\label{A_eq1}
\end{eqnarray}
Near the supersymmetric $AdS_5$ vacuum, we find
\begin{equation}
\Sigma\sim -\left(\frac{g}{\sqrt{2}g_1}\right)^{\frac{1}{3}}+ e^{-\frac{2r}{L}},\qquad 2\phi_3+\phi_5\sim e^{\frac{2r}{L}},\qquad \phi_3-\phi_5\sim e^{-\frac{4r}{L}}
\end{equation}
which implies that $\phi_3$ and $\phi_5$ are dual to two different linear combinations of operators of dimensions $\Delta=6$ and $\Delta=4$. Setting $\phi_5=\phi_3$, we recover the solutions considered in the previous case. However, in this case, the irrelevant deformation corresponding to $2\phi_3+\phi_5$ cannot be consistently truncated out, so the $N=2$ SCFT must be an IR fixed point of an RG flow from a certain field theory in the UV.
\\
\indent Combing \eqref{phi3_eq1} and \eqref{phi5_eq11}, we find
\begin{equation}
\frac{d\phi_5}{d\phi_3}=\frac{1-2e^{2\phi_3}+e^{2\phi_5}}{1-e^{2\phi_5}}
\end{equation}
which can be solved by
\begin{equation}
e^{\phi_3+\phi_5}\sqrt{1-e^{2\phi_3-2\phi_5}}+\sin^{-1}e^{\phi_3-\phi_5}=C
\end{equation} 
with an integration constant $C$. We can further simplify this expression by defining   
\begin{equation}
\varphi_1=\phi_3+\phi_5\qquad \textrm{and}\qquad \varphi_2=\phi_3-\phi_5\label{varphi12_def}
\end{equation}
which results in
\begin{equation}
e^{\varphi_1}=\frac{C-\sin^{-1}e^{\varphi_2}}{\sqrt{1-e^{2\varphi_2}}}\, .\label{varphi1_sol}
\end{equation}
To make the solution approach the $AdS_5$ vacuum with $\varphi_1=\varphi_2=0$, we need to choose
\begin{equation}
C=\frac{\pi}{2}\, .
\end{equation}
We can now find the solutions for $A$ and $\Sigma$ as finctions of $\varphi_2$. It is then useful to note the BPS equation for $\varphi_2$ which takes the form
\begin{equation}
\varphi'_2=g\Sigma^{-1}e^{-\frac{1}{2}(\varphi_1-\varphi_2)}(e^{-2\varphi_2}-1).
\end{equation} 
Using \eqref{varphi12_def} and \eqref{varphi1_sol}, we can combine this equation with \eqref{Sigma_eq11} and \eqref{A_eq1} to obtain 
\begin{eqnarray}
\Sigma^{-3}&=&\frac{g_1e^{-\frac{\varphi_2}{2}}\left[\sin^{-1}e^{\varphi_2}-C-e^{\varphi_2}\sqrt{1-e^{2\varphi_2}}\right]}{\sqrt{2}g(1-e^{2\varphi_2})^{\frac{1}{4}}\sqrt{C-\sin^{-1}e^{\varphi_2}}}+\frac{\Sigma_0e^{-\frac{\varphi_2}{2}}(1-e^{2\varphi_2})^{\frac{3}{4}}}{C-\sin^{-1}e^{\varphi_2}},\\
A&=&\frac{1}{4}\varphi_2-\frac{1}{2}\ln \Sigma-\frac{3}{8}\ln (1-e^{2\varphi_2})+\frac{1}{4}\ln\left(\sin^{-1}e^{\varphi_2}-C\right)
\end{eqnarray}
with $\Sigma_0$ being another integration constant. As in the previous case, we have neglected an additive integration constant for $A$. In order to make the solution for $\Sigma$ becomes $\Sigma^{-3}=-\frac{\sqrt{2}g_1}{g}$ for $\varphi_2=0$ at the $AdS_5$ vacuum, we need to set $\Sigma_0=0$. 
\\
\indent Finally, using all these results, we can solve for $\varphi_2$ as
\begin{equation}
2g(\rho-\rho_0)=\ln(1+e^{\frac{\varphi_2}{2}})-\ln(1-e^{\frac{\varphi_2}{2}})-2\tan^{-1}e^{\frac{\varphi_2}{2}}
\end{equation}
with the new radial coordinate $\rho$ defined by $\frac{d\rho}{dr}=\frac{e^{-\frac{\varphi_1}{2}}}{\Sigma}$ and $\rho_0$ being an integration constant. As already mentioned, there is a deformation by a dimension-6 operator in the presence of vacuum expectation values of operators of dimensions $\Delta=4$ and $\Delta=2$. Therefore, the solution holographically describes an RG flow to an $N=2$ conformal fixed point in the IR as in the previous case. To determine a possible UV field theory, we again consider the uplifted eleven-dimensional metric.  
\\
\indent In this case, we choose the $S^4$ coordinates to be
\begin{eqnarray}
\mu^1&=&\cos\xi\cos\zeta\cos\vartheta,\qquad \mu^2=\cos\xi\cos\zeta\sin\vartheta,\qquad \mu^3=\cos\xi\sin\zeta\cos\theta,\nonumber \\
\mu^4&=&\cos\xi\sin\zeta\sin\theta,\qquad \mu^5=\sin\xi\, .
\end{eqnarray}
At $\rho\rightarrow \rho_0$, the five-dimensional solution is singular with
\begin{eqnarray}
& &\varphi_2\sim \frac{2}{3}\ln(\rho_0-\rho),\qquad \varphi_1\sim \textrm{constant},\nonumber \\
& &\Sigma\sim (\rho_0-\rho)^{\frac{1}{9}},\qquad A\sim \frac{1}{9}\ln(\rho_0-\rho).
\end{eqnarray}
Using the formulae given in the appendix, we find 
\begin{equation}
\phi\sim \frac{1}{45}\ln(\rho_0-\rho),\qquad \lambda\sim -\frac{2}{45}\ln(\rho_0-\rho),\qquad w\sim \frac{4}{9}\ln(\rho_0-\rho)
\end{equation}
which lead to the eleven-dimensional metric of the form
\begin{eqnarray}
ds^2_{11}&\sim& (\rho_0-\rho)^{-\frac{2}{9}}[dx^2_{1,3}+d\rho^2+ds^2(H_2)]+(\rho_0-\rho)^{\frac{4}{9}}\left[\sin^2\xi\cos^2\zeta d\xi^2 \right.\nonumber \\
& &\left.+\cos^2\xi \sin^2\zeta d\zeta^2+\cos^2\xi\cos^2\zeta (d\vartheta-\omega_H)^2+\sin2\xi \cos\zeta\sin\zeta d\xi d\zeta\right]\nonumber \\
& &+(\rho_0-\rho)^{\frac{4}{3}}\left(\sin^2\xi\cos^2\zeta d \xi^2+\cos^2\xi\cos^2\zeta d\zeta^2+\cos^2\xi\sin^2\zeta d\theta^2 \right.\nonumber \\
& &\left.-\sin2\xi\cos\zeta\sin\zeta d\xi d \zeta\right). \label{UV2_FT}
\end{eqnarray}
As in the previous case, we see that the two compact directions along $H^2$ become large, and we expect the UV field theory to be the six-dimensional field theory on the world-volume of M5-branes wrapped on $H^2$. We also note that the $SO(2)\times SO(2)$ symmetry of the solution corresponds to the $S^1\times S^1$ isometry along $\vartheta$ and $\theta$.

\subsection{Holographic RG flows with $SO(2)_{\textrm{diag}}$ symmetry}
\begin{figure}
         \centering
               \begin{subfigure}[b]{0.32\textwidth}
                 \includegraphics[width=\textwidth]{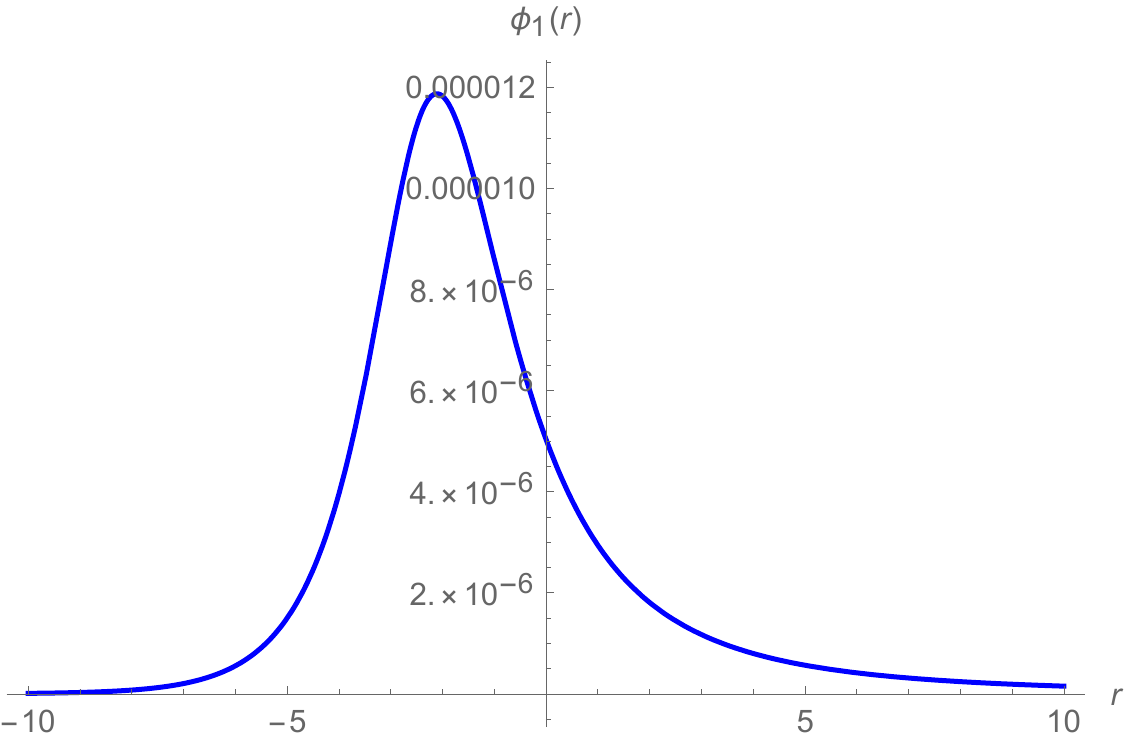}
                 \caption{Solution for $\phi_1(r)$}
         \end{subfigure}
         \begin{subfigure}[b]{0.32\textwidth}
                 \includegraphics[width=\textwidth]{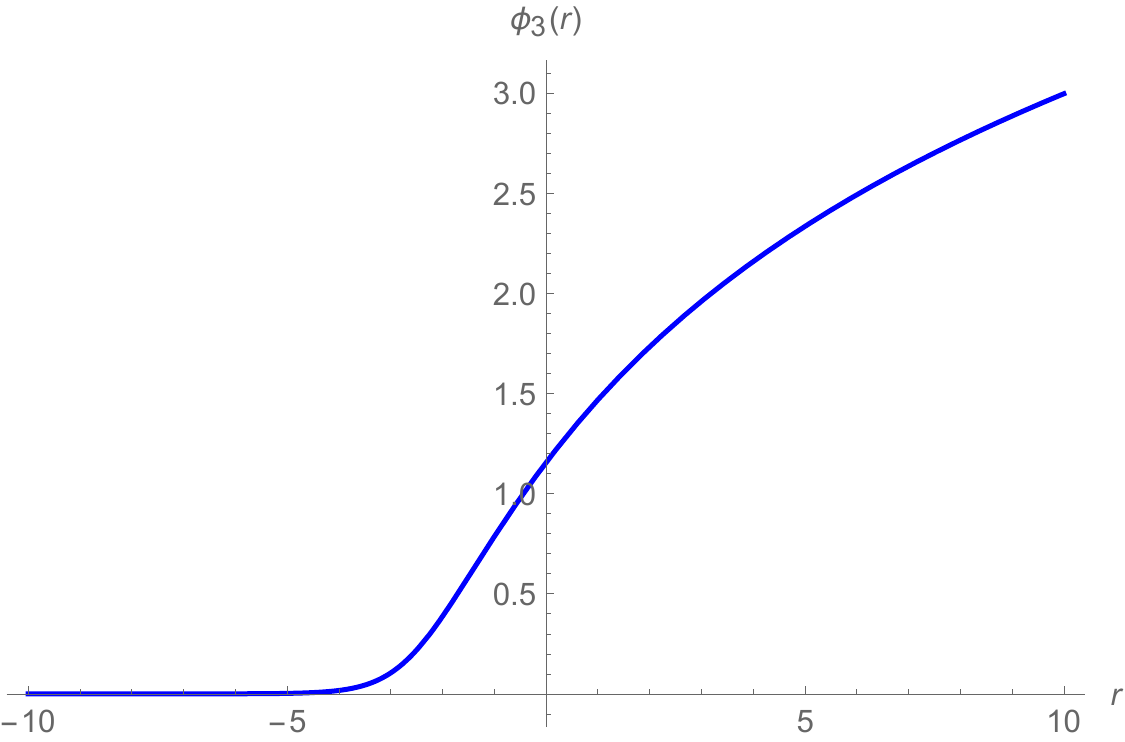}
                 \caption{Solution for $\phi_3(r)$}
         \end{subfigure}
         \begin{subfigure}[b]{0.32\textwidth}
                 \includegraphics[width=\textwidth]{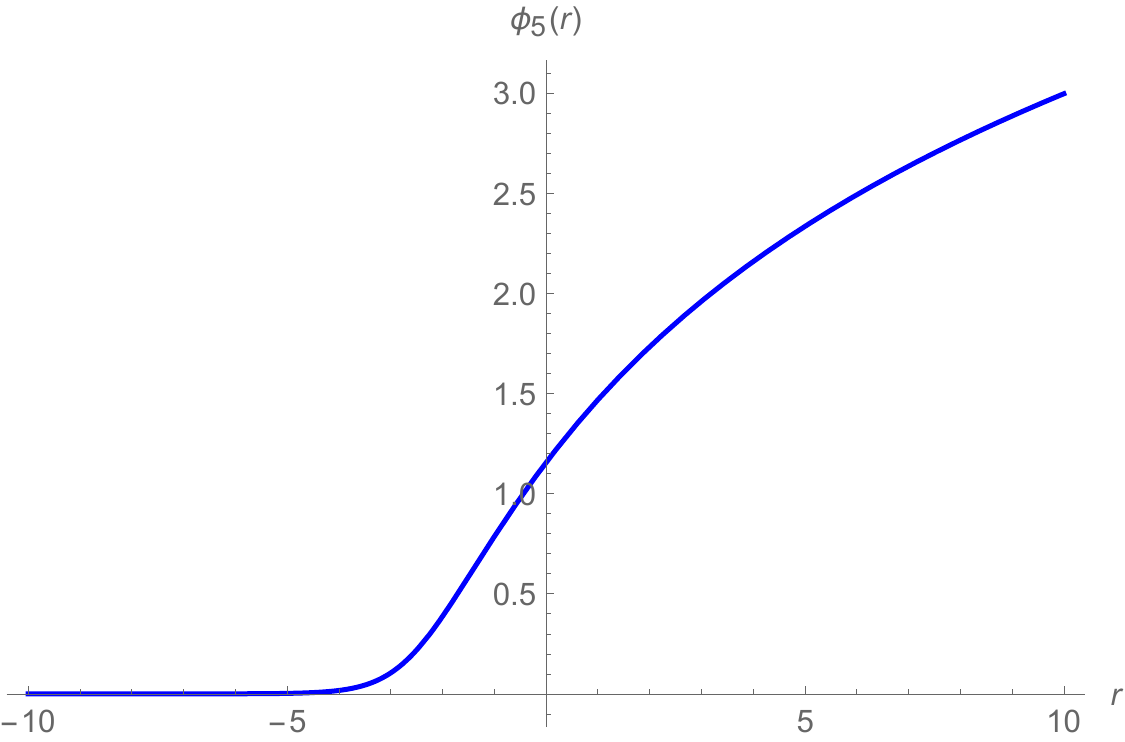}
                 \caption{Solution for $\phi_5(r)$}
         \end{subfigure}\\
               \begin{subfigure}[b]{0.32\textwidth}
                 \includegraphics[width=\textwidth]{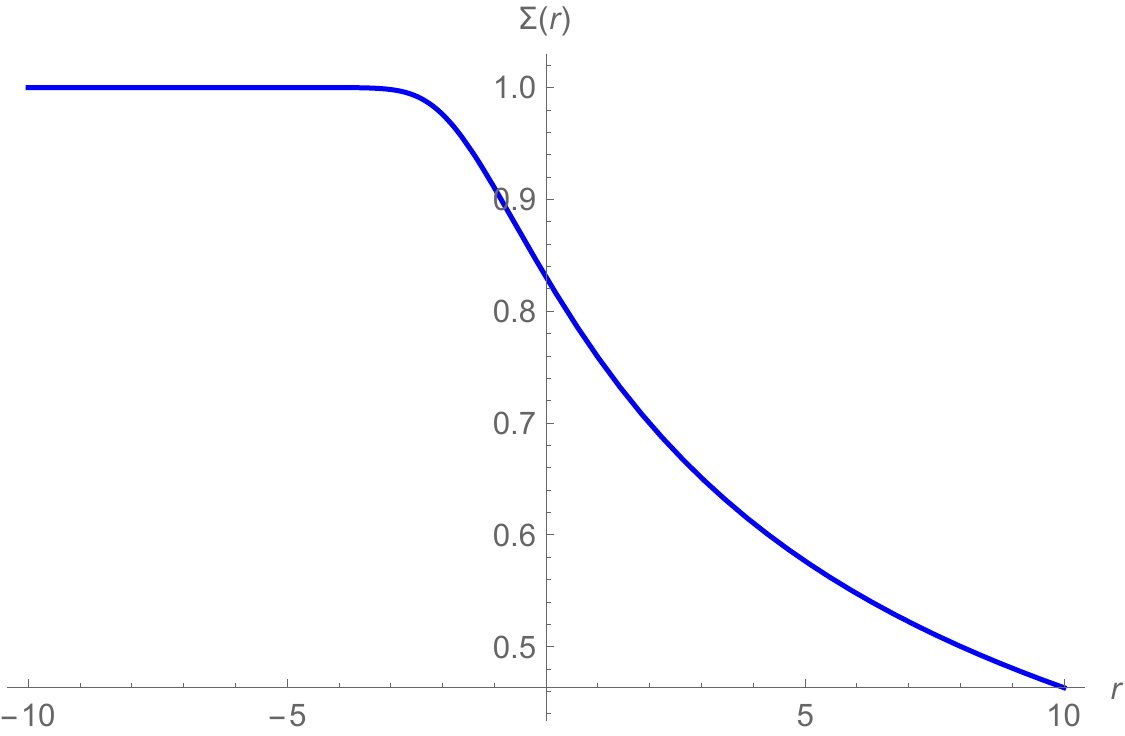}
                 \caption{Solution for $\Sigma(r)$}
         \end{subfigure}
         \begin{subfigure}[b]{0.32\textwidth}
                 \includegraphics[width=\textwidth]{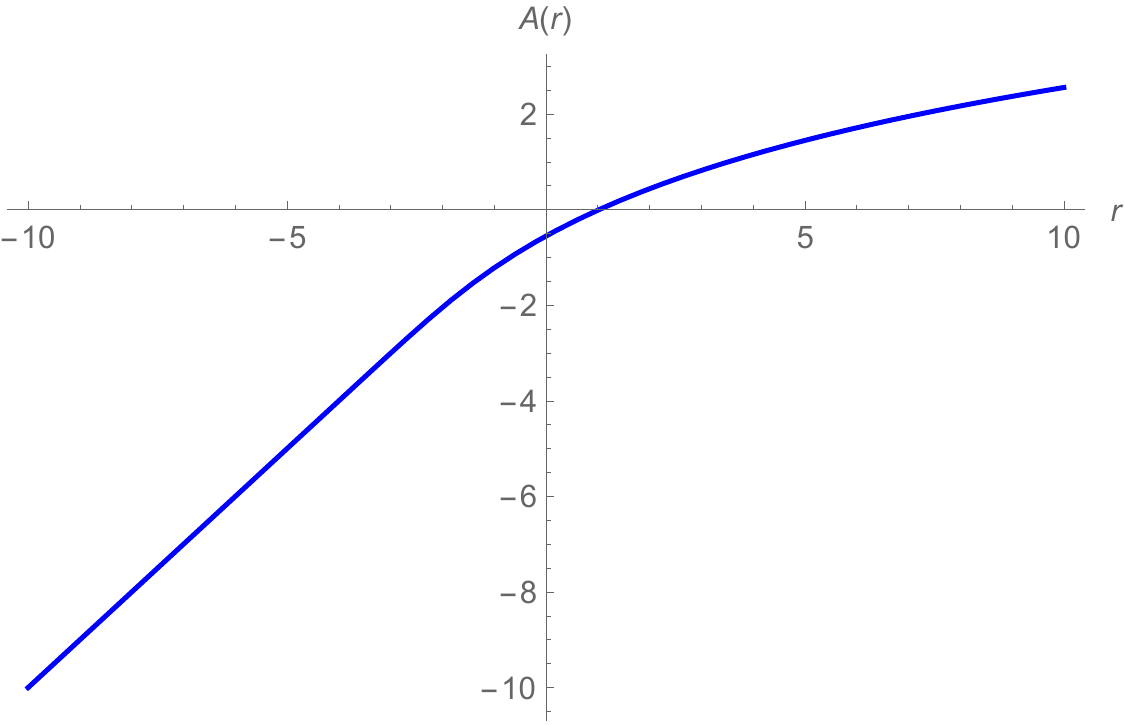}
                 \caption{Solution for $A(r)$}
         \end{subfigure}
         \begin{subfigure}[b]{0.32\textwidth}
                 \includegraphics[width=\textwidth]{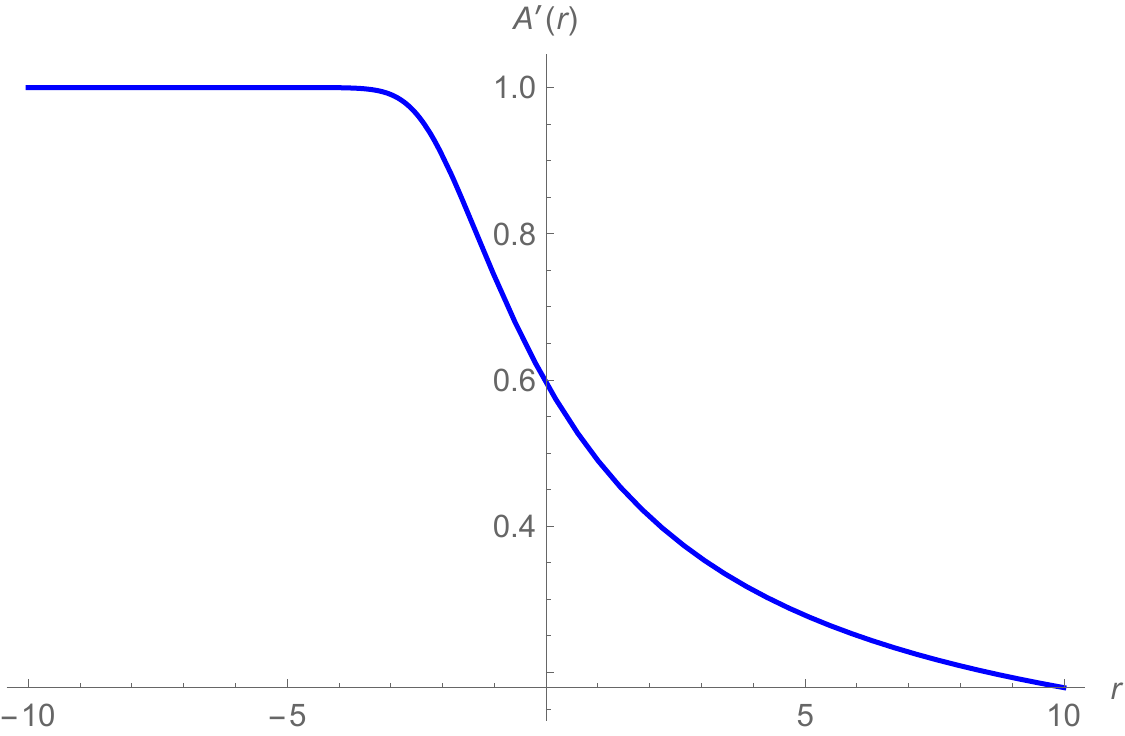}
                 \caption{Solution for $A'(r)$}
         \end{subfigure} 
         \caption{An $N=1$ supersymmetric RG flow from a UV field theory to the IR $N=2$ SCFT dual to $N=4$ $AdS_5$ vacuum for $g=2$.}\label{fig7}
 \end{figure}

For non-vanishing $\phi_1$, the solutions will preserve only $N=2$ supersymmetry and $SO(2)_{\textrm{diag}}$ symmetry. In this case, the BPS equations read 
\begin{eqnarray}
\phi'_1&=&\frac{\sinh\phi_1}{2\Sigma}\left[2g\sinh\phi_5\tanh\phi_3+\cosh\phi_1\{2g(\cosh\phi_5-2\sinh\phi_5)+\sqrt{2}g_1\Sigma^3\}\right],\nonumber \\ 
& &\\
\phi_3'&=&\frac{1}{8}\Sigma^{-1}e^{-\phi_5}\left[\sinh2\phi_3\{g+3g\cosh2\phi_1-ge^{2\phi_5}(3+\cosh2\phi_1)\right.\nonumber \\
& &\left.+2\sqrt{2}g_1\Sigma^3e^{\phi_5}\sinh^2\phi_1\}+4g(e^{2\phi_5}-1)\cosh\phi_1\cosh2\phi_3\right],\\
\phi'_5&=&\frac{1}{8}g\Sigma^{-1}e^{-\phi_5}\left[11-\cosh2\phi_3-4e^{2\phi_5}(\sinh\phi_3-\cosh\phi_1\cosh\phi_3)^2 \right.\nonumber \\
& &\left.-6\cosh2\phi_1\cosh^2\phi_3+4\cosh\phi_1\sinh2\phi_3 \right],
\end{eqnarray}
\begin{eqnarray}
\Sigma'&=&\frac{1}{24}\left[ge^{-\phi_5}\{11-6\cosh2\phi_1\cosh^2\phi_3+4\cosh\phi_1\sinh2\phi_3-\cosh2\phi_3 \right.\nonumber \\
& & +4e^{2\phi_5}(\sinh\phi_3-\cosh\phi_1\cosh\phi_3)^2\}+2\sqrt{2}g_1\Sigma^3(3+\cosh2\phi_1\nonumber \\
& &\left.+2\cosh2\phi_3\sinh^2\phi_1) \right],\\
A'&=&\frac{1}{24}\Sigma^{-1}\left[ge^{-\phi_5}\{11-\cosh2\phi_3+4e^{2\phi_5}(\sinh\phi_3-\cosh\phi_1\cosh\phi_3)^2\right. \nonumber \\ 
& &-6\cosh2\phi_1\cosh^2\phi_3 +4\cosh\phi_1\sinh2\phi_3\}-\sqrt{2}g_1\Sigma^3(3+\cosh2\phi_1\nonumber \\
& &\left.+2\cosh2\phi_3\sinh^2\phi_1)\right].
\end{eqnarray}
We first consider the asymptotic behavior near the $N=4$ $AdS_5$ vacuum given by  
\begin{eqnarray}
\Sigma\sim e^{-\frac{2r}{L}},\qquad \phi_1\sim  e^{\frac{r}{L}},\qquad 2\phi_3+\phi_5\sim e^{\frac{2r}{L}},\qquad \phi_3-\phi_5\sim e^{-\frac{4r}{L}}\, .
\end{eqnarray}
In addition to the dual operators of dimensions $2$, $4$ and $6$ appearing in the previous case, there is a source term for an irrelevant operator of dimension $\Delta=5$ dual to $\phi_1$. As in the previous case, the scalars corresponding to irrelevant deformations cannot be consistently truncated out. We also point out that due to non-vanishing $\phi_1$, the solutions preserve only four supercharges or $N=1$ supersymmetry in four dimensions. Therefore, in this case, the $N=2$ SCFT must be an IR fixed point of an RG flow involving a supersymmetry breaking deformation dual to $\phi_1$ in the UV field theory. 
\\
\indent In this case, the BPS equations are much more complicated, and we are not able to find analytic solutions. We will then look for numerical solutions. An example of numerical solutions with $g=2$ is shown in figure \ref{fig7}. We also note that although the RG flow preserves only four supercharges due to non-vanishing $\phi_1$, the non-conformal phase in the UV preserves eight supercharges since $\phi_1=0$ near the UV limit. In this case, the solution is only obtained numerically, and it is more complicated to find the eleven-dimensional metric. However, we also expect the UV field theory of the RG flow dual to this solution to be a six-dimensional field theory on the world-volume of M5-branes wrapped on an $H^2$ as in the previous cases. In particular, due to the vanishing of $\phi_1$ in the UV limit, the holographic dual of the UV field theory would be described by an eleven-dimensional metric similar to that given in \eqref{UV2_FT}.    

\section{Supersymmetric Janus solutions}\label{Janus_solutions}
\begin{figure}
         \centering
         \begin{subfigure}[b]{0.32\textwidth}
                 \includegraphics[width=\textwidth]{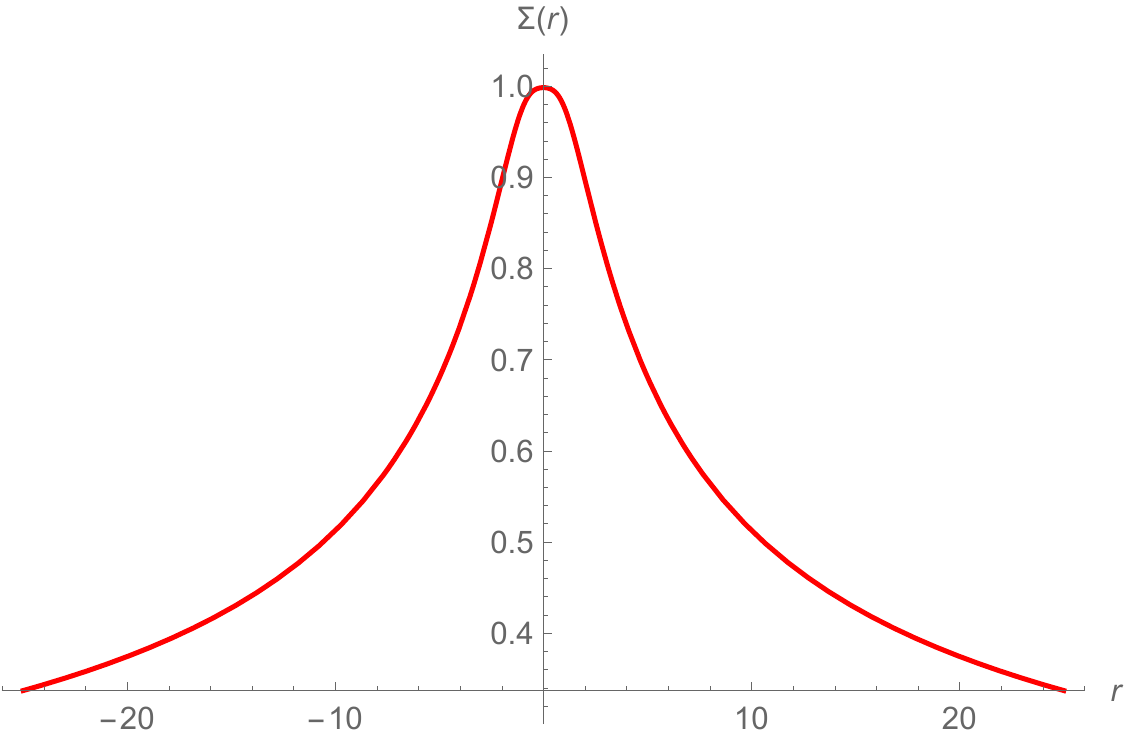}
                 \caption{Solution for $\Sigma(r)$}
         \end{subfigure}
\begin{subfigure}[b]{0.32\textwidth}
                 \includegraphics[width=\textwidth]{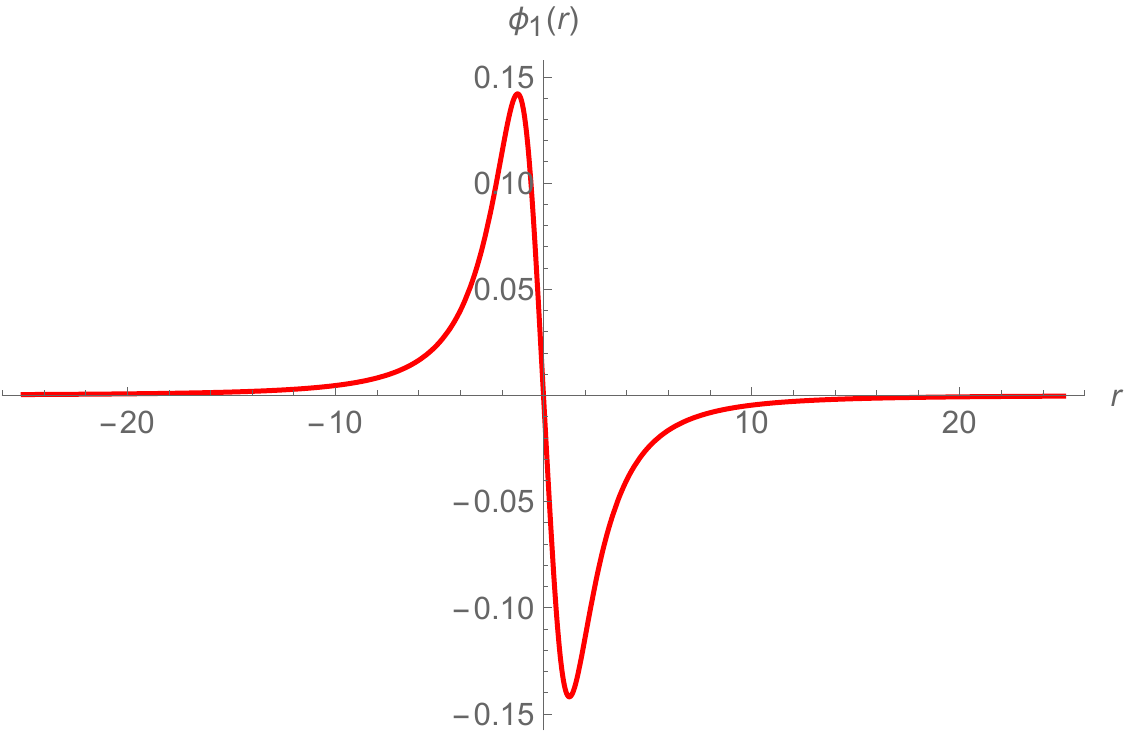}
                 \caption{Solution for $\phi_1(r)$}
         \end{subfigure}
         \begin{subfigure}[b]{0.32\textwidth}
                 \includegraphics[width=\textwidth]{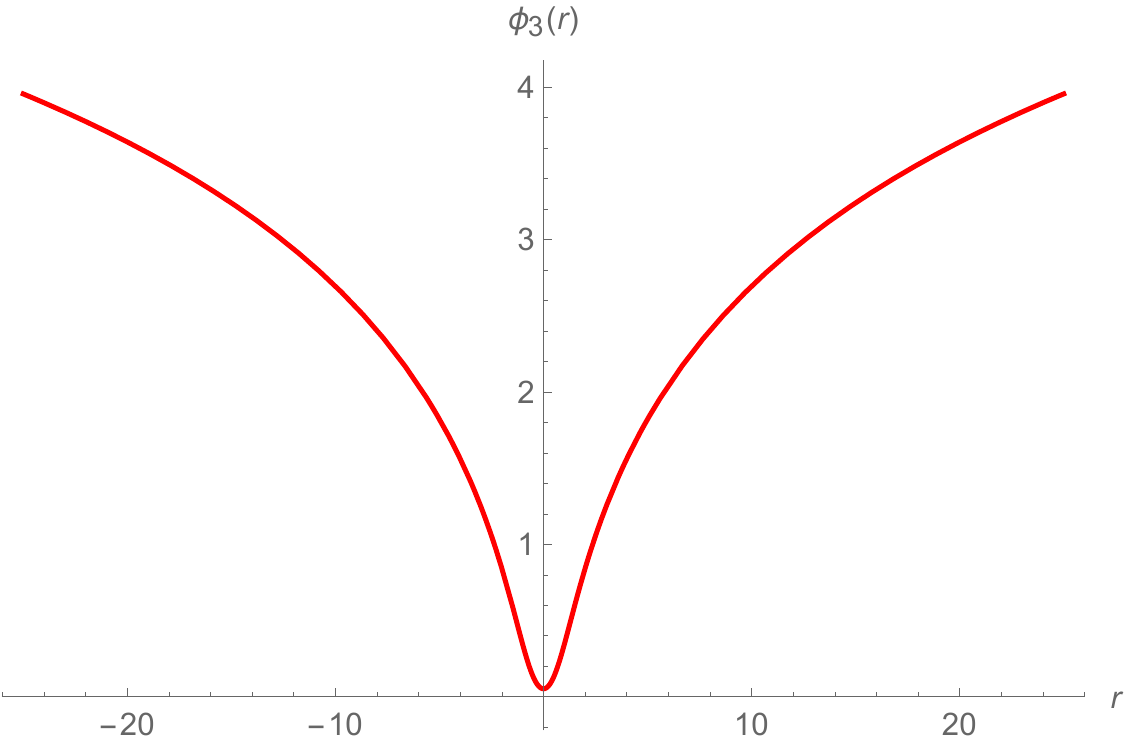}
                 \caption{Solution for $\phi_3(r)$}
         \end{subfigure}\\ 
         \begin{subfigure}[b]{0.35\textwidth}
                 \includegraphics[width=\textwidth]{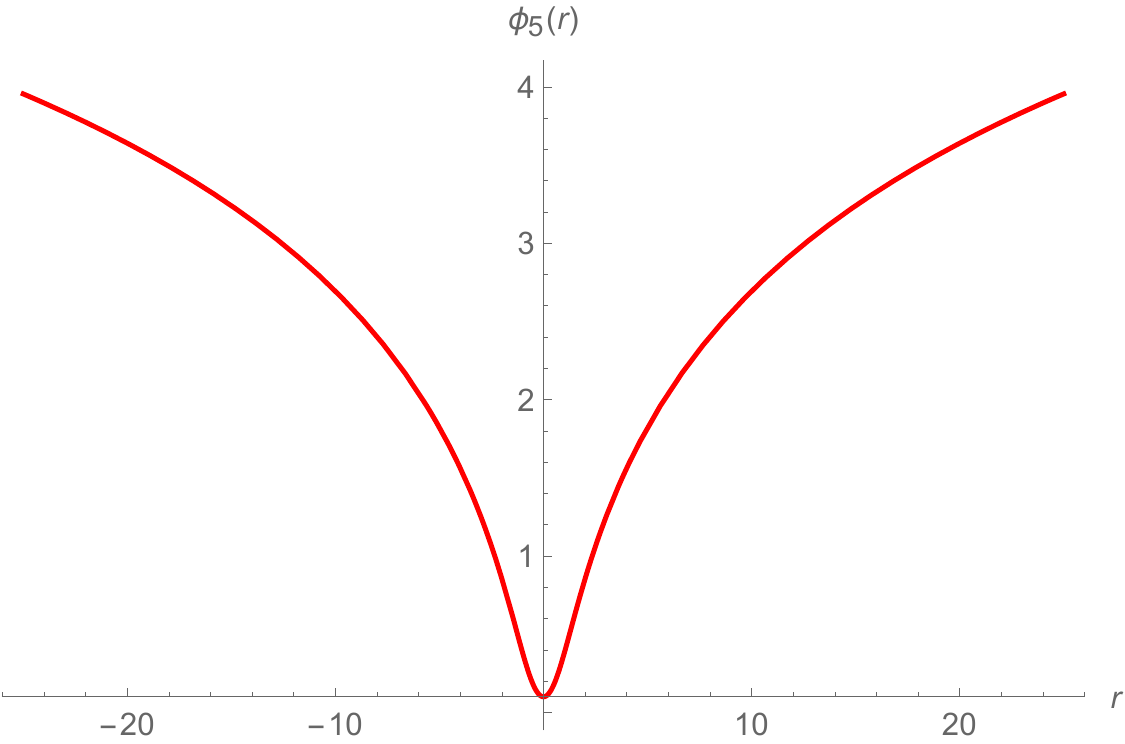}
                 \caption{Solution for $\phi_5(r)$}
         \end{subfigure}
         \begin{subfigure}[b]{0.35\textwidth}
                 \includegraphics[width=\textwidth]{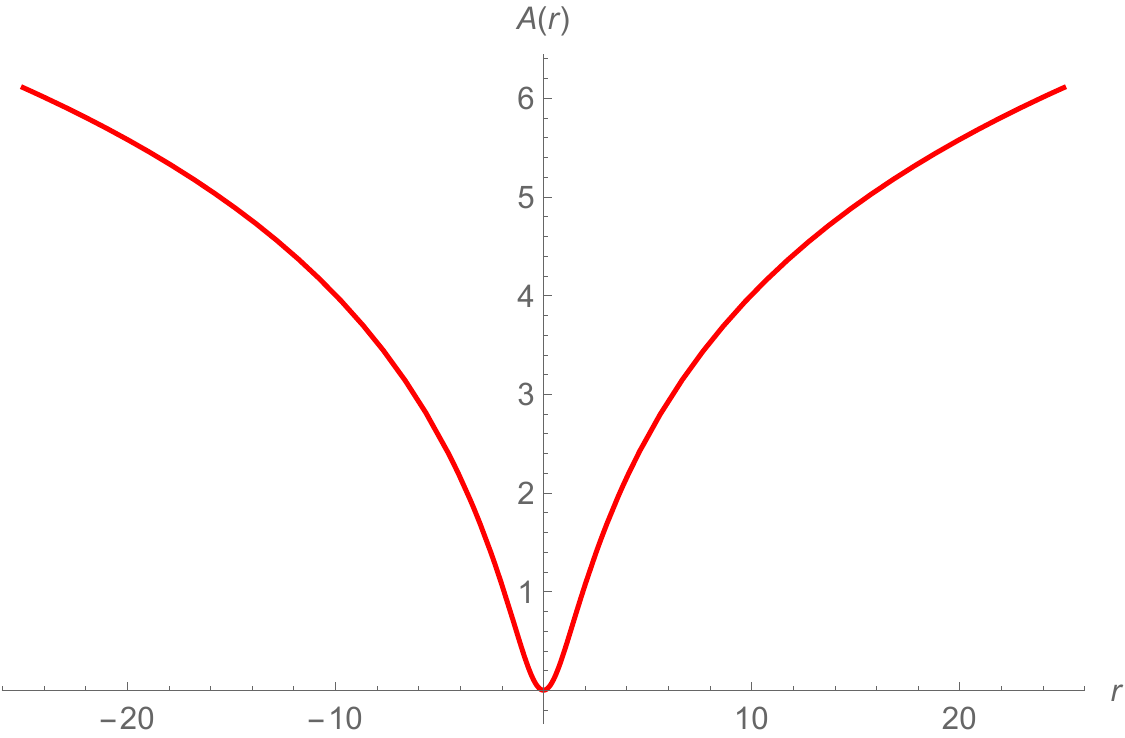}
                 \caption{Solution for $A(r)$}
         \end{subfigure} 
         \caption{An example of Janus solutions interpolating between $N=2$ non-conformal phases with $\ell=1$, $\kappa=-1$ and $g=2$.}\label{fig1}
 \end{figure}

In this section, we look for supersymmetric Janus solutions describing three-dimensional conformal interfaces within $N=2$ field theories in four dimensions. To preserve $SO(2,3)$ conformal symmetry in three dimensions, we take the metric ansatz to be an $AdS_4$-sliced domain wall
\begin{equation} 
ds^2=e^{2A(r)}ds^2_{AdS_4}+dr^2
\end{equation}
with $ds^2_{AdS_4}$ being the metric on $AdS_4$ with radius $\ell$. To find the relevant BPS equations for Janus solutions, we will closely follow the recent analysis in \cite{5D_N4_Janus}. We first note that the structure of $A^{ij}_1$ tensor given in \eqref{A1} is very similar to that of \cite{5D_N4_Janus}. In particular, there are two real and two complex eigenvalues.
\\
\indent As pointed out in \cite{5D_N4_Janus}, the real eigenvalues cannot lead to Janus solutions in the form of curved domain walls given above. Equivalently, the real eigenvalues can only support the flat domain walls describing holographic RG flows studied in the previous section. Accordingly, in this section, we will consider the complex eigenvalue $\alpha$ and take the Killing spinors of the unbroken supersymmetry to be $\epsilon_1$ and $\epsilon_3$. We also point out that as in \cite{5D_N4_Janus}, $\alpha$ does not give rise to a viable superpotential in terms of which the scalar potential can be written. 
\\
\indent We begin the anlysis of the BPS equations by considering the variations $\delta\chi_i$ which give 
\begin{equation}
\Sigma'\gamma_{\hat{r}}\epsilon_1=\mc{A}\epsilon_3\qquad \textrm{and}\qquad \Sigma'\gamma_{\hat{r}}\epsilon_3=\mc{A}^*\epsilon_1\label{Sigma_eq1}
\end{equation}
with
\begin{eqnarray}
\mc{A}&=&\frac{1}{3}\left[2g\sinh\phi_5(\cosh\phi_1\sinh\phi_3+i\sinh\phi_1-\cosh\phi_3)(i\cosh\phi_1\sinh\phi_3\right.\nonumber \\
& &\left.-\sinh\phi_1) +i(\cosh\phi_1-i\sinh\phi_1\sinh\phi_3)^2(g\cosh\phi_5+\sqrt{2}g_1\Sigma^3)\right].
\end{eqnarray}
Following \cite{5D_N4_Janus}, we find that the two equations in \eqref{Sigma_eq1} lead to the BPS equation for $\Sigma$ of the form
\begin{equation}
\Sigma'=\eta |\mc{A}|\label{Sigma_eq2}
\end{equation}  
and a projector
\begin{equation}
\gamma_{\hat{r}}\epsilon_1=\eta \frac{\mc{A}}{|\mc{A}|}\epsilon_3\qquad \textrm{and}\qquad \gamma_{\hat{r}}\epsilon_3=\eta \frac{\mc{A}^*}{|\mc{A}|}\epsilon_1\, .\label{gamma_r_proJanus}
\end{equation}
In these equations, we have introduced a sign factor $\eta=\pm 1$. 
\\
\indent From $\delta\lambda^a_i$, we find two sets of equations of the form
\begin{equation}
\phi'_5\gamma_{\hat{r}}\epsilon_3=\mc{B}^*\epsilon_1\qquad \textrm{and}\qquad \phi'_5\gamma_{\hat{r}}\epsilon_1=\mc{B}\epsilon_3\label{phi5_eq1}
\end{equation}
and 
\begin{equation}
(\phi'_3-i\cosh\phi_3\phi'_1)\gamma_{\hat{r}}\epsilon_1=\mc{C}^*\epsilon_3\qquad \textrm{and}\qquad (\phi'_3+i\cosh\phi_3\phi'_1)\gamma_{\hat{r}}\epsilon_3=\mc{C}\epsilon_1\, .\label{phi13_eq1}
\end{equation}
In these equations, the functions $\mc{B}$ and $\mc{C}$ are given by
\begin{eqnarray}
\mc{B}&=&g\Sigma^{-1}(\cosh\phi_3-\cosh\phi_1\sinh\phi_3-i\sinh\phi_1)\left[\sinh\phi_5(\sinh\phi_1-i\cosh\phi_3 \right.\nonumber \\
& &\left. -i\cosh\phi_1\sinh\phi_3)-2\cosh\phi_5(\sinh\phi_1-i\cosh\phi_1\sinh\phi_3)\right],\\
\mc{C}&=&-\frac{1}{4}\Sigma^{-1}\left[2g\cosh\phi_3\sinh2\phi_1(\cosh\phi_5-2\sinh\phi_5)+2gi\cosh\phi_5\sinh^2\phi_1\times \right.\nonumber \\
& &\times \sinh2\phi_3+4g(\sinh\phi_1\sinh\phi_3+i\cosh\phi_1\cosh2\phi_3-i\cosh^2\phi_1\sinh2\phi_3)\times \nonumber \\
& &\left.\times \sinh\phi_5 -\sqrt{2}g_1\Sigma^{3}(\cosh\phi_3\sinh2\phi_1+i\sinh^2\phi_1\sinh2\phi_3)\right].
\end{eqnarray}
Using the $\gamma_{\hat{r}}$ projection given in \eqref{gamma_r_proJanus}, we find that equation \eqref{phi5_eq1} leads to BPS equations for $\phi_5$
\begin{equation}
\phi'_5=\eta\frac{\mc{A}^*\mc{B}}{|\mc{A}|}=\eta\frac{\mc{A}\mc{B}^*}{|\mc{A}|}
\end{equation}
giving rise to an algebraic constraint for consistency of these two equations
\begin{equation}
\mc{A}^*\mc{B}=\mc{A}\mc{B}^*\, .
\end{equation}
The explicit form of this constraint is remarkably simple
\begin{equation}
\Sigma^3=-\frac{g}{\sqrt{2}g_1}\textrm{sech}\phi_5=\textrm{sech}\phi_5\label{constraint1}
\end{equation}
in which we have used $g=-\sqrt{2}g_1$ in the last equality.
\\
\indent Repeating the same procedure in \eqref{phi13_eq1}, we find additional two BPS equations for $\phi_1$ and $\phi_3$ of the form
\begin{equation}
\phi_3'=\frac{\eta}{|\mc{A}|}\textrm{Re}\, (\mc{C}\mc{A})\qquad \textrm{and}\qquad \cosh\phi_3\phi_1'=\frac{\eta}{|\mc{A}|}\textrm{Im}\, (\mc{C}\mc{A}).
\end{equation}
\indent We now consider the gravitino variations along $AdS_4$ directions with coordinates $x^\alpha$ for $\alpha=0,1,2,3$. The five-dimensional coordinates will be split as $x^\mu=(x^\alpha,r)$. As in \cite{Bobev_5D_Janus2}, using the Killing spinor equations for $AdS_4$ of the form
\begin{equation}
\widetilde{\nabla}_\alpha\epsilon_i=\frac{i}{2\ell}\kappa_i\gamma_{r}\gamma_{\alpha}\epsilon_i
\end{equation}
with $\kappa_i=\pm 1$, we find 
\begin{eqnarray}
\left(A'-\frac{i}{\ell}\kappa_1 e^{-A}\right)\gamma_{\hat{r}}\epsilon_1=\mc{W}\epsilon_3\qquad 
\textrm{and}\qquad & & \left(A'-\frac{i}{\ell}\kappa_3 e^{-A}\right)\gamma_{\hat{r}}\epsilon_3=\mc{W}^*\epsilon_1\quad\label{gravitino_eq1}
\end{eqnarray}
with 
\begin{eqnarray}
\mc{W}&=&-\frac{1}{6}\Sigma^{-1}\left[i(2g\cosh\phi_5-\sqrt{2}g_1\Sigma^3)(\cosh\phi_1\sinh\phi_3+\cosh\phi_3+i\sinh\phi_1)\right.\nonumber \\
& &\left.+4g\sinh\phi_5(\sinh\phi_1 -i\cosh\phi_1\sinh\phi_3)\right](\cosh\phi_1\sinh\phi_3\nonumber \\
& &-\cosh\phi_3+i\sinh\phi_1).
\end{eqnarray}
In obtaining the two equations in \eqref{gravitino_eq1}, we have rewritten the covariant derivative in terms of the covariant derivative $\widetilde{\nabla}_\alpha$ on $AdS_4$ according to the relation 
\begin{equation}
D_\alpha\epsilon_i=\widetilde{\nabla}_\alpha\epsilon_i-\frac{1}{2}A'\gamma_{r}\gamma_{\alpha}\epsilon_i\, .
\end{equation}
with the chirality matrix on $AdS_4$ given by $\gamma_r=i\gamma_{\hat{0}}\gamma_{\hat{1}}\gamma_{\hat{2}}\gamma_{\hat{3}}$.
Consistency between the two equations in \eqref{gravitino_eq1} implies $\kappa_3=-\kappa_1$. 
\\
\indent Using the $\gamma_{\hat{r}}$ projector given in \eqref{gamma_r_proJanus} and writing $\kappa=\kappa_1=-\kappa_3$, we find the BPS equation for $A$ and another algebraic constraint
\begin{equation}
A'=\eta \frac{\textrm{Re}(i\mc{W}\mc{A}^*)}{|\mc{A}|}\qquad \textrm{and}\qquad \frac{\kappa}{\ell}e^{-A}=-\eta\frac{\textrm{Im}(\mc{W}\mc{A}^*)}{|\mc{A}|}\, .\label{BPS_gravi}
\end{equation}
It can also be verified that the two algebraic constraints in \eqref{constraint1} and \eqref{BPS_gravi} are compatible with all the remaining BPS equations. Furthermore, all the BPS equations and these constraints also imply the second-ordered field equations. We also note that the two equations in \eqref{gravitino_eq1} also imply the relation
\begin{equation}
{A'}^2+\frac{1}{\ell^2}e^{-2A}=|\mc{W}|^2\, .
\end{equation}
Finally, the remaining condition $\delta\psi_{\hat{r}i}$ determines the radial dependence of the Killing spinors.
\\
\indent The algebraic constraint given in \eqref{BPS_gravi} takes the form
\begin{equation}
\frac{\kappa}{\ell}e^{-A}=\eta\frac{\sqrt{2}g g_1}{3|\mc{A}|}\cosh^3\phi_3\sinh\phi_1\sinh\phi_5\Sigma^2(\cosh\phi_1-\tanh\phi_3)^2\, .\label{constraint}
\end{equation}
We readily see that for either $\phi_1=0$ or $\phi_5=0$, the constraint implies that the $AdS_4$ radius $\ell\rightarrow \infty$ resulting in a flat domain wall. From this constraint, it might appear that a further simplification with $\phi_3=0$ could still give curved domain wall solutions. However, this is not compatible with the BPS equation for $\phi_3$ since $\phi_3'\neq 0$ for $\phi_3=0$ unless $\phi_5=0$. 
\\
\indent The above BPS equations can not be analytically solved. Therefore, we will look for numerical Janus solutions. Since the Killing spinors are given by only $\epsilon_1$ and $\epsilon_3$ subject to the projector \eqref{gamma_r_proJanus}, the solutions preserve only four supercharges. For regular Janus solutions, the solutions are asymptotically $AdS_5$ geometry on both sides of the interfaces. In particular, this implies that the metric function $A(r)$ has a turning point at a particular value of $r=r_0$ namely $A'(r_0)=0$. As $r\rightarrow \pm \infty$, the asymptotic behavior of $A(r)$ is given by $A\sim \frac{r}{L}$ with $L$ being the $AdS_5$ radius. In this case, $A(r)$ has a minimum at $r_0$. 
\\
\indent From the BPS equation for $\Sigma$ given in \eqref{Sigma_eq2}, we have
\begin{equation}    
\Sigma'=\eta\sqrt{\mc{A}_1^2+\mc{A}_2^2}
\end{equation}
with $\mc{A}_1$ and $\mc{A}_2$ being real and imaginary parts of $\mc{A}$. Follow the smoothness analysis in \cite{6D_Janus}, we need to smoothly sewn the two branches of the solution with $\eta=1$ and $\eta=-1$ at some value of $r=\tilde{r}_0$ in order to obtain regular Janus solutions. In particular, this requires $\Sigma'(\tilde{r}_0)=0$ or equivalently $\mc{A}_1=\mc{A}_2=0$ at $r=\tilde{r}_0$. This also implies that $\Sigma$ attains a minimum or a maximum at $r=\tilde{r}_0$. For convenience, we will also choose $\tilde{r}_0=r_0$ such that both $A(r)$ and $\Sigma(r)$ have a turning point at the same value of $r=r_0$. We will also require $\Sigma(r_0)$ to be a maximum at the turning point. 
\\
\indent For $\mc{A}_1=0$ condition, we have
\begin{eqnarray}
\sinh\phi_1\left[g\cosh\phi_1\sinh\phi_3(\cosh\phi_5-2\sinh\phi_5)+g\cosh\phi_3\sinh\phi_5\phantom{\sqrt{2}}\right.& &\nonumber \\
\left.+\sqrt{2}g_1\cosh\phi_1\sinh\phi_3
\Sigma^3\right]=0\, .& &\quad\label{con1}
\end{eqnarray}   
To satisfy this condition, the simplest possibility is to set 
\begin{equation}
\phi_1(r_0)=0\, .\label{boundary_con0}
\end{equation}
Using this result in $\mc{A}_2=0$ condition together with \eqref{constraint1}, we find
\begin{eqnarray}
\Sigma(r_0)^3&=&-\frac{g}{\sqrt{2}g_1}\textrm{sech}\phi_5(r_0)\nonumber \\ 
\textrm{and}\qquad \phi_3(r_0)&=&\frac{1}{2}\ln\left[\cosh\phi_5(r_0)(\cosh\phi_5(r_0)+\sinh\phi_5(r_0))\right].\label{boundary_con1}
\end{eqnarray}       
With these results, the second algebraic constraint given in \eqref{constraint} can be used to determine the value of $A(r_0)$. Therefore, we can determine the values of all the fields at the turning point in terms of a free parameter $\phi_5(r_0)$. It turns out that for any value of $\phi_5(r_0)$,
\begin{equation}
A'(r_0)=\Sigma'(r_0)=\phi'_1(r_0)=\phi'_3(r_0)=\phi'_5(r_0)=0\, .
\end{equation}
However, from the constraint \eqref{constraint1}, we find that the maximal value of $\Sigma$ is $1$ at $\phi_5=0$. All these results would imply that $\Sigma=1$ identically. This also leads to $\phi_5=0$ identically resulting in a flat domain wall solution.                     
\\
\indent Another possibility of setting the bracket in \eqref{con1} to zero leads to either
\begin{equation}
\phi_1(r_0)=\cosh^{-1}\tanh\phi_3(r_0)\qquad \textrm{and}\qquad \phi_5(r_0)=\frac{1}{2}\ln[\cosh2\phi_3(r_0)-2]
\end{equation}
or
\begin{equation}
\phi_5(r_0)=0\qquad \textrm{and}\qquad \phi_1(r_0)=\cosh^{-1}\left[\frac{1}{2}\coth\phi_3(r_0)\right].
\end{equation}
The fomer has no real solutions while the latter leads to $\Sigma(r)=1$ identically as in the previous case. Therefore, there do not seem to exist any supersymmetric regular Janus solutions interpolating between the supersymmetric $AdS_5$ vacuum on each side of the interface. It should be emphasized that this is a consequence of requiring both $\Sigma(r)$ and $A(r)$ to have a turning point at $r_0$. There could be regular Janus solutions with $A(r)$ attaining a turning point at a different value of $r\neq r_0$. In this case, the solution space is generically parametrized by two parameters rather than one independent parameter, $\phi_3(r_0)$ or $\phi_5(r_0)$ considered above. This is due to the fact that there are five BPS equations with two algebraic constraints given in \eqref{constraint1} and \eqref{constraint}. However, from a numerical search, we have not found solutions of this type.
\\
\indent The numerical analysis shows that there exist Janus solutions interpolating between non-conformal phases of $N=2$ SCFT dual to the $N=4$ $AdS_5$ vacuum. An example of these solutions is given in figure \ref{fig1}. To find this numerical solution, we have chosen the turning point $r_0=0$ and used \eqref{boundary_con0} and \eqref{boundary_con1} with $\phi_5(0)=0.1$.  Both sides of the interface correspond to a non-conformal phase of $N=2$ SCFT in four dimensions. We also note that, on both sides of the interface, we have $\phi_1=0$ implying the enhancement of supersymmetry to eight supercharges. This $N=2$ non-conformal phase of the $N=2$ SCFT should be identified with the one appearing in the RG flow solution shown in figure \ref{fig7}. Therefore, we expect the Janus solution in figure \ref{fig1} to describe conformal interfaces within an $N=2$ non-conformal field theory in four dimensions arising from a decompactification limit of a six-dimensional field theory on M5-branes wrapped on $H^2$ as mentioned in the previous section. On the field theory side, this conformal interface could be induced by position-dependent deformations or position-dependent vacuum expectation values of the dual operators. Similar solutions are also given in \cite{Minwoo_4DN8_Janus} and \cite{ISO7_Janus} in which supersymmetric Janus solutions in $ISO(7)$ maximal gauged supergravity in four dimensions have been found. In that case, the solutions are also attracted to the non-conformal phases rather than the conformal fixed points.

\begin{figure}
         \centering
         \begin{subfigure}[b]{0.32\textwidth}
                 \includegraphics[width=\textwidth]{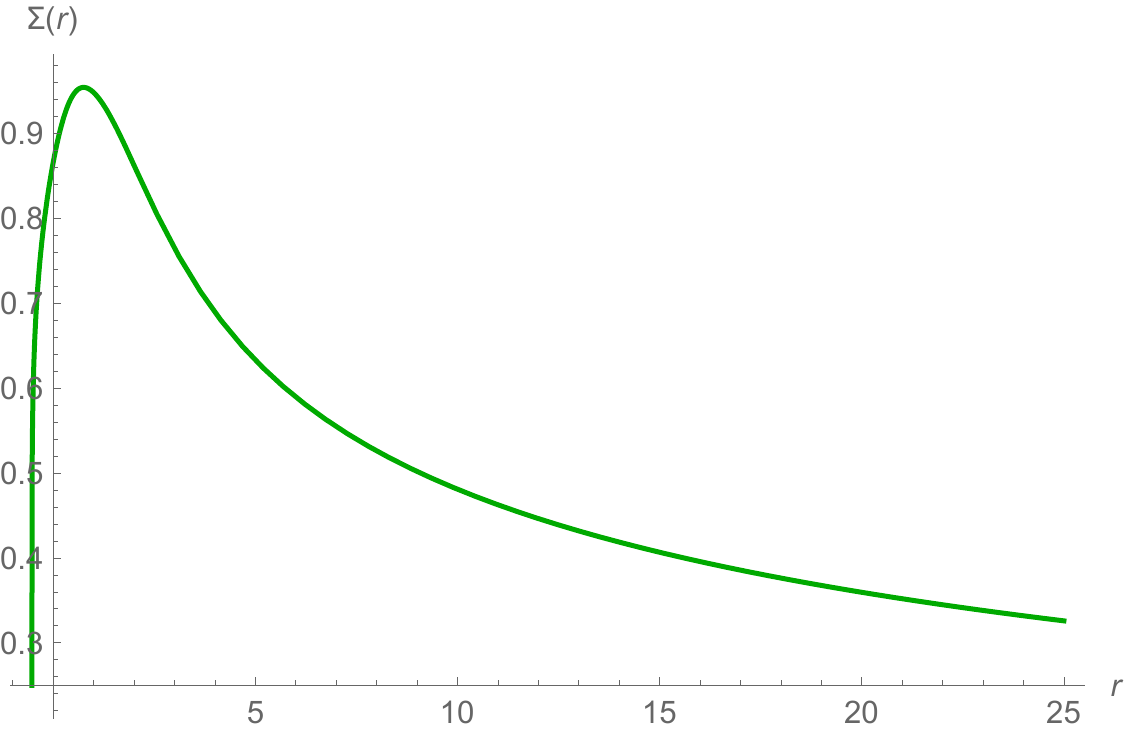}
                 \caption{Solution for $\Sigma(r)$}
         \end{subfigure}
\begin{subfigure}[b]{0.32\textwidth}
                 \includegraphics[width=\textwidth]{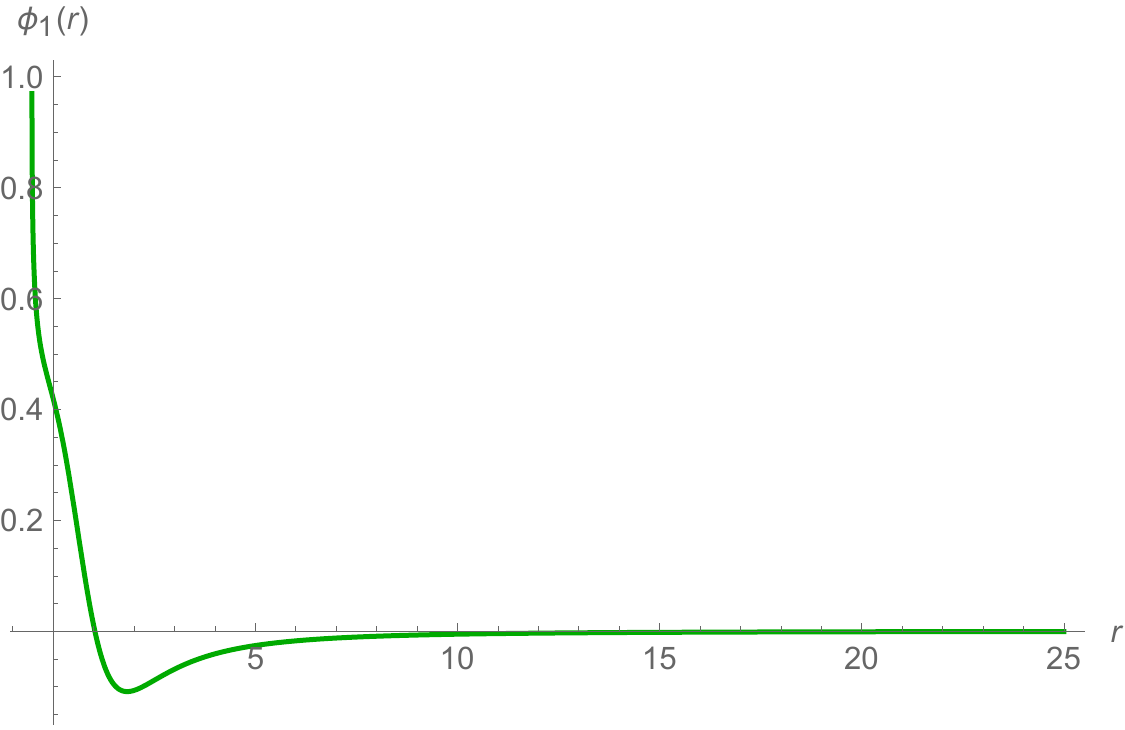}
                 \caption{Solution for $\phi_1(r)$}
         \end{subfigure}
         \begin{subfigure}[b]{0.32\textwidth}
                 \includegraphics[width=\textwidth]{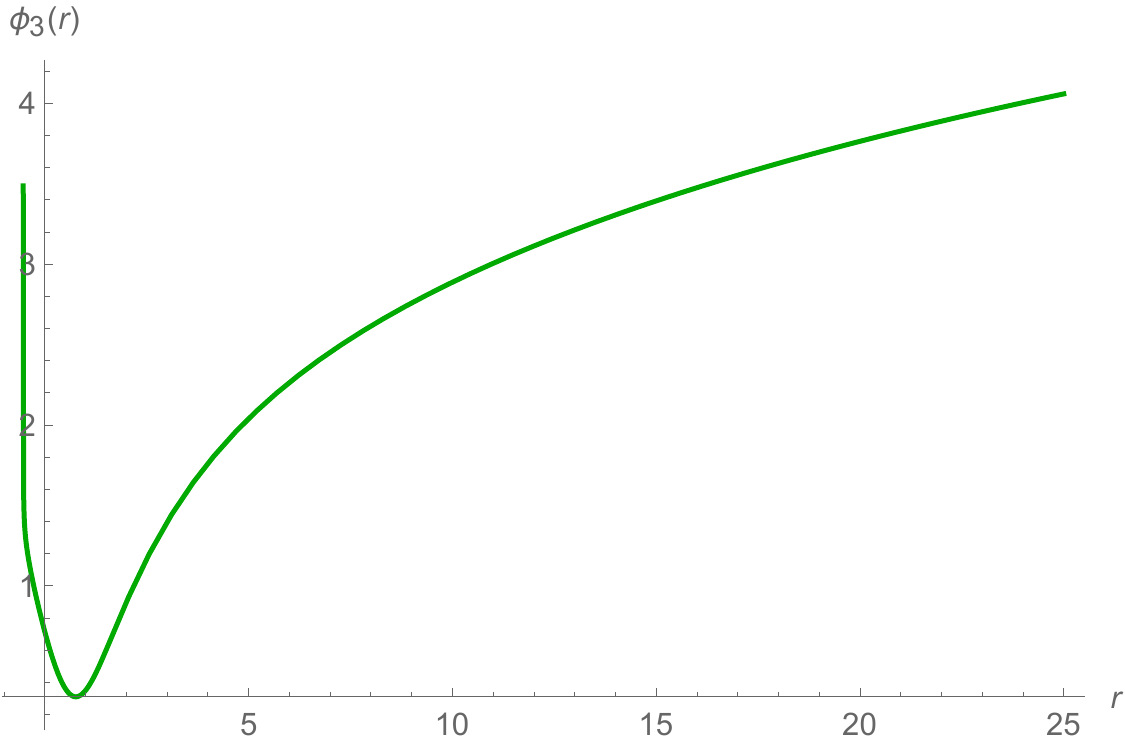}
                 \caption{Solution for $\phi_3(r)$}
         \end{subfigure}\\ 
         \begin{subfigure}[b]{0.35\textwidth}
                 \includegraphics[width=\textwidth]{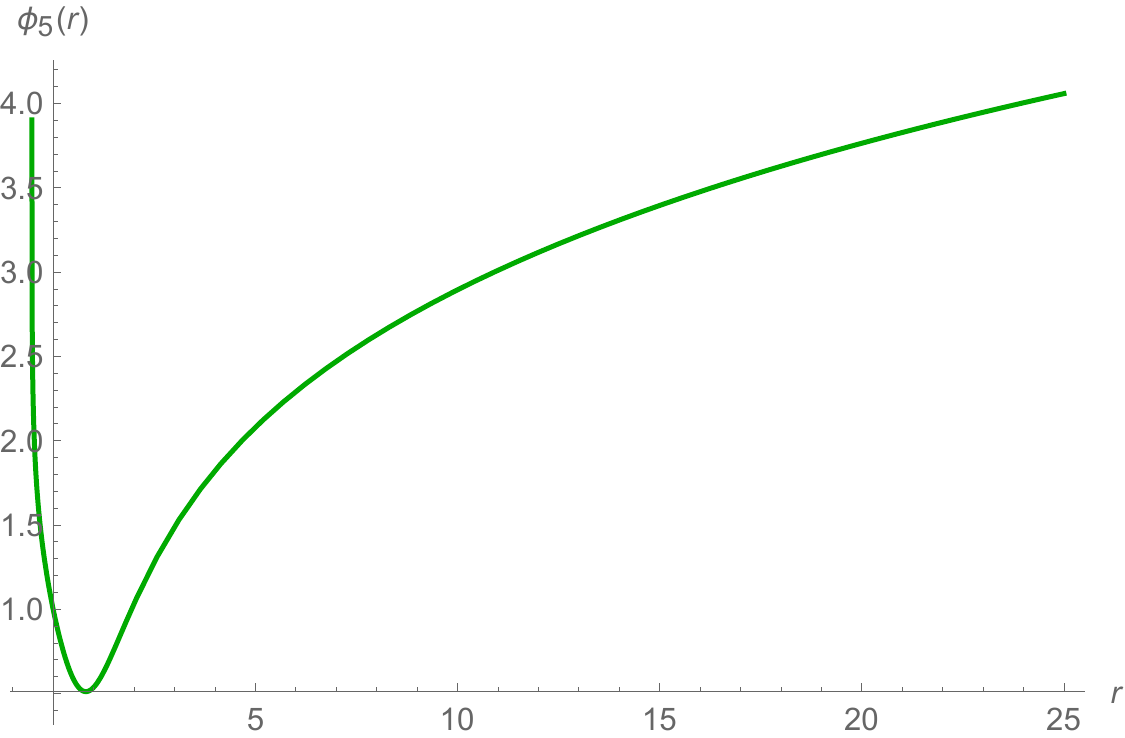}
                 \caption{Solution for $\phi_5(r)$}
         \end{subfigure}
         \begin{subfigure}[b]{0.35\textwidth}
                 \includegraphics[width=\textwidth]{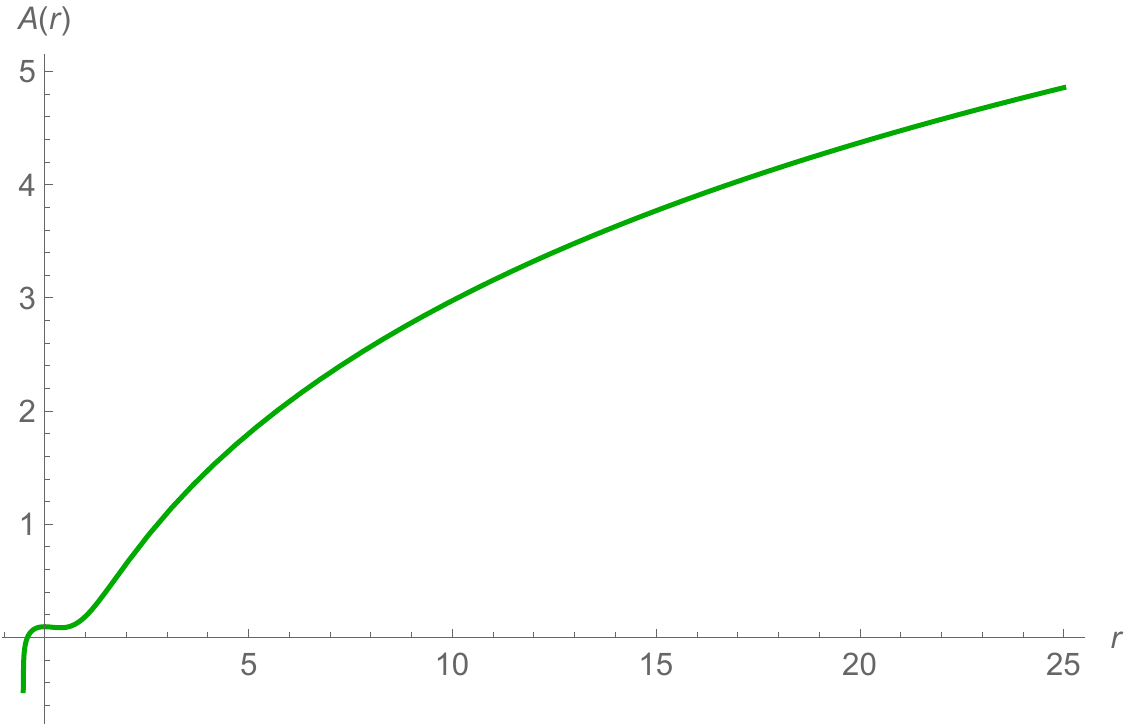}
                 \caption{Solution for $A(r)$}
         \end{subfigure} 
         \caption{An example of Janus solutions interpolating between an $N=2$ non-conformal phase and a singularity with $\ell=1$, $\kappa=-1$ and $g=2$.}\label{fig3}
 \end{figure}

\begin{figure}
         \centering
         \begin{subfigure}[b]{0.32\textwidth}
                 \includegraphics[width=\textwidth]{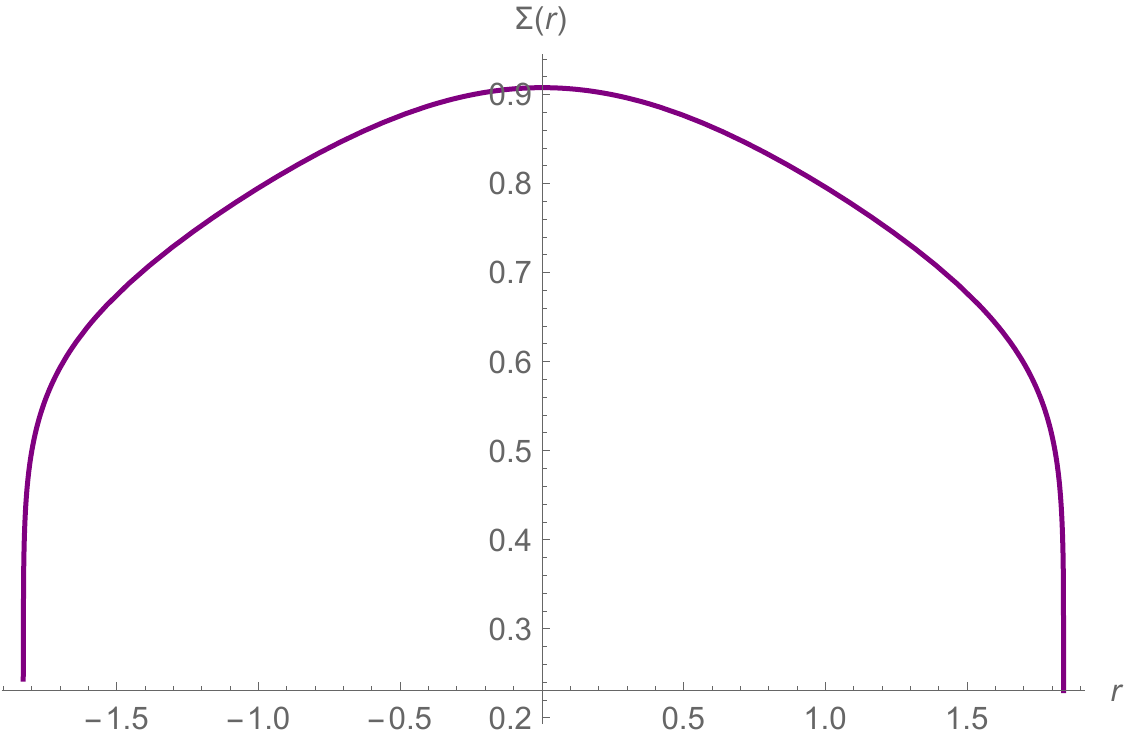}
                 \caption{Solution for $\Sigma(r)$}
         \end{subfigure}
\begin{subfigure}[b]{0.32\textwidth}
                 \includegraphics[width=\textwidth]{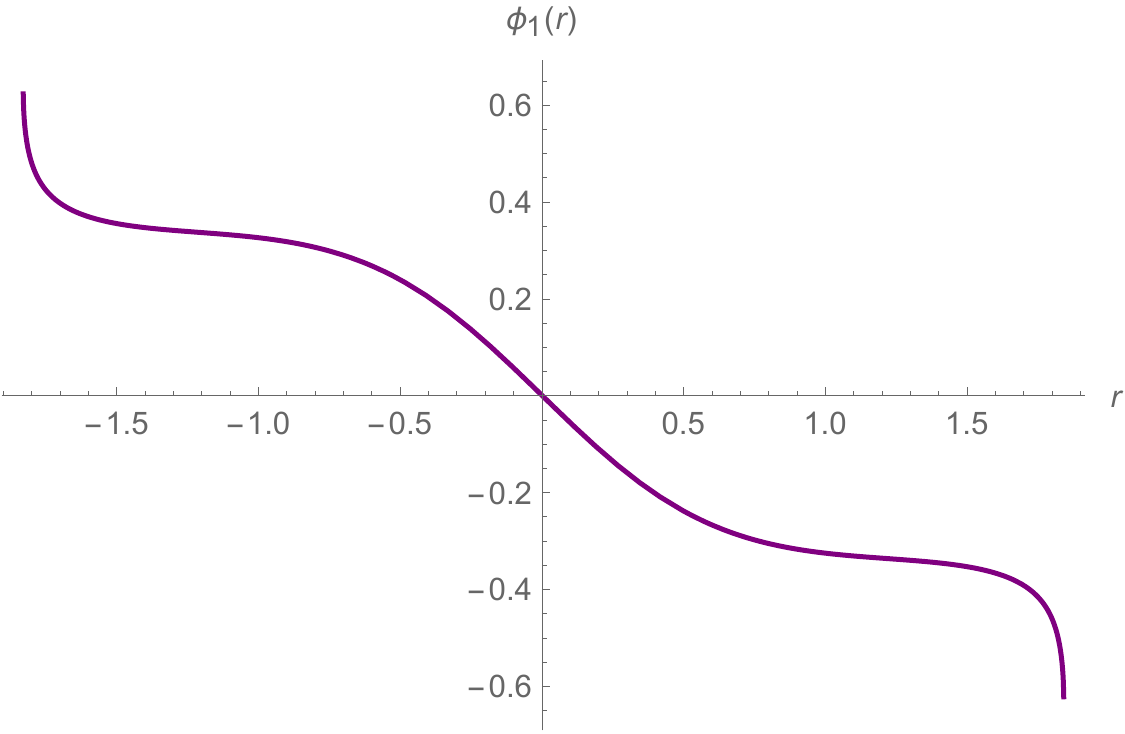}
                 \caption{Solution for $\phi_1(r)$}
         \end{subfigure}
         \begin{subfigure}[b]{0.32\textwidth}
                 \includegraphics[width=\textwidth]{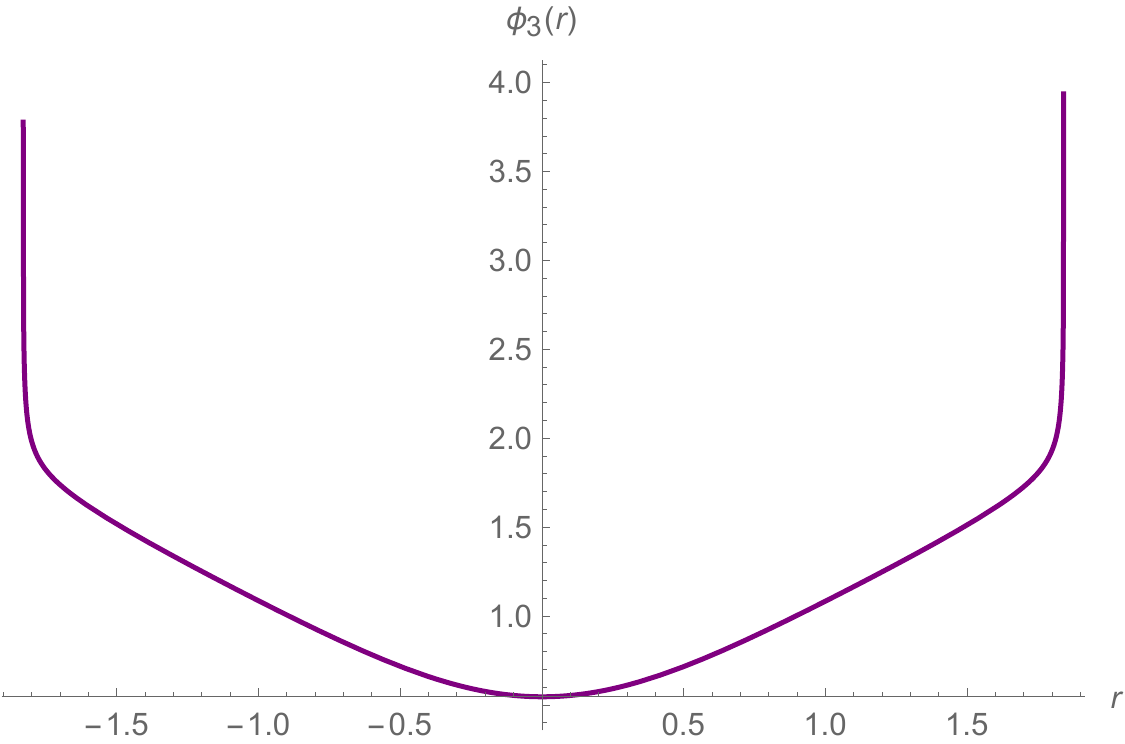}
                 \caption{Solution for $\phi_3(r)$}
         \end{subfigure}\\ 
         \begin{subfigure}[b]{0.35\textwidth}
                 \includegraphics[width=\textwidth]{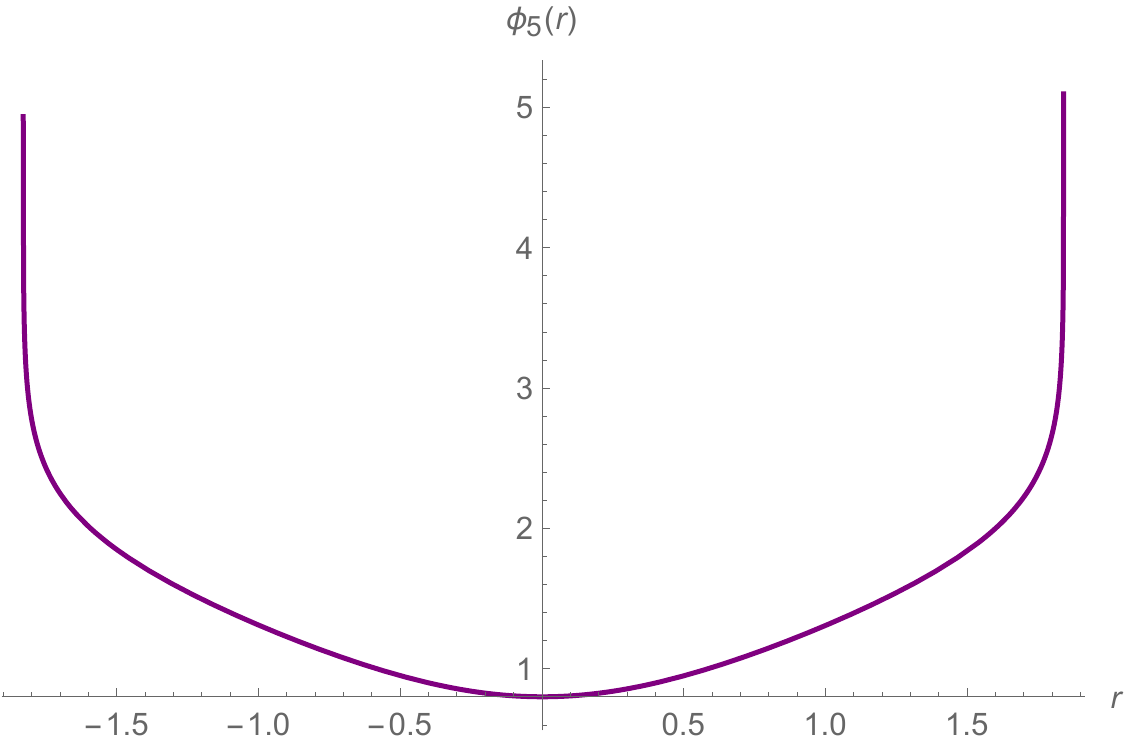}
                 \caption{Solution for $\phi_5(r)$}
         \end{subfigure}
         \begin{subfigure}[b]{0.35\textwidth}
                 \includegraphics[width=\textwidth]{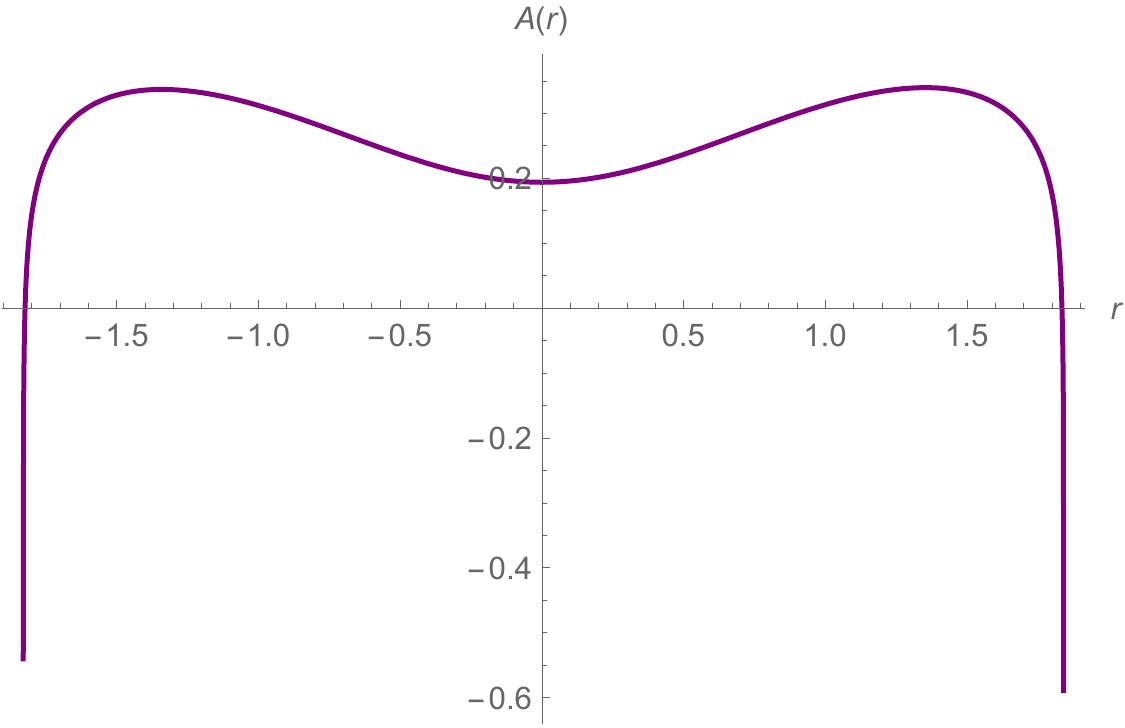}
                 \caption{Solution for $A(r)$}
         \end{subfigure} 
         \caption{An example of Janus solutions interpolating between singularities on both sides with $\ell=1$, $\kappa=-1$ and $g=2$.}\label{fig4}
 \end{figure}

\indent For larger values of $\phi_5(0)$, one side of the solutions becomes singular. An example of these solutions with $\phi_5(0)=1$ is shown in figure \ref{fig3}. This should describe a conformal boundary within the $N=2$ non-conformal field theory arising from M5-branes wrapped on $H^2$ as pointed out in \cite{BCFT_Gutperle}. Depending on the boundary conditions, there are also solutions that are singular on both sides of the interfaces. An example of these solutions is shown in figure \ref{fig4}. A similar solution has also been obtained in four-dimensional $N=4$ gauged supergravity arising from a truncation of eleven-dimensional supergravity on a tri-sasakian manifold \cite{tri-sasakian-flow}.

\section{Supersymmetric $AdS_5$ black strings}\label{string}
In this section, we consider solutions interpolating between the $N=4$ supersymmetric $AdS_5$ vacuum and an $AdS_3\times \Sigma$ geometry with $\Sigma$ being a Riemann surface. These solutions describe supersymmetric black strings in asymptotically $AdS_5$ space. Holographically, the solutions describe RG flows across dimensions from the $N=2$ SCFT in four dimensions to two-dimensional SCFTs in the IR. The latter arises from twisted compactifications of the former on $\Sigma$.
\\
\indent The ansatz for the metric is given by
\begin{equation}
ds^2=e^{2f(r)}dx^2_{1,1}+dr^2+e^{2h(r)}(d\theta^2+f^2_\kappa(\theta)d\phi^2)
\end{equation}
with 
\begin{equation}
f_\kappa(\theta)=\begin{cases}
  \sin\theta,  & \kappa=1\phantom{-}\quad \textrm{for}\quad \Sigma^2=S^2 \\
  \theta,  & \kappa=0\phantom{-}\quad \textrm{for}\quad \Sigma^2=T^2\\
  \sinh\theta,  & \kappa=-1\quad \textrm{for}\quad \Sigma^2=H^2
\end{cases}.\label{f_def}
\end{equation}
We will split the five-dimensional coordinates as $x^\mu=(x^\alpha,r,\theta,\phi)$ with $\alpha=0,1$. With an obvious choice of vielbein
\begin{equation}
e^{\hat{\alpha}}=e^fdx^\alpha,\qquad e^{\hat{r}}=dr,\qquad e^{\hat{\theta}}=e^hd\theta,\qquad e^{\hat{\phi}}=e^hf_\kappa(\theta)d\phi,
\end{equation}
non-vanishing components of the spin connection are given by
\begin{eqnarray}
{\omega^{\hat{a}}}_{\hat{r}}=f'e^{\hat{a}},\qquad {\omega^{\hat{\theta}}}_{\hat{r}}=h'e^{\hat{\theta}},\qquad {\omega^{\hat{\phi}}}_{\hat{r}}=h'e^{\hat{\phi}},\qquad {\omega^{\hat{\phi}}}_{\hat{\theta}}=\frac{f_\kappa'(\theta)}{f_\kappa(\theta)}e^{-h}e^{\hat{\phi}}\, .
\end{eqnarray} 
with $f_\kappa'(\theta)=\frac{df_\kappa(\theta)}{d\theta}$. 
\\
\indent In order to preseve some amount of supersymmetry, we will follow the standard procedure of performing a topological twist by turning on some gauge fields to cancel ${\omega^{\hat{\phi}}}_{\hat{\theta}}$ component of the spin connention. We will consider a twist achieved by turning on $SO(2)\times SO(2)$ gauge fields $A^0$ and $A^3$. There are four $SO(2)\times SO(2)$ singlet scalars consisting of the dilaton and other three scalars from $SO(5,3)/SO(5)\times SO(3)$. The coset representative for the latter can be obtained by setting $\phi_1=\phi_2=0$ in \eqref{SO2d_coset}. Relevant components of the composite connection are given by
\begin{equation}
{Q_i}^j=\frac{i}{2}\left[g_1 A^0\delta_i^k-gA^3{(\mathbf{I}_2\otimes \sigma_3)_i}^k\right]{(\sigma_2\otimes \sigma_3)_k}^j\, .\label{Q_string}
\end{equation}    
We then turn on $SO(2)\times SO(2)$ gauge fields of the form 
\begin{equation}
A^0=a_0f_\kappa'(\theta)d\phi\qquad \textrm{and}\qquad A^3=a_3f'_\kappa (\theta)d\phi\, .
\end{equation}
The corresponding field strength tensors are given by
\begin{equation}
F^0=dA^0=-\kappa a_0e^{-2h}e^{\hat{\theta}}\wedge e^{\hat{\phi}}\quad \textrm{and}\quad F^3=dA^3=-\kappa a_3e^{-2h}e^{\hat{\theta}}\wedge e^{\hat{\phi}}
\end{equation}
in which we have used the relation $f''_\kappa(\theta)=-\kappa f_\kappa(\theta)$. Finally, for solutions with $r$-dependent scalar fields, we need to impose the $\gamma_{\hat{r}}$ projector of the form
\begin{equation}
 \gamma_{\hat{r}}\epsilon_i=-{(\sigma_2\otimes \sigma_3)_i}^j\epsilon_j \label{gamma_r_string}
\end{equation} 
in which we have chosen a definite sign choice in order to make the $AdS_5$ vacuum appear in the limit $r\rightarrow \infty$.   

\subsection{$AdS_5$ black strings preserving four supercharges}
\begin{figure}
         \centering
               \begin{subfigure}[b]{0.32\textwidth}
                 \includegraphics[width=\textwidth]{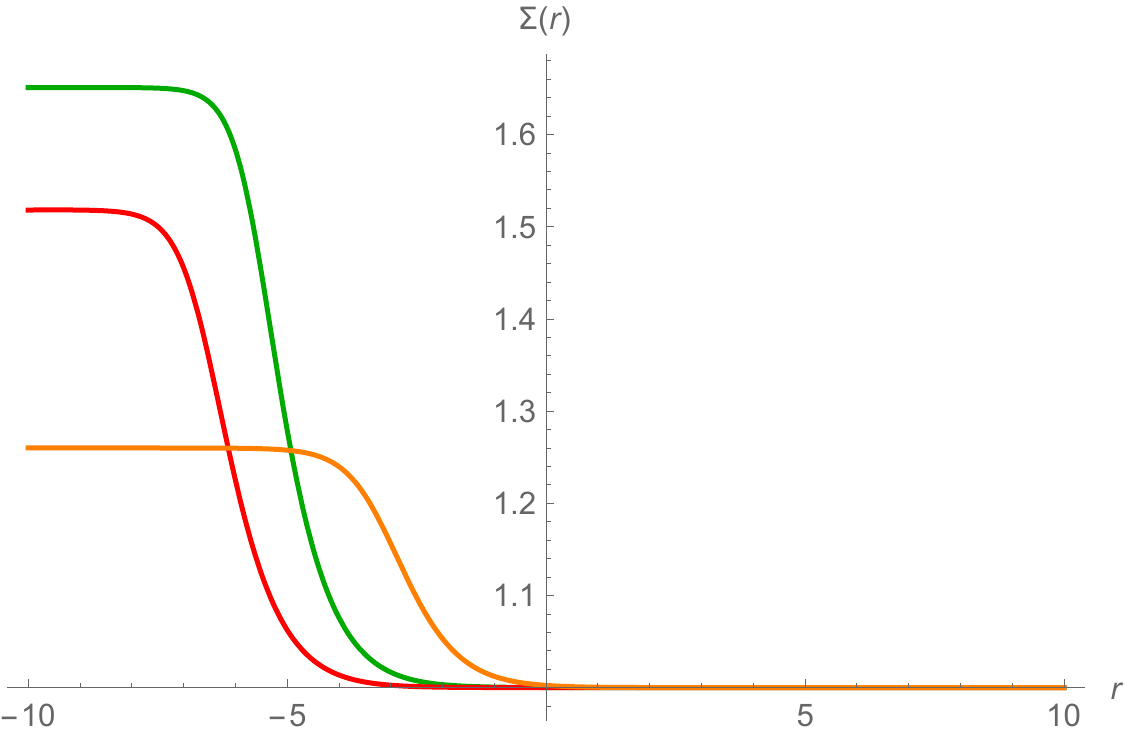}
                 \caption{Solutions for $\Sigma(r)$}
         \end{subfigure}
         \begin{subfigure}[b]{0.32\textwidth}
                 \includegraphics[width=\textwidth]{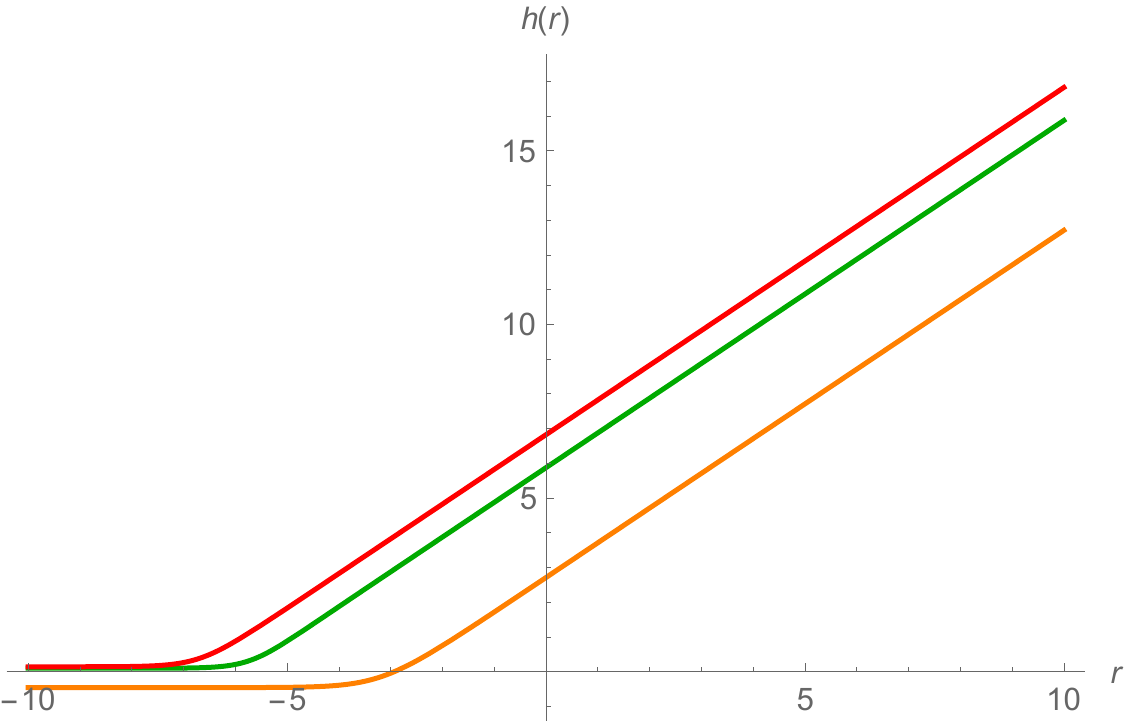}
                 \caption{Solutions for $h(r)$}
         \end{subfigure}
          \begin{subfigure}[b]{0.32\textwidth}
                 \includegraphics[width=\textwidth]{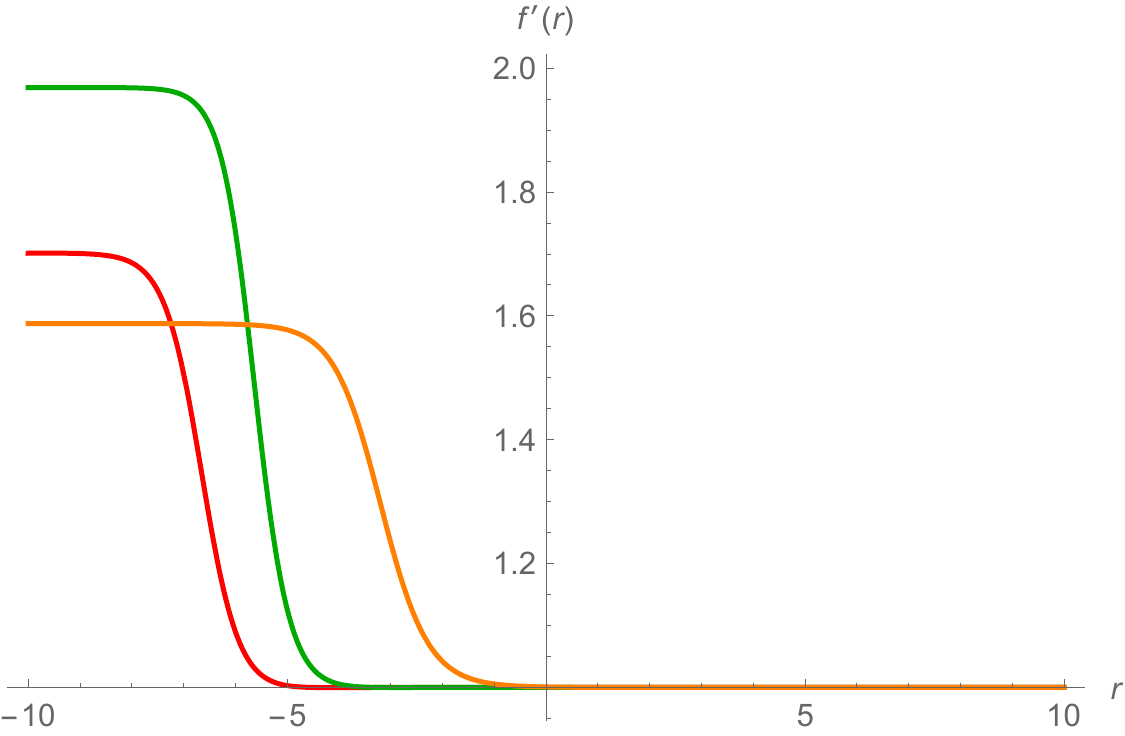}
                 \caption{Solutions for $f'(r)$}
         \end{subfigure}
         \caption{Examples of supersymmetric $AdS_5$ black string solutions. The orange line represents the solution interpolating between $N=4$ $AdS_5$ vacuum and $AdS_3\times H^2$ geometry preserving $8$ supercharges. The red (green) line corresponds to a solution interpolating between $AdS_5$ vacuum and $AdS_3\times H^2$ ($AdS_3\times S^2$) preserving $4$ supercharges.}\label{fig5}
 \end{figure}

We begin with supersymmetric $AdS_5$ black strings preserving four supercharges. These solutions can be obtained by performing a twist using an $SO(2)$ gauge field. From the composite connection given in \eqref{Q_string}, we find that
\begin{equation}
\delta \psi_{i\hat{\phi}}=\frac{1}{2}\frac{f_\kappa'(\theta)}{f_\kappa(\theta)}e^{-h}\gamma_{\hat{\phi}\hat{\theta}}\epsilon_i+\frac{i}{2}\left[g_1a_0-ga_3(\mathbf{I}_2\otimes \sigma_3)\right]\frac{f_\kappa'(\theta)}{f_\kappa(\theta)}e^{-h}{(\sigma_2\otimes \sigma_3)_i}^j\epsilon_j+\ldots\label{twist_string}
\end{equation}
with $\ldots$ denoting other terms in the variation of $\delta\psi_{i\hat{\phi}}$. The topological twist amounts to the cancellation between the two terms appearing in \eqref{twist_string}. There are two possibilities to achieve this by turning on only one $SO(2)$ gauge field.
\begin{itemize}
\item $A^0$-twist: \\
We can set $A^3=0$ and turn on $A^0$ to cancel the spin connection on $\Sigma$. This is achieved by imposing the following projector
\begin{equation}
\gamma_{\hat{\phi}\hat{\theta}}\epsilon_i=-i{(\sigma_2\otimes \sigma_3)_i}^j\epsilon_j
\end{equation}
together with a twist condition
\begin{equation}
g_1a_0=1\, .
\end{equation}
\item $A^3$-twist:\\
In this case, we set $A^0=0$ and imposing the projector
\begin{equation}
\gamma_{\hat{\phi}\hat{\theta}}\epsilon_i=i{(\sigma_2\otimes \mathbf{I}_2)_i}^j\epsilon_j\label{string_N8_pro}
\end{equation}
as well as a twist condition
\begin{equation}
ga_3=1\, .
\end{equation}
\end{itemize}
\indent The $A^0$-twist leads to the following BPS equations
\begin{eqnarray}
\phi_3'&=&g\Sigma^{-1}e^{-2\phi_3}\sinh\phi_5,\\
\phi_5'&=&-\frac{1}{2}\Sigma^{-1}ge^{-2\phi_3-\phi_5}(1-2e^{2\phi_3}+e^{2\phi_5}),\\
\Sigma'&=&\frac{1}{6}\left[ge^{-2\phi_3-\phi_5}(2e^{2\phi_3}+e^{2\phi_5}-1)+2\sqrt{2}g_1\Sigma^3-2\sqrt{2}\kappa a_0\Sigma^{-1}e^{-2h}\right],\\
h'&=&-\frac{1}{6}\Sigma^{-2}\left[\sqrt{2}g_1\Sigma^4-g\Sigma e^{-2\phi_3-\phi_5}(2e^{2\phi_3}+e^{2\phi_5}-1)+2\sqrt{2}\kappa a_0e^{-2h}\right],\quad \\
f'&=&-\frac{1}{6}\Sigma^{-2}\left[\sqrt{2}g_1\Sigma^4-g\Sigma e^{-2\phi_3-\phi_5}(2e^{2\phi_3}+e^{2\phi_5}-1)-\sqrt{2}\kappa a_0e^{-2h}\right].
\end{eqnarray}
We also note that compatibility between the BPS equations and the field equations requires $\phi_4=0$. In addition, it can be verified that the two-form fields can be consistently set to zero. We now look for $AdS_3\times \Sigma$ fixed point at which $\phi'_3=\phi_5'=\Sigma'=h'=0$ and $f'=\frac{1}{L_3}$ with $L_3$ being the $AdS_3$ radius. It turns out that the BPS equations do not admit any $AdS_3\times \Sigma$ fixed point solutions. The IR geometry is singular with
\begin{equation} 
\phi_3=\phi_5\sim0,\quad \Sigma\sim \frac{1}{\sqrt{r-r_0}},\quad h\sim \ln (r-r_0),\quad f\sim \frac{1}{4}\ln(r-r_0)
\end{equation}
for a finite value of $r=r_0$. Uplifting to eleven dimensions, we find that the metric component
\begin{equation}
\hat{g}_{00}\sim \textrm{constant},
\end{equation}
so the singularity is physically acceptable. We then expect the solution to describe an RG flow from four-dimensional $N=2$ SCFT to non-conformal field theory in two dimensions.
\\
\indent For the $A^3$-twist, a similar analysis leads to the following BPS equations
\begin{eqnarray}
\phi_3'&=&g\Sigma^{-1}e^{-2\phi_3}\sinh\phi_5,\\
\phi_5'&=&-\frac{1}{2}\Sigma^{-1}e^{-\phi_5}\left[ge^{-2\phi_3}(1-2e^{2\phi_3}+e^{2\phi_5})+\kappa a_3\Sigma^2e^{-2h}(e^{2\phi_5}-1)\right],\\
\Sigma'&=&\frac{1}{6}\left[ge^{-2\phi_3-\phi_5}(e^{2\phi_5}+2e^{2\phi_3}-1)+2\sqrt{2}g_1\Sigma^3-2\kappa a_3e^{-2h}\Sigma^2\cosh\phi_5\right],\\
h'&=&\frac{1}{6}\Sigma^{-1}\left[ge^{-2\phi_3-\phi_5}(e^{2\phi_5}+2e^{2\phi_3}-1)-\sqrt{2}g_1\Sigma^3+4\kappa a_3\Sigma^2e^{-2h}\cosh\phi_5\right],\qquad \\
f'&=&\frac{1}{6}\Sigma^{-1}\left[ge^{-2\phi_3-\phi_5}(e^{2\phi_5}+2e^{2\phi_3}-1)-\sqrt{2}g_1\Sigma^3-2\kappa a_3\Sigma^2e^{-2h}\cosh\phi_5\right].\qquad\,\,
\end{eqnarray}
These equations admit one supersymmetric $AdS_3\times \Sigma$ fixed point given by
\begin{eqnarray}
& &\phi_3=\phi_5=0,\qquad \Sigma=-\left(\frac{\sqrt{2}g}{g_1}\right)^{\frac{1}{3}},\nonumber \\
& &h=\frac{1}{2}\ln\left[-\kappa a_3\left(\frac{2}{gg_1^2}\right)^{\frac{1}{3}}\right],\qquad L_{3}=-\left(\frac{\sqrt{2}}{g_1g^2}\right)^{\frac{1}{3}}\, .
\end{eqnarray}
This solution gives a real warp factor $h$ only for $\kappa=-1$, so in this case, there is only an $AdS_3\times H^2$ fixed point. 
\\
\indent We also note that the fixed point preserves eight supercharges due to the projector \eqref{string_N8_pro}. Recall that the supersymmetry parameters $\epsilon_i$ transforming under $SO(1,3)\times SO(5)_R$ as $(\mathbf{4},\mathbf{4})$. Following the analysis in \cite{MN_nogo}, we decompose this representation under the subgroup $SO(1,1)\times SO(2)_\Sigma\times SO(2)\times SO(2)_R$ in which $SO(2)\times SO(2)_R\subset SO(2)\times SO(3)_R\subset SO(5)_R$ and $SO(1,1)\times SO(2)_\Sigma\subset SO(1,3)$. The $SO(2)_R\subset SO(3)_R\sim SO(3)\subset ISO(3)$ corresponds to the $A^3$ gauge field that participates in the twist. Since the twist in performed by identifying $SO(2)_\Sigma$ with $SO(2)_R$, the unbroken supersymmetry corresponds to the twisted Killing spinors in the representations with opposite charges under $SO(2)_\Sigma$ and $SO(2)_R$; $(+,\pm,+\mp)$, $(+,\pm,-\mp)$, $(-,\pm,+\mp)$ and $(-,\pm,-\mp)$. This leads to $N=(2,2)$ superconformal symmetry in two dimensions. However, due to an extra $\gamma_{\hat{r}}$-projector given in \eqref{gamma_r_string}, the flow solutions interpolating between the $AdS_5$ vacuum and this $AdS_3\times H^2$ geometry preserve only four supercharges corresponding to $N=(2,2)$ Poincare supersymmetry in two dimensions. This can also be explicitly seen as follows. Recall that in five dimensions, $i\gamma^{\hat{0}}\gamma^{\hat{1}}\gamma^{\hat{r}}\gamma^{\hat{\theta}}\gamma^{\hat{\phi}}=\pm\mathbf{I}_4$, the two projectors given in \eqref{gamma_r_string} and \eqref{string_N8_pro} imply that
\begin{equation}
\gamma^{\hat{0}}\gamma^{\hat{1}}\epsilon_i=\pm{(\mathbf{I}_2\otimes \sigma_3)_i}^j\epsilon_j\, .
\end{equation}  
This indicates that half of the unbroken supersymmetry has opposite two-dimensional chirality to the other half leading to $N=(2,2)$ supersymmetry.    
\\
\indent It should be noted that since all scalars from vector multiplets vanish, this $AdS_3\times H^2$ fixed point is the same as that found in pure $N=4$ gauged supergravity with $U(1)\times SU(2)$ gauge group \cite{Romans_5DN4}. The uplifted eleven-dimensional solution has been discussed in \cite{Gauntlett_pure_5DN4_from_11D}. An example of numerical solutions for these interpolating solutions is shown by the orange line in figure \ref{fig5}. In this solution, we have set $\phi_3=\phi_5=0$ and $g=2$ corresponding to a unit $AdS_5$ radius. The solution describes a black string in asymptotically $AdS_5$ space with a near horizon geometry given by $AdS_3\times H^2$. Upon uplifted to eleven dimensions, this leads to a supersymmetric $AdS_3\times H^2\times H^2\times S^4$ geometry preserving eight supercharges \cite{Gauntlett_pure_5DN4_from_11D}. Holographically, this solution describes an RG flow from $N=2$ SCFT in four dimensions to $N=(2,2)$ two-dimensional SCFT in the IR. 
\\
\indent We can also compute the central charge of the dual two-dimensional $N=(2,2)$ SCFT in the IR by the standard formula
\begin{equation}
c=\frac{3L_3}{2G_N^{(3)}}\, .
\end{equation}  
The Newton's constant in three dimensions is related to that in five dimensions by
\begin{equation} 
\frac{1}{G_N^{(3)}}=\frac{e^{2h_0}\textrm{vol}(\Sigma)}{G_N^{(5)}}
\end{equation}
with $h_0$ being the value of $h(r)$ at the $AdS_3\times \Sigma$ fixed point. $G_N^{(5)}$ is in turn obtained by a truncation of eleven-dimensional supergravity on $\tilde{H}^2\times S^4$ as
\begin{equation}
\frac{1}{G_N^{(5)}}=\frac{\Sigma_0^{-\frac{3}{5}}\textrm{vol}(\tilde{H}^2)\textrm{vol}(S^4)R_{S^4}^4}{G_N^{(11)}}
\end{equation}
in which $G_N^{(11)}=16\pi^7\ell_p^9$ with $\ell_p$ being eleven-dimensional Plank's length. We also recall that the $S^4$ truncation of eleven-dimensional supergravity leads to an $AdS_7\times S^4$ geometry with
\begin{equation}
L_7^2=4(\pi N)^{\frac{2}{3}}\ell_p^3=\frac{4}{m^2}\qquad \textrm{and}\qquad R_{S^4}=\frac{1}{m}
\end{equation}
with $m$ being the gauge coupling constant in seven-dimensional gauged supergravity. 
\\
\indent Following \cite{MN_nogo}, we will work with a unit such that the $AdS_7$ radius $L_7=1$ or $m=2$. This leads to 
\begin{equation}
G_N^{(11)}=\frac{\pi^5}{4N^2}\qquad \textrm{and}\qquad R_{S^4}=\frac{1}{2}\, .
\end{equation}
With all these, we eventually find $G_N^{(5)}$ of the form
\begin{eqnarray}
\frac{1}{G_N^{(5)}}&=&\frac{\Sigma_0^{-\frac{3}{5}}\textrm{vol}(\tilde{H}^2)\textrm{vol}(S^4)N^2}{4\pi^5}\nonumber \\
&=&\frac{\Sigma_0^{-\frac{3}{5}}|\tilde{g}-1|N^2}{2\pi^2}\, .\label{GN_5}
\end{eqnarray} 
In the second line, we have used the volume of a unit $S^4$, $\textrm{vol}(S^4)=\frac{\pi^2}{2}$ and the volume of a genus $g\neq 1$ Riemann surface
\begin{equation}
\textrm{vol}(\Sigma)=4\pi|g-1|\, .
\end{equation}
Finally, we can determine the central charge of the dual two-dimensional SCFT 
\begin{equation}
c=\frac{3}{\pi}L_3N^2\Sigma_0^{-\frac{3}{5}}e^{2h_0}|\tilde{g}-1||\hat{g}-1|\label{2D_cen}
\end{equation}
with $\tilde{g}$ and $\hat{g}$ denoting the genera of $\tilde{H}^2$ and $\Sigma$, respectively. For the $AdS_3\times H^2$ fixed point given above, we find
\begin{equation}
c=\frac{3(2^{\frac{4}{5}})N^2a_3|\tilde{g}-1||\hat{g}-1|}{\pi g^2}\, .
\end{equation}

\begin{figure}
         \centering
         \begin{subfigure}[b]{0.32\textwidth}
                 \includegraphics[width=\textwidth]{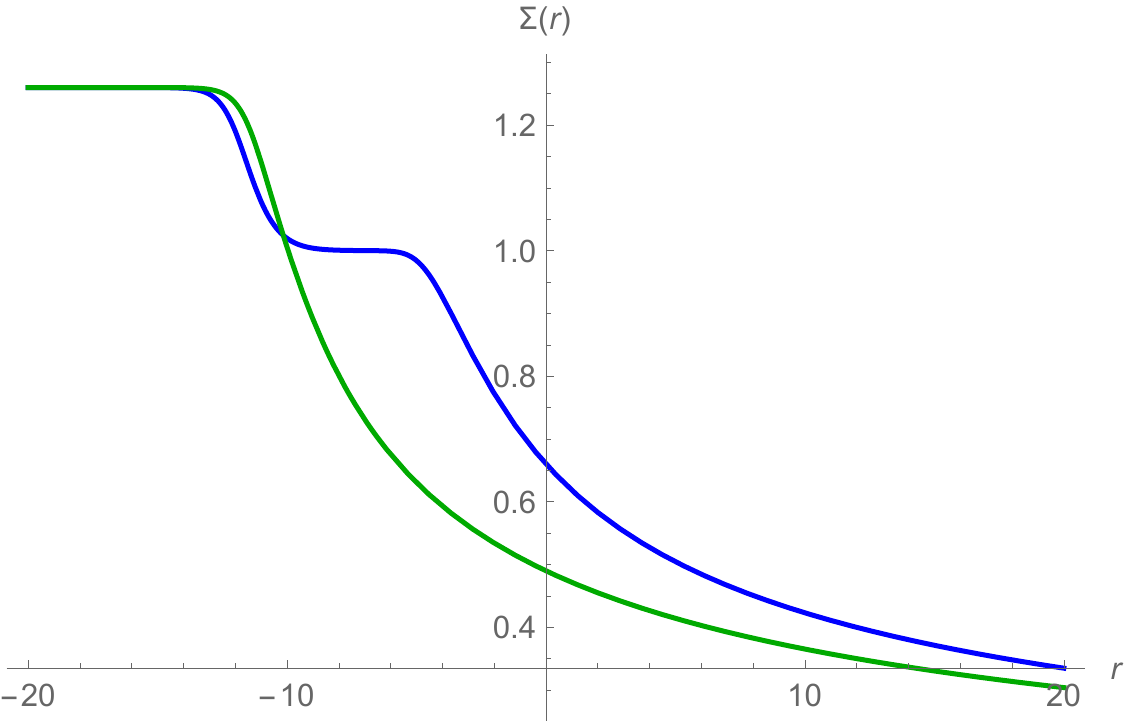}
                 \caption{Solutions for $\Sigma(r)$}
         \end{subfigure}
         \begin{subfigure}[b]{0.32\textwidth}
                 \includegraphics[width=\textwidth]{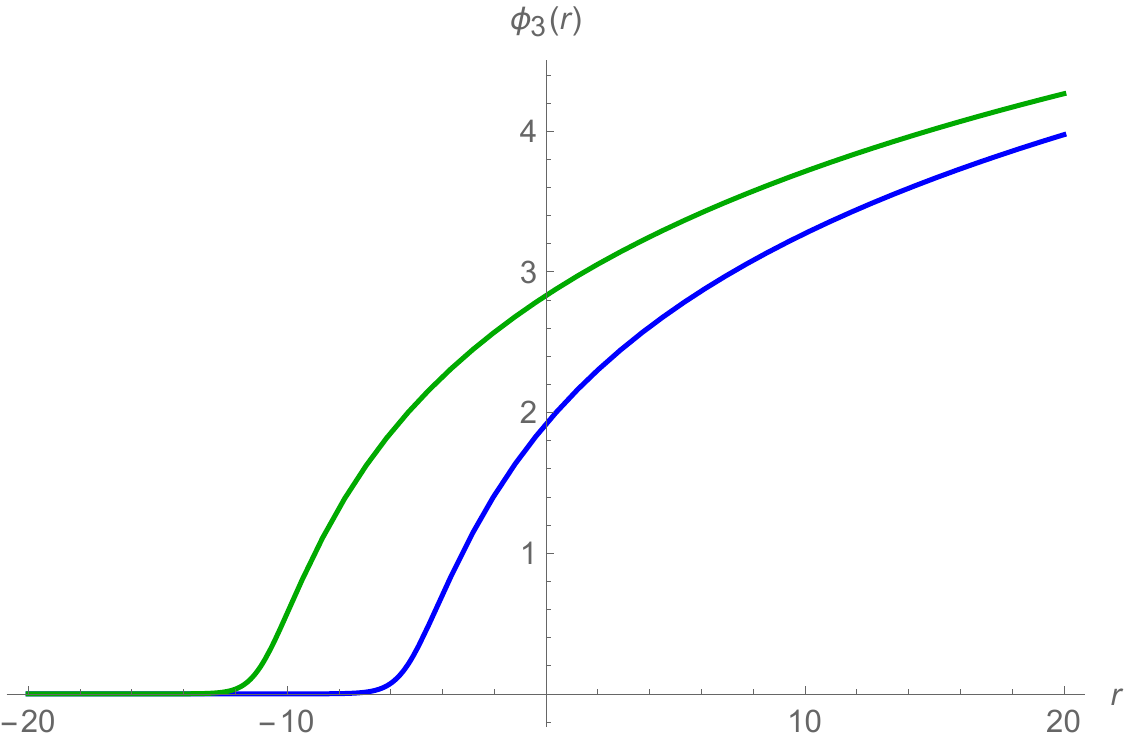}
                 \caption{Solutions for $\phi_3(r)$}
         \end{subfigure}
          \begin{subfigure}[b]{0.32\textwidth}
                 \includegraphics[width=\textwidth]{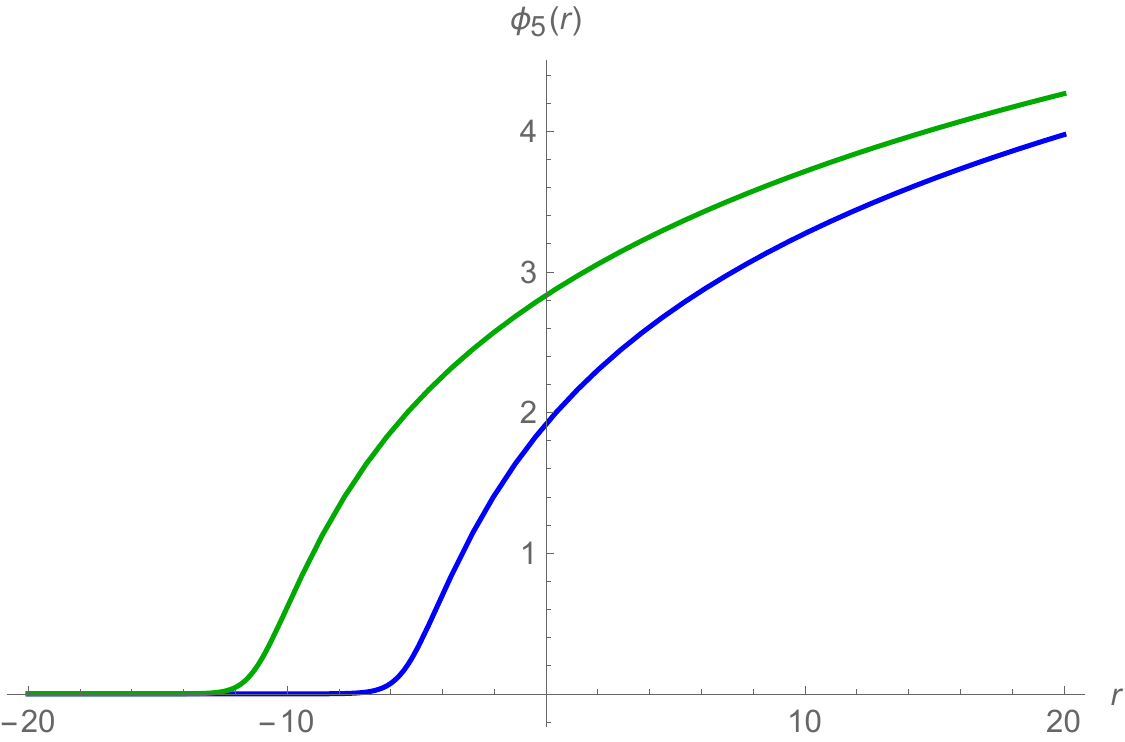}
                 \caption{Solutions for $\phi_5(r)$}
         \end{subfigure}\\
               \begin{subfigure}[b]{0.32\textwidth}
                 \includegraphics[width=\textwidth]{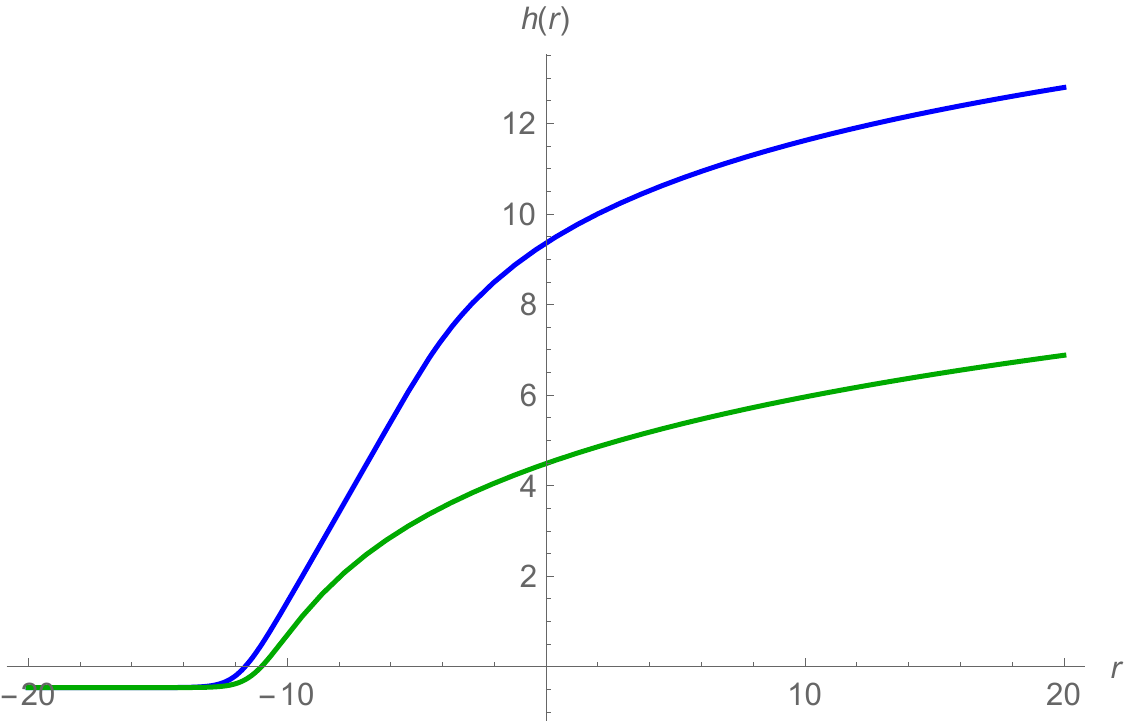}
                 \caption{Solutions for $h(r)$}
         \end{subfigure}
         \begin{subfigure}[b]{0.32\textwidth}
                 \includegraphics[width=\textwidth]{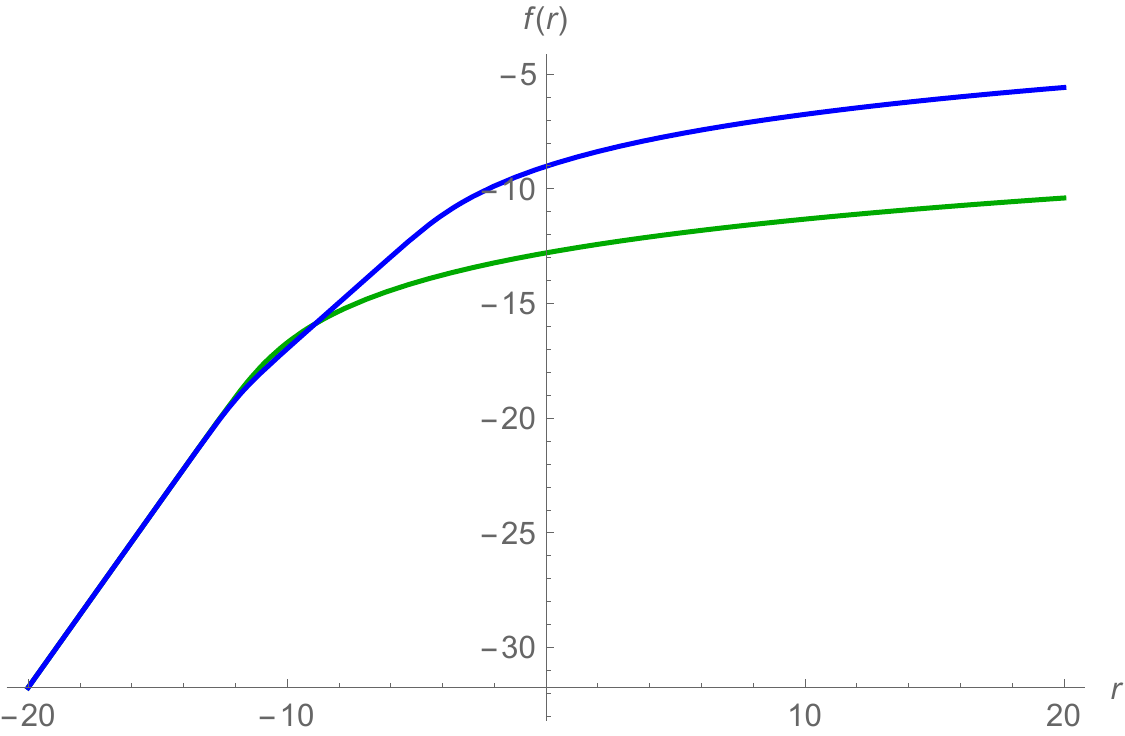}
                 \caption{Solutions for $f(r)$}
         \end{subfigure}
          \begin{subfigure}[b]{0.32\textwidth}
                 \includegraphics[width=\textwidth]{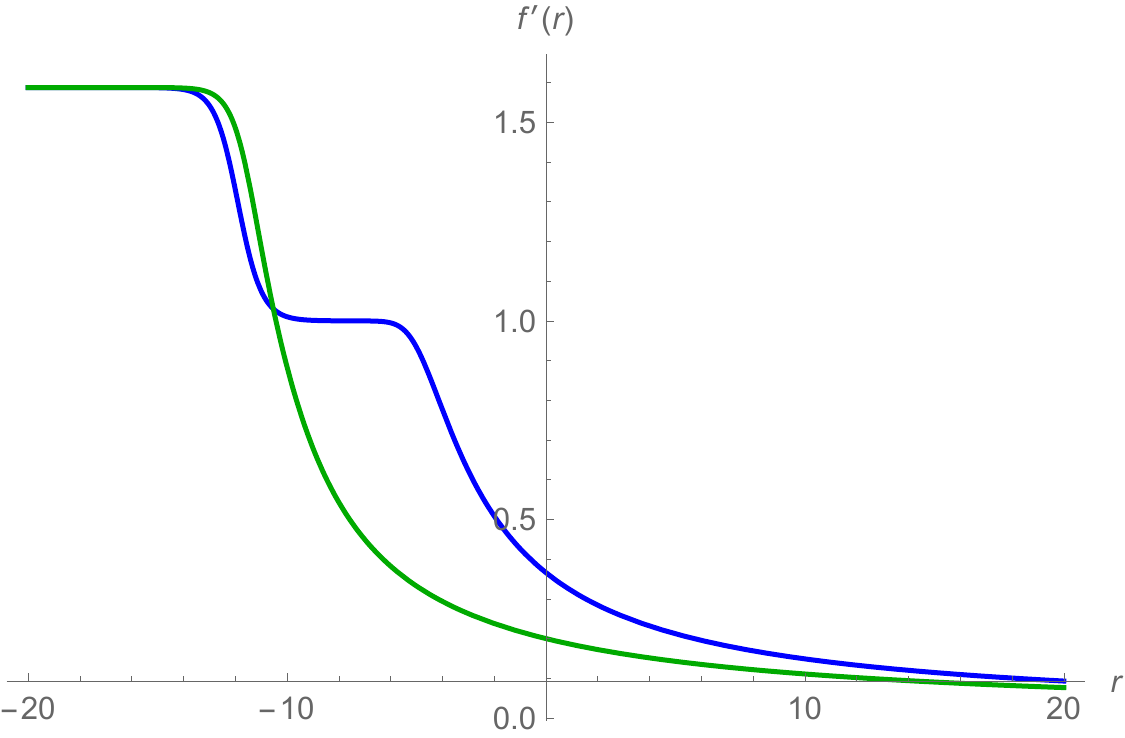}
                 \caption{Solutions for $f'(r)$}
         \end{subfigure}
         \caption{Examples of RG flows from an $N=2$ non-conformal field theory to $N=(2,2)$ SCFT in two dimensions (green) and to four-dimensional $N=2$ SCFT and $N=(2,2)$ SCFT in two dimensions (blue) with $g=2$.}\label{fig51}
 \end{figure}

By turning on scalar fields from vector multiplets, we obtain black string solutions that are not solutions of pure $N=4$ gauged supergravity although the $AdS_3\times H^2$ fixed point is the solution of both pure and matter-coupled $N=4$ gauged supergravities. The numerical analysis shows that there exist solutions describing holographic RG flows from a domain wall, dual to an $N=2$ non-conformal phase of the $N=2$ SCFT, to the $N=(2,2)$ SCFT in two dimensions. An example of these solutions is shown by the green line in figure \ref{fig51}. By tuning the boundary condition very close to the $AdS_3\times H^2$ fixed point, we also find solutions interpolating between domain wall and $AdS_3\times H^2$ fixed point with an intermediate $AdS_5$ geometry. These solutions would holographically describe RG flows from $N=2$ four-dimensional field theory obtained from M5-branes wrapped on $H^2$ to $N=2$ SCFT in four dimensions and then to $N=(2,2)$ SCFT in two dimensions. An example of these solutions is shown by the blue line in figure \ref{fig51}. Both types of these solutions shown in figure \ref{fig51} also describe black strings in asymptotically domain wall space-time.  
 
\subsection{$AdS_5$ black strings preserving two supercharges} 
\begin{figure}
         \centering
         \begin{subfigure}[b]{0.32\textwidth}
                 \includegraphics[width=\textwidth]{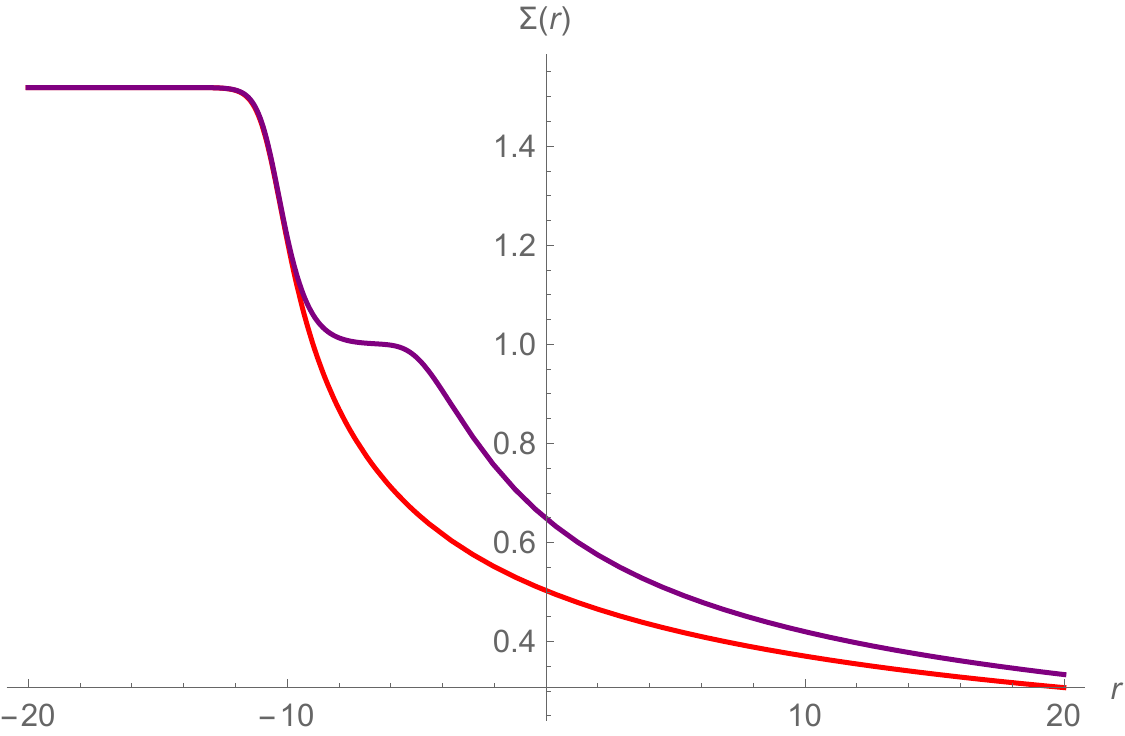}
                 \caption{Solutions for $\Sigma(r)$}
         \end{subfigure}
         \begin{subfigure}[b]{0.32\textwidth}
                 \includegraphics[width=\textwidth]{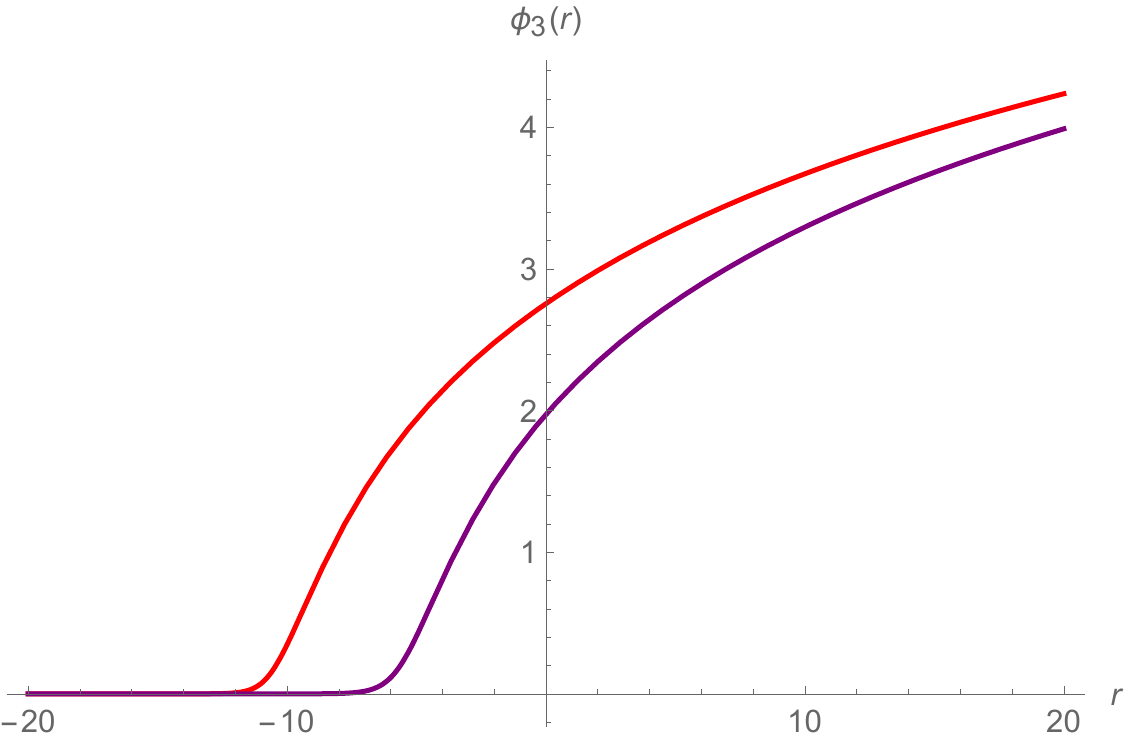}
                 \caption{Solutions for $\phi_3(r)$}
         \end{subfigure}
          \begin{subfigure}[b]{0.32\textwidth}
                 \includegraphics[width=\textwidth]{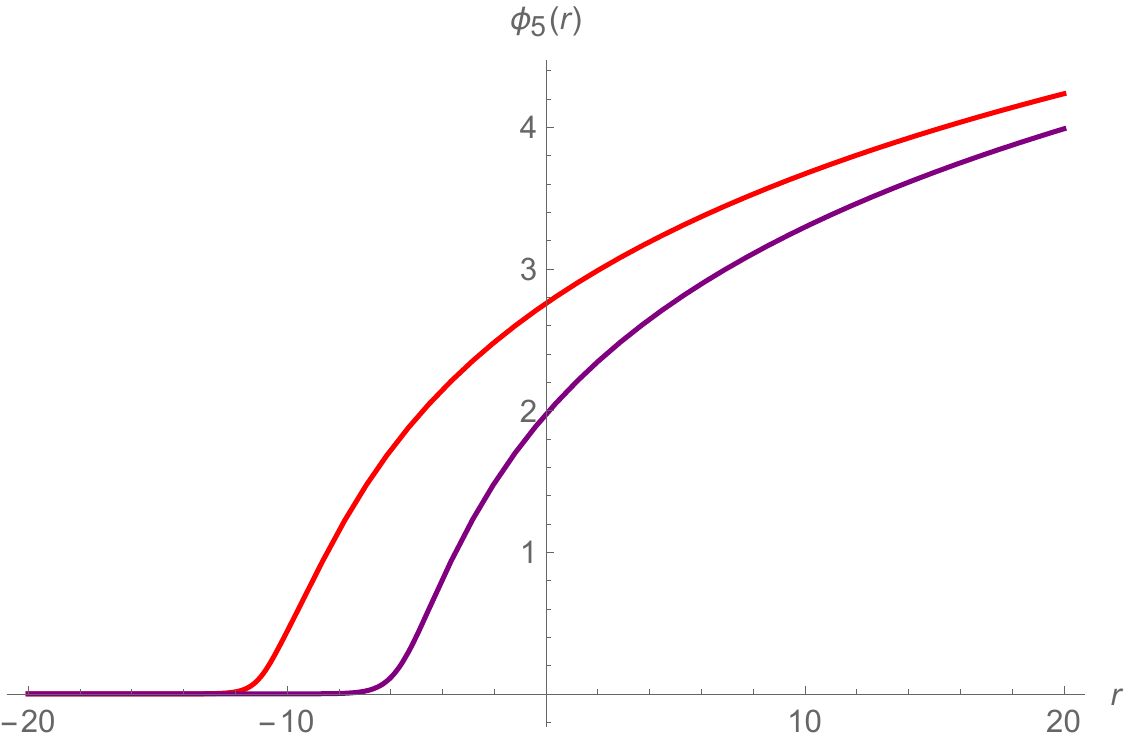}
                 \caption{Solutions for $\phi_5(r)$}
         \end{subfigure}\\
               \begin{subfigure}[b]{0.32\textwidth}
                 \includegraphics[width=\textwidth]{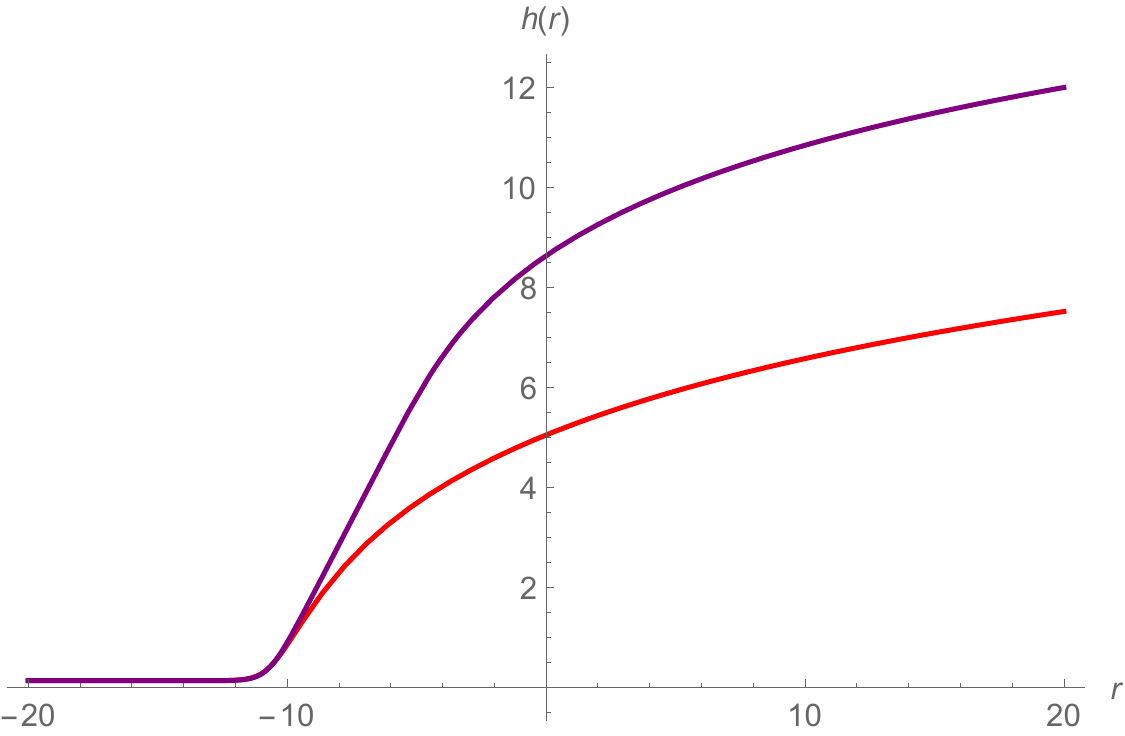}
                 \caption{Solutions for $h(r)$}
         \end{subfigure}
         \begin{subfigure}[b]{0.32\textwidth}
                 \includegraphics[width=\textwidth]{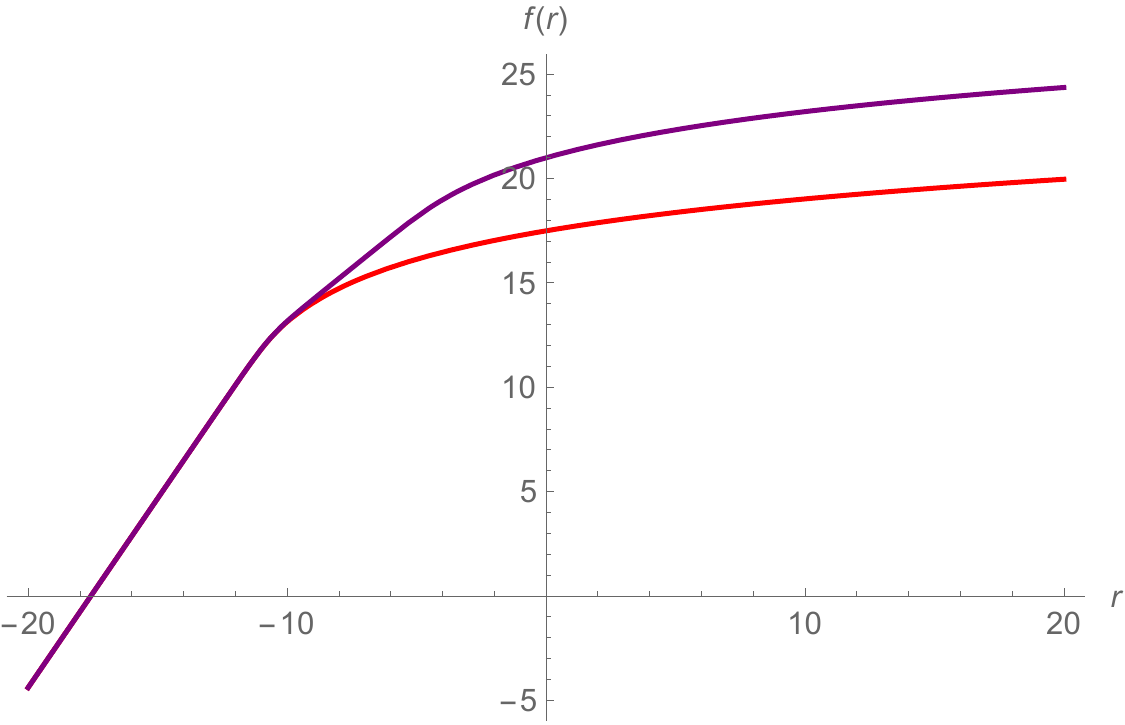}
                 \caption{Solutions for $f(r)$}
         \end{subfigure}
          \begin{subfigure}[b]{0.32\textwidth}
                 \includegraphics[width=\textwidth]{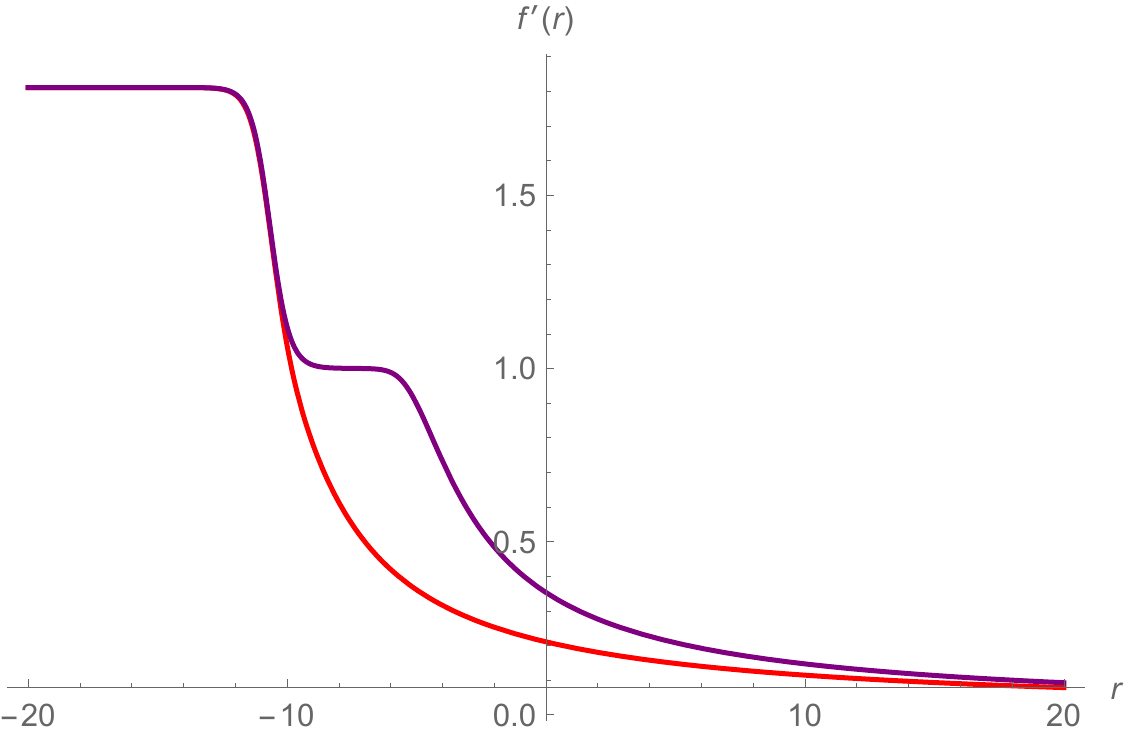}
                 \caption{Solutions for $f'(r)$}
         \end{subfigure}
         \caption{Examples of RG flows from an $N=2$ non-conformal field theory to $N=(0,2)$ SCFT in two dimensions dual to $AdS_3\times H^2$ geometry (red) and to four-dimensional $N=2$ SCFT and $N=(0,2)$ SCFT in two dimensions (purple) with $g=2$.}\label{fig52}
 \end{figure}

\begin{figure}
         \centering
         \begin{subfigure}[b]{0.32\textwidth}
                 \includegraphics[width=\textwidth]{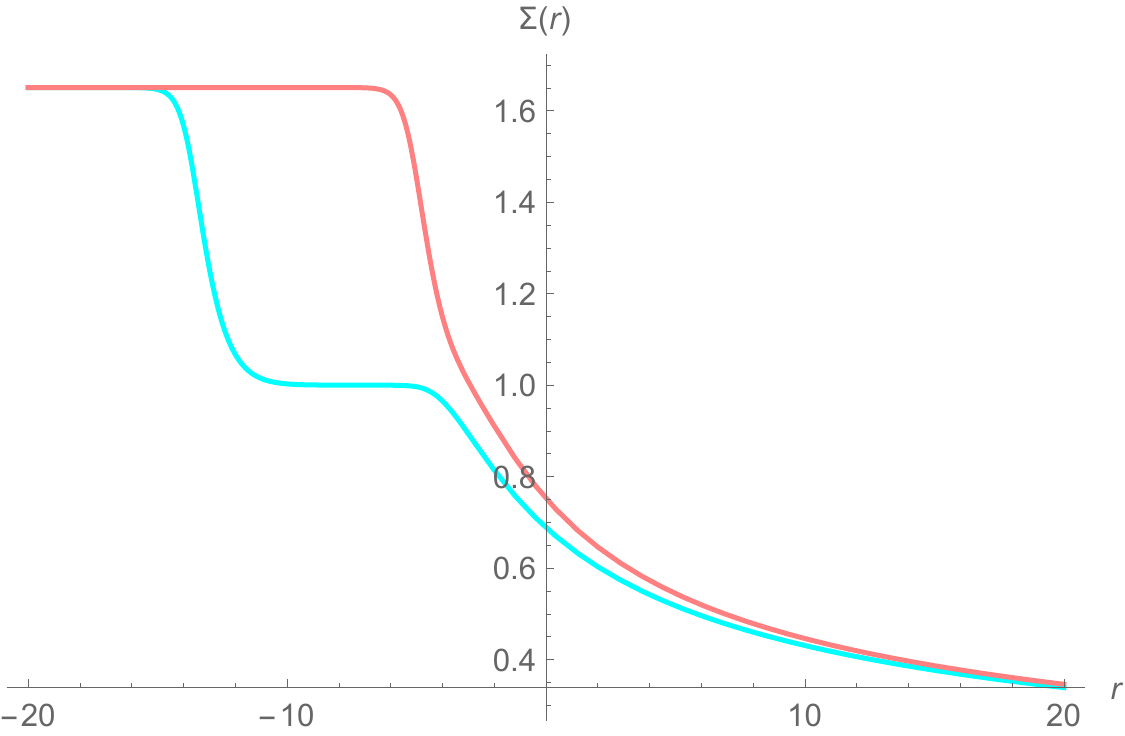}
                 \caption{Solutions for $\Sigma(r)$}
         \end{subfigure}
         \begin{subfigure}[b]{0.32\textwidth}
                 \includegraphics[width=\textwidth]{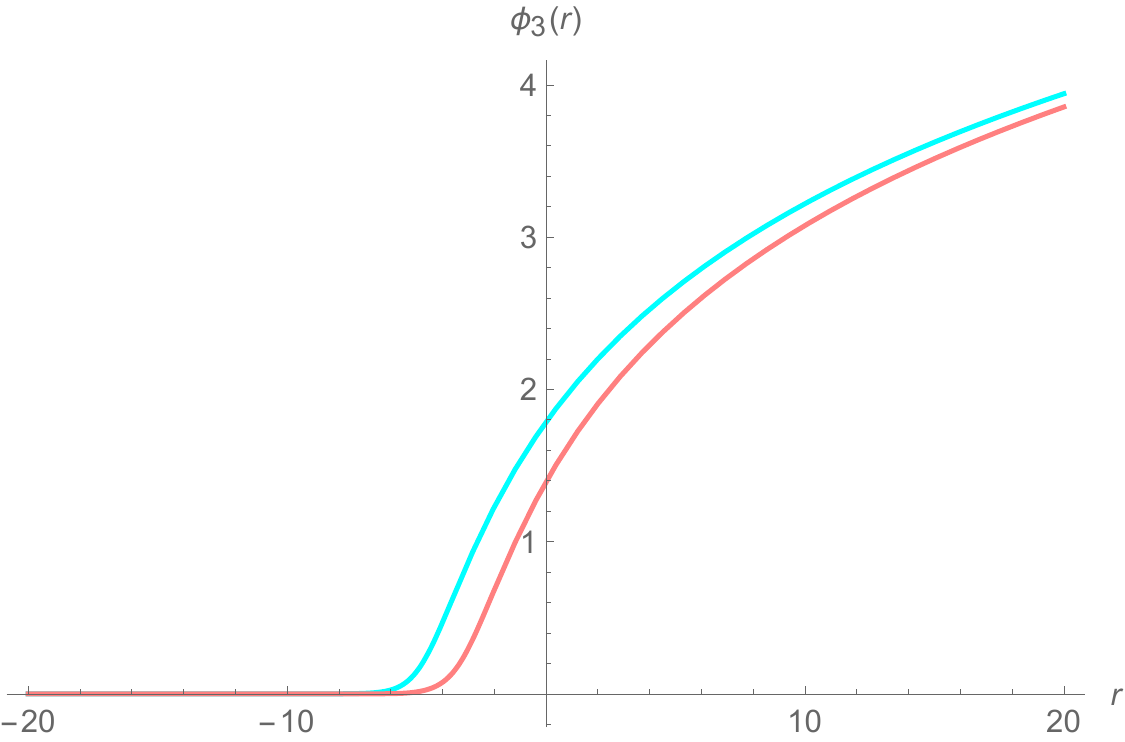}
                 \caption{Solutions for $\phi_3(r)$}
         \end{subfigure}
          \begin{subfigure}[b]{0.32\textwidth}
                 \includegraphics[width=\textwidth]{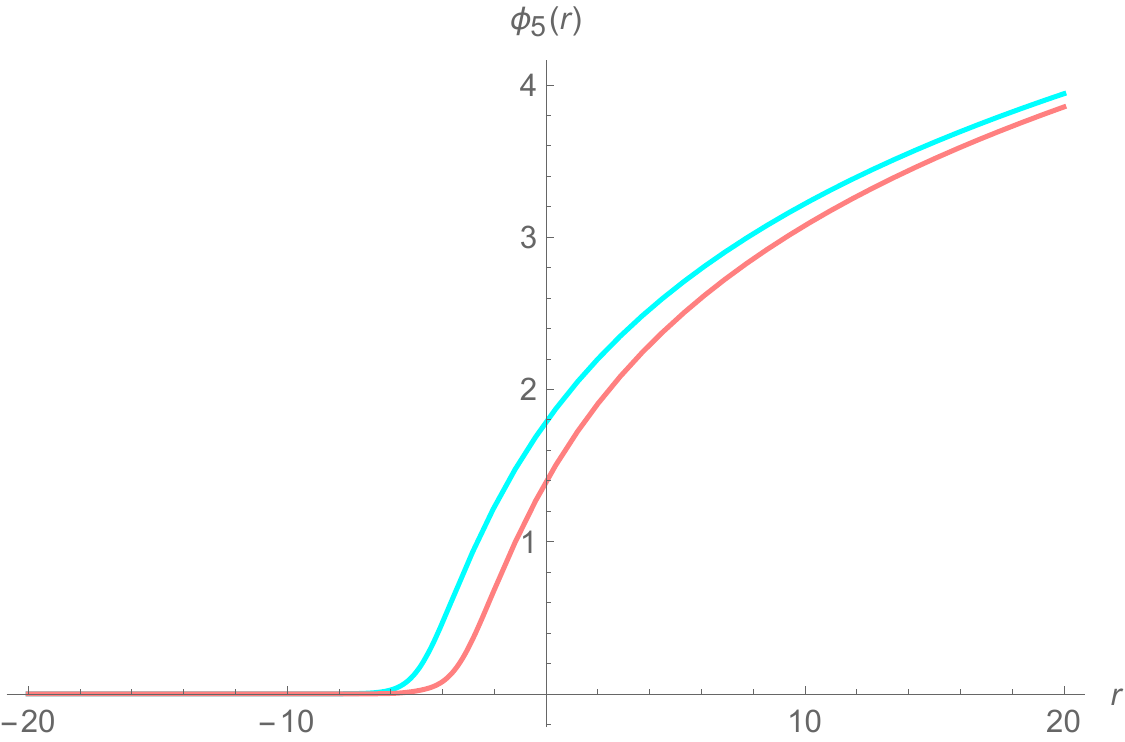}
                 \caption{Solutions for $\phi_5(r)$}
         \end{subfigure}\\
               \begin{subfigure}[b]{0.32\textwidth}
                 \includegraphics[width=\textwidth]{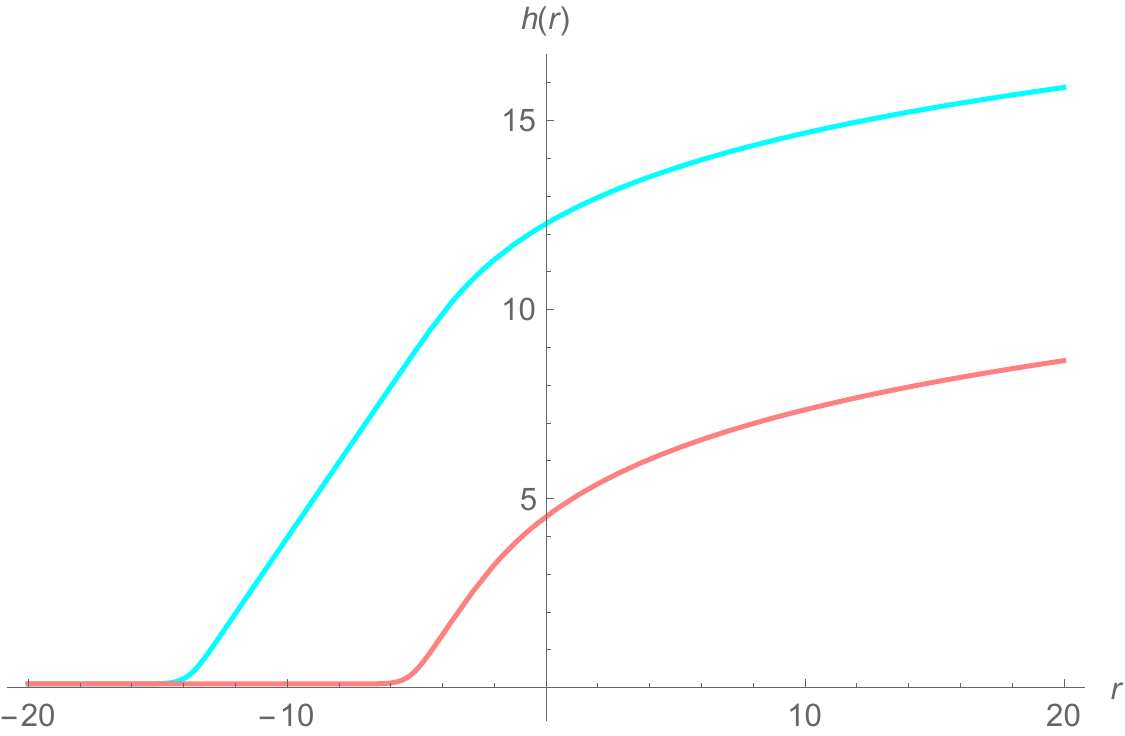}
                 \caption{Solutions for $h(r)$}
         \end{subfigure}
         \begin{subfigure}[b]{0.32\textwidth}
                 \includegraphics[width=\textwidth]{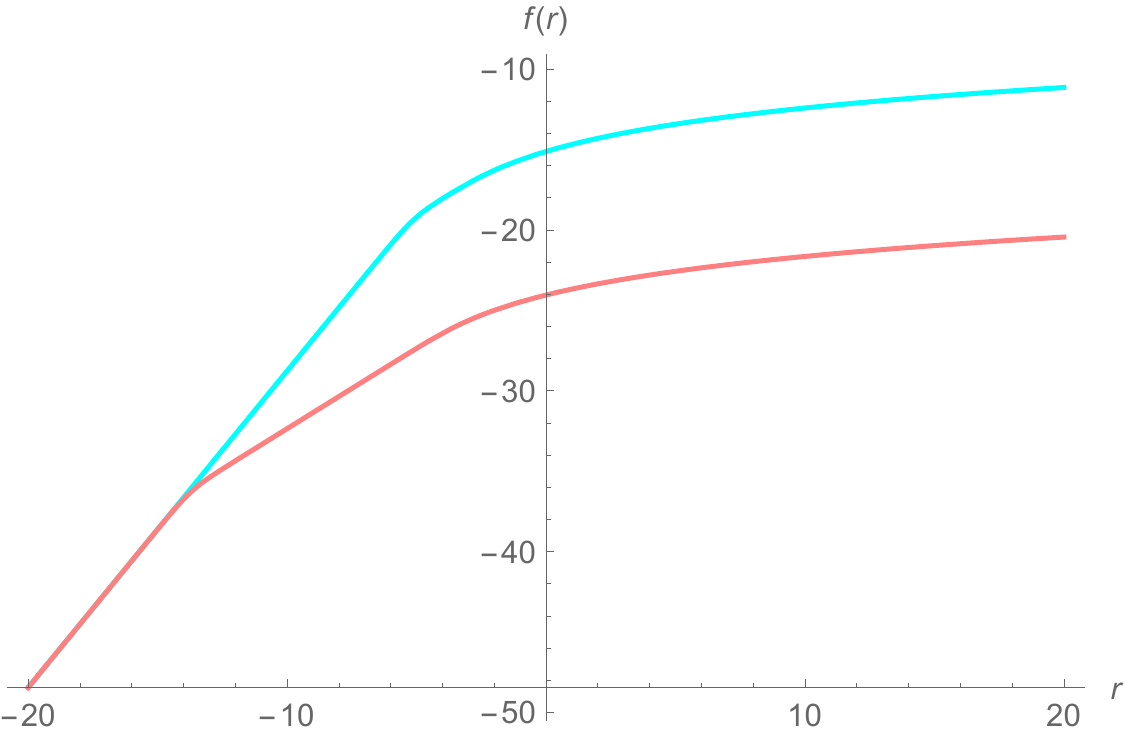}
                 \caption{Solutions for $f(r)$}
         \end{subfigure}
          \begin{subfigure}[b]{0.32\textwidth}
                 \includegraphics[width=\textwidth]{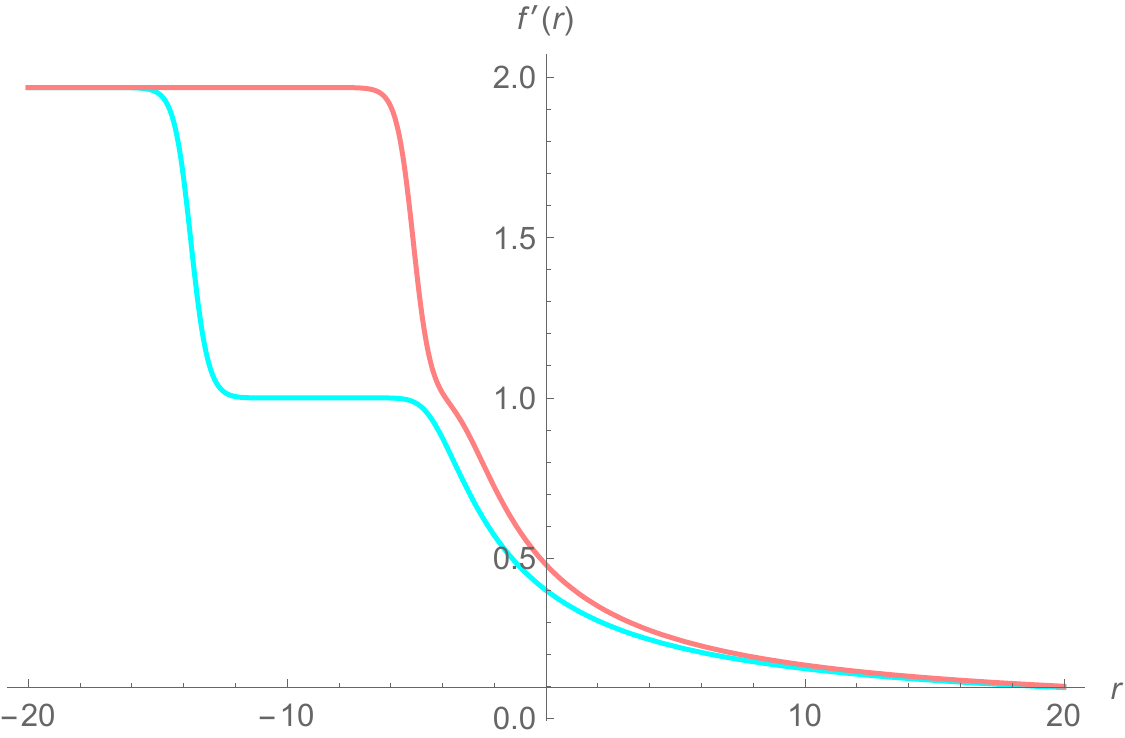}
                 \caption{Solutions for $f'(r)$}
         \end{subfigure}
         \caption{Examples of RG flows from an $N=2$ non-conformal field theory to $N=(0,2)$ SCFT in two dimensions dual to $AdS_3\times S^2$ geometry (pink) and to four-dimensional $N=2$ SCFT and $N=(0,2)$ SCFT in two dimensions (cyan) with $g=2$.}\label{fig53}
 \end{figure}

We now consider a twist on $\Sigma$ by $SO(2)\times SO(2)$ gauge fields. To cancel the spin connection on $\Sigma$ as shown in \eqref{twist_string}, we impose the following projectors
\begin{equation}
{(\mathbf{I}_2\otimes \sigma_3)_i}^j\epsilon_j=-\epsilon_i\qquad \textrm{and}\qquad \gamma_{\hat{\phi}\hat{\theta}}\epsilon_i=-i{(\sigma_2\otimes \sigma_3)_i}^j\epsilon_j\label{string_proN2} 
\end{equation} 
together with a twist condition
\begin{equation}
g_1a_0+ga_3=1\, .
\end{equation}      
Using the $\gamma_{\hat{r}}$-projector given in \eqref{gamma_r_string}, we find the following BPS equations
\begin{eqnarray}
\phi_3'&=&g\Sigma^{-1}e^{-2\phi_3}\sinh\phi_5,\\
\phi_5'&=&-\frac{1}{2}\Sigma^{-1}e^{-\phi_5}\left[ge^{-2\phi_3}(1-2e^{2\phi_3}+e^{2\phi_5})+\kappa a_3\Sigma^2e^{-2h}(e^{2\phi_5}-1)\right],\\
\Sigma'&=&\frac{1}{6}\left[ge^{-2\phi_3-\phi_5}(e^{2\phi_5}+2e^{2\phi_3}-1)-2\kappa \Sigma^{-1}(\sqrt{2}a_0+a_3\Sigma^3\cosh\phi_5)\right.\nonumber \\
& &\left.+2\sqrt{2}g_1\Sigma^3 \right],\\
h'&=&\frac{1}{6}\Sigma^{-2}\left[g\Sigma e^{-2\phi_3-\phi_5}(e^{2\phi_5}+2e^{2\phi_3}-1)-\sqrt{2}g_1\Sigma^4-2\sqrt{2}\kappa a_0e^{-2h}\right.\nonumber \\
& &\left.+4\kappa a_3\Sigma^3e^{-2h}\cosh\phi_5\right],\\
f'&=&\frac{1}{6}\Sigma^{-2}\left[g\Sigma e^{-2\phi_3-\phi_5}(e^{2\phi_5}+2e^{2\phi_3}-1)-\sqrt{2}g_1\Sigma^4+\sqrt{2}\kappa a_0e^{-2h}\right.\nonumber \\
& &\left.-2\kappa a_3\Sigma^3e^{-2h}\cosh\phi_5\right].
\end{eqnarray}
As in the previous case, consistency with all the field equations requires $\phi_4=0$. In this case, due to an extra projector in \eqref{string_proN2}, $AdS_3\times \Sigma$ fixed points preserve four supercharges while the full interpolating RG flow solutions preserve only two supercharges. In this case, the first projector in \eqref{string_proN2} implies that $\epsilon_1=\epsilon_3=0$. Using the relation $i\gamma^{\hat{0}}\gamma^{\hat{1}}\gamma^{\hat{r}}\gamma^{\hat{\theta}}\gamma^{\hat{\phi}}=\pm\mathbf{I}_4$ again, we find
\begin{equation}
\gamma^{\hat{0}}\gamma^{\hat{1}}\epsilon_{2,4}=\mp \epsilon_{2,4}
\end{equation}
which implies that the unbroken supercharges have a definite two-dimensional chirality corresponding to $N=(0,2)$ or $N=(2,0)$ supersymmetry.       
\\
\indent From the BPS equations, we find an $AdS_3\times \Sigma$ fixed point given by  
\begin{eqnarray}
& & \Sigma=\left[\frac{\sqrt{2}(a_0g_1-a_3g)}{g_1a_3}\right]^{\frac{1}{3}},\qquad h=\frac{1}{6}\ln\left[\frac{2\kappa a_3^4}{g_1^2(g_1a_0-ga_3)}\right],\nonumber \\
& &\phi_5=\phi_3=0,\qquad L_3=\frac{2}{(2a_3g-a_0g_1)}\left[\frac{\sqrt{2}a_3^2(a_0g_1-a_3g)}{g_1}\right]^{\frac{1}{3}}.\qquad 
\end{eqnarray}
Unlike the previous case of $SO(2)$ twist, there can be both $AdS_3\times H^2$ and $AdS_3\times S^2$ solutions depending on the values of $a_3$ and $g$. We again find that this is the same as the $AdS_3\times \Sigma$ solutions obtained in pure $N=4$ gauged supergravity. An example of numerical solutions interpolating between the $AdS_5$ vacuum and an $AdS_3\times H^2$ geometry is shown by the red line in figure \ref{fig5}. In this solution, we have chosen the following numerical values of various parameters
\begin{equation}
g=2,\qquad \kappa=-1,\qquad a_3=2\, .
\end{equation}
For solutions interpolating between the $AdS_5$ vacuum and an $AdS_3\times S^2$ geometry, a numerical solution is shown by the green line in figure \ref{fig5} with
\begin{equation}
g=2,\qquad \kappa=1,\qquad a_3=-2\, .
\end{equation}  
Both of these solutions should describe holographic RG flows from $N=2$ SCFT to two-dimensional $N=(0,2)$ SCFTs. The latter arise from twisted compactifications of the former on $H^2$ and $S^2$, respectively. 
\\
\indent Equivalently, these solutions correspond to black strings in asymptotically $AdS_5$ space with near horizon geometries given by $AdS_3\times H^2$ and $AdS_3\times S^2$. These geometries give rise to $AdS_3\times H^2\times H^2\times S^4$ and $AdS_3\times S^2\times H^2\times S^4$ solutions in eleven-dimensional supergravity via a consistent truncation on $H^2\times S^4$ \cite{Gauntlett_pure_5DN4_from_11D}. As in the previous case, using \eqref{2D_cen}, we can compute the central charge of the dual two-dimensional $N=(0,2)$ SCFT 
\begin{equation}
c=-\frac{24}{\pi}\frac{\kappa N^2a_3^2|\tilde{g}-1||\hat{g}-1|}{g^2(\sqrt{2}a_0+4a_3)}\left(\frac{a_3}{\sqrt{2}a_0+2a_3}\right)^{\frac{1}{5}}.
\end{equation}
\indent As in the previous case, by turning on scalars from vector multiplets, we find RG flow solutions interpolating between $AdS_3\times \Sigma$ fixed points and an $N=2$ non-conformal phase of $N=2$ SCFT in four dimensions as shown in figures \ref{fig52} and \ref{fig53}. These solutions describe RG flows from an $N=2$ field theory in four dimensions, arising from M5-branes wrapped on $H^2$, to two-dimensional $N=(0,2)$ SCFTs dual to $AdS_3\times H^2$ or $AdS_3\times S^2$ geometries in the IR. The solutions represented by red and pink lines correspond to RG flows directly from the $N=2$ non-conformal phase to two-dimensional $N=(0,2)$ SCFTs while solutions shown by purple and cyan lines describe RG flows to two-dimensional $N=(0,2)$ SCFTs via $N=2$ SCFT in four dimensions. As in the previous case, these solutions correspond to black strings in asymptotically domain wall space-time.

We end this section by some comments on the solutions with $SO(2)_{\textrm{diag}}$ twist. In this case, the scalar coset representative is given by \eqref{SO2d_coset}. The two $SO(2)$ gauge fields $A^0$ and $A^3$ are related by $g_1A^0=gA^3$. As in the previous section, non-vanishing $\phi_1$ scalar breaks supersymmetry corresponding to the Killing spinors $\epsilon_1$ and $\epsilon_3$. The topological twist is achieved by imposing the projectors
\begin{equation}
\gamma_{\hat{\phi}\hat{\theta}}\epsilon_2=\epsilon_4\qquad \textrm{and}\qquad \gamma_{\hat{\phi}\hat{\theta}}\epsilon_4=-\epsilon_2
\end{equation}
and a twist condition $ga_3=\frac{1}{2}$. However, in this case, there are no $AdS_3\times \Sigma$ fixed points from the resulting BPS equations. We refrain from giving the detail on this analysis here. 
\section{Supersymmetric $AdS_5$ black holes}\label{BH}
\begin{figure}
         \centering
               \begin{subfigure}[b]{0.32\textwidth}
                 \includegraphics[width=\textwidth]{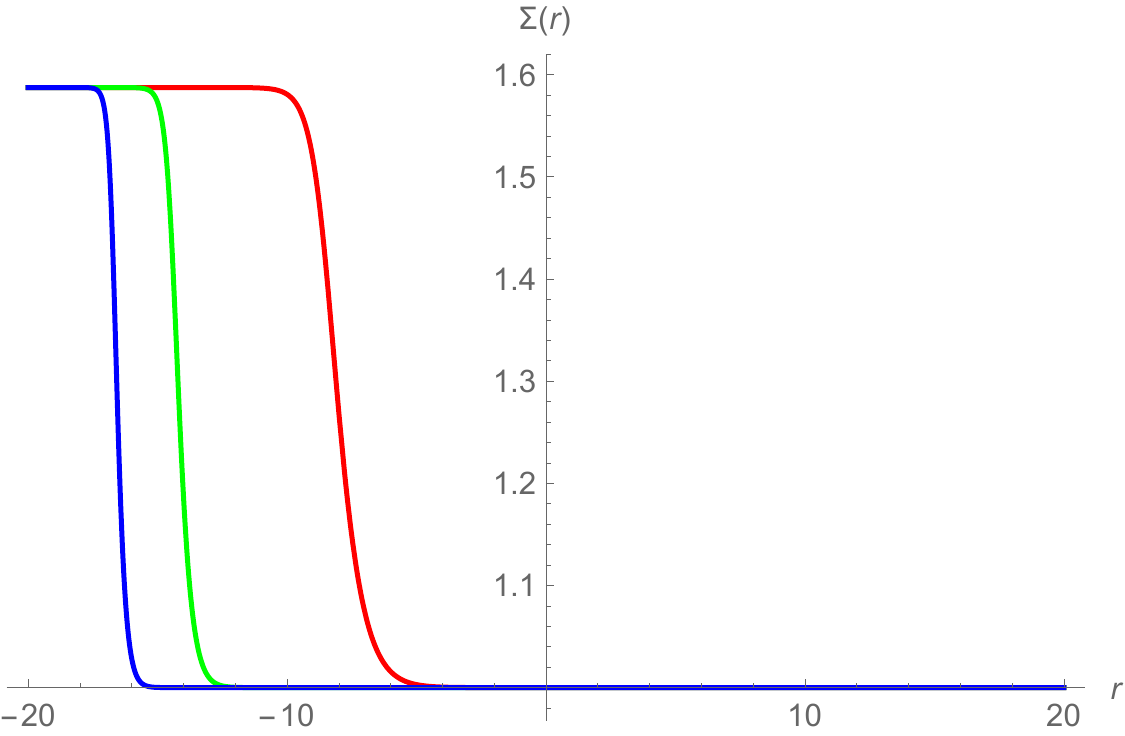}
                 \caption{Solution for $\Sigma(r)$}
         \end{subfigure}
         \begin{subfigure}[b]{0.32\textwidth}
                 \includegraphics[width=\textwidth]{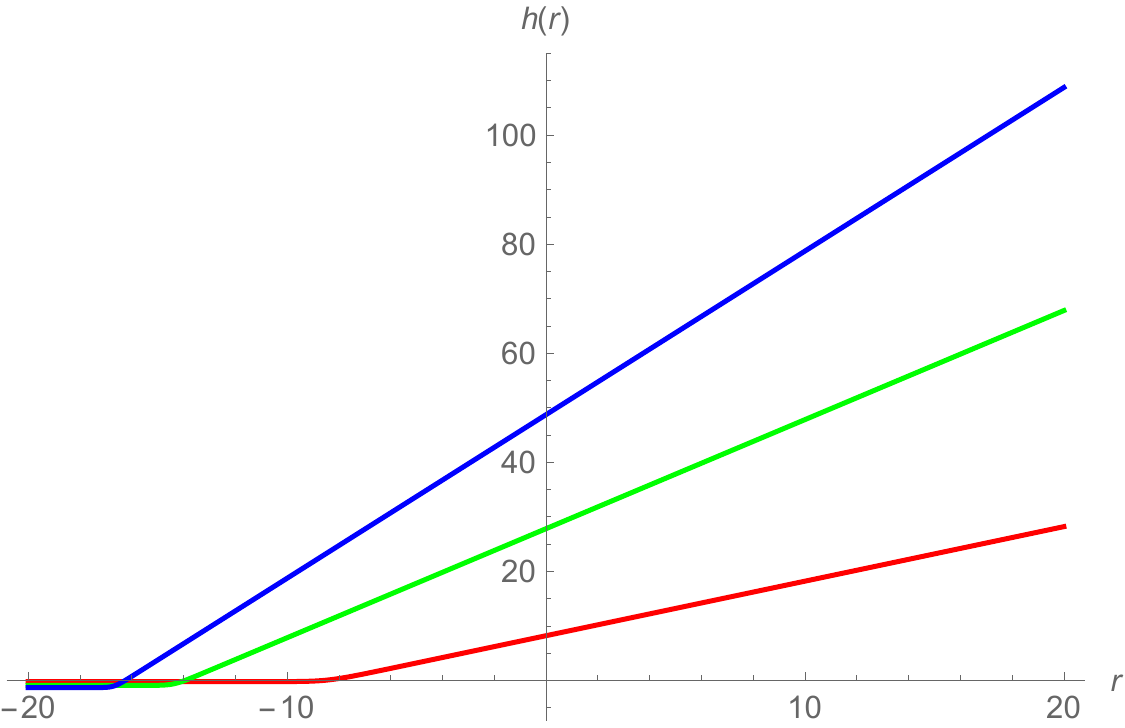}
                 \caption{Solution for $h(r)$}
         \end{subfigure}
         \begin{subfigure}[b]{0.32\textwidth}
                 \includegraphics[width=\textwidth]{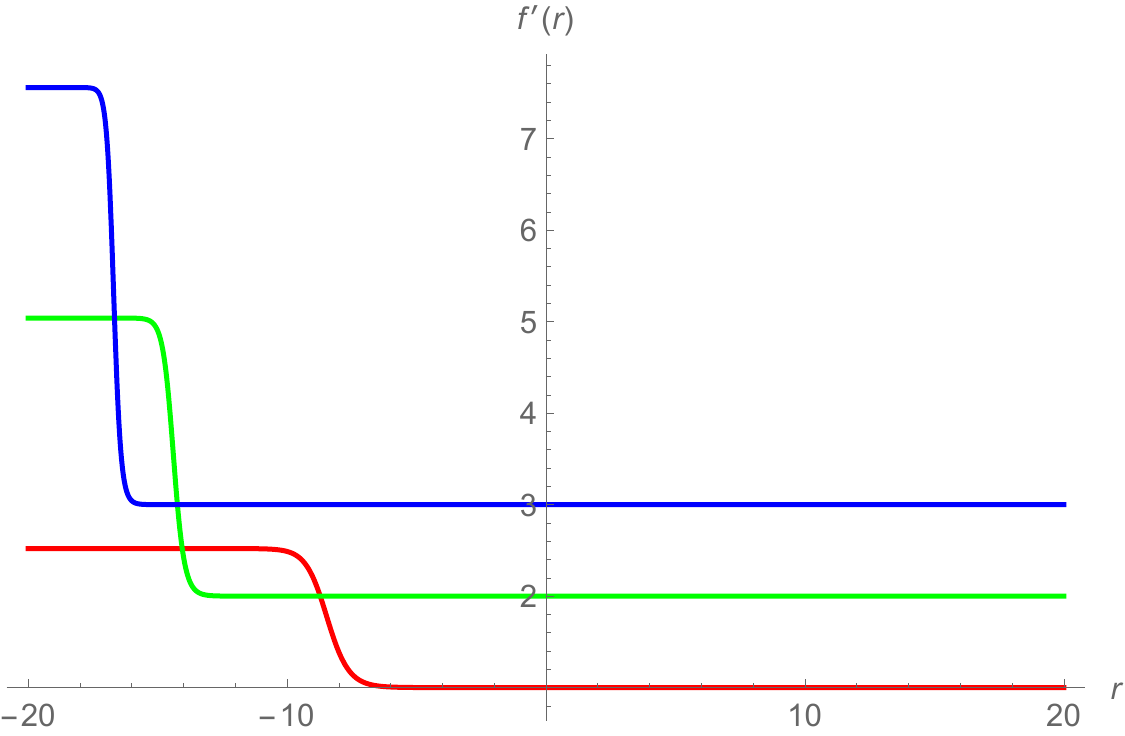}
                 \caption{Solution for $f'(r)$}
         \end{subfigure}
         \caption{Supersymmetric $AdS_5$ black hole solutions with $AdS_2\times H^3$ near horizon geometry for $g=2$ (red), $g=4$ (green) and $g=6$ (blue).}\label{fig6}
 \end{figure}

\begin{figure}
         \centering
               \begin{subfigure}[b]{0.35\textwidth}
                 \includegraphics[width=\textwidth]{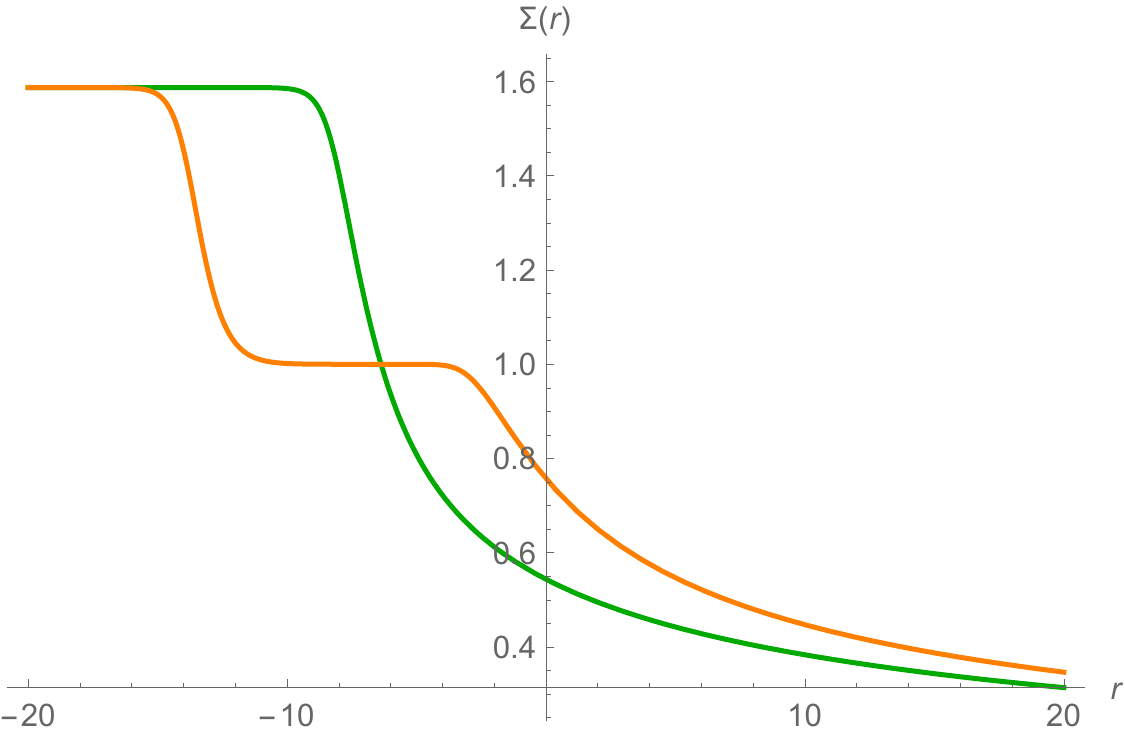}
                 \caption{Solution for $\Sigma(r)$}
         \end{subfigure}
                        \begin{subfigure}[b]{0.35\textwidth}
                 \includegraphics[width=\textwidth]{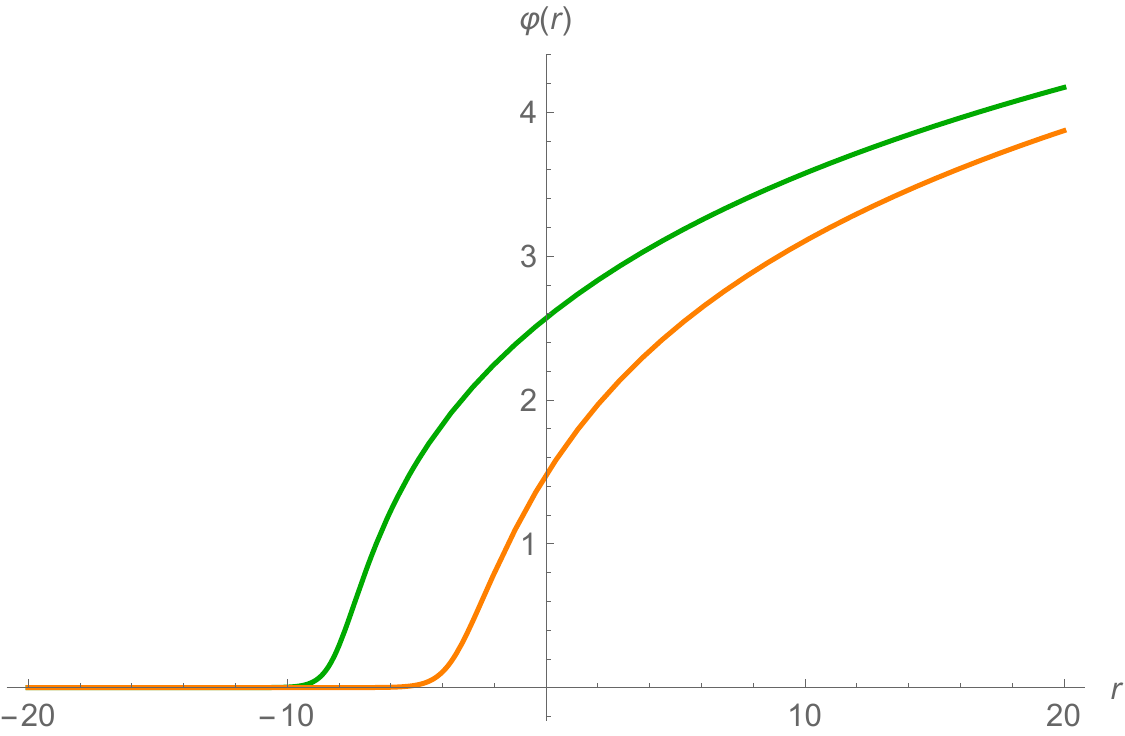}
                 \caption{Solution for $\varphi(r)$}
         \end{subfigure}\\
         \begin{subfigure}[b]{0.35\textwidth}
                 \includegraphics[width=\textwidth]{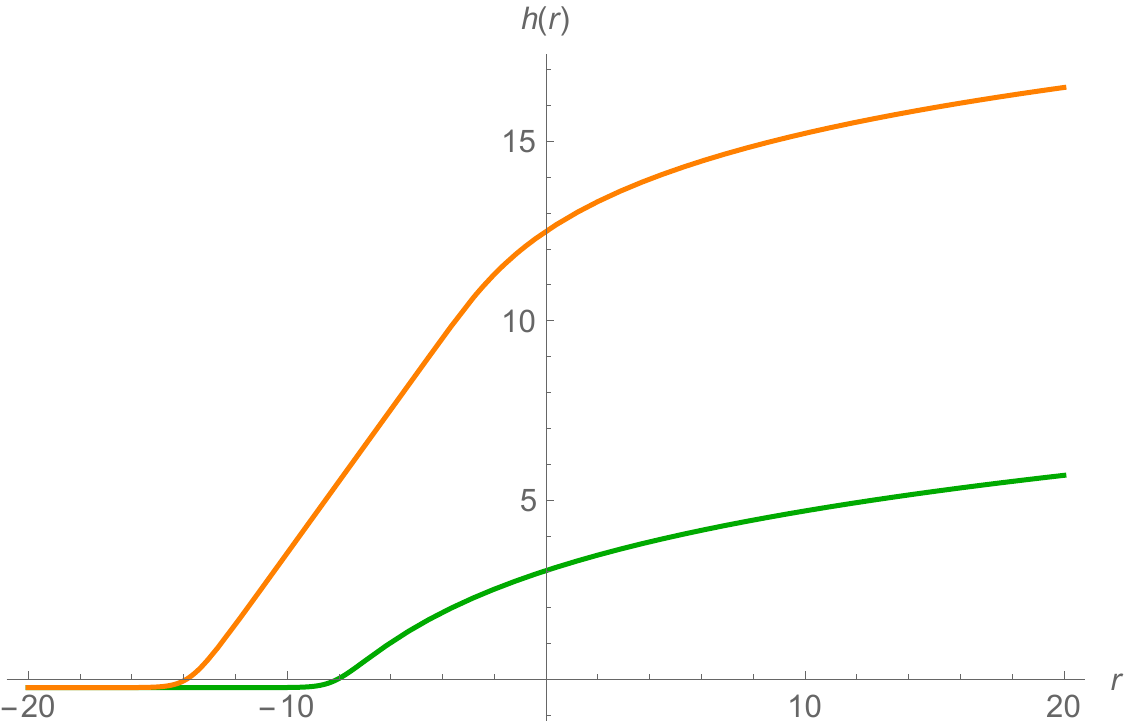}
                 \caption{Solution for $h(r)$}
         \end{subfigure}
         \begin{subfigure}[b]{0.35\textwidth}
                 \includegraphics[width=\textwidth]{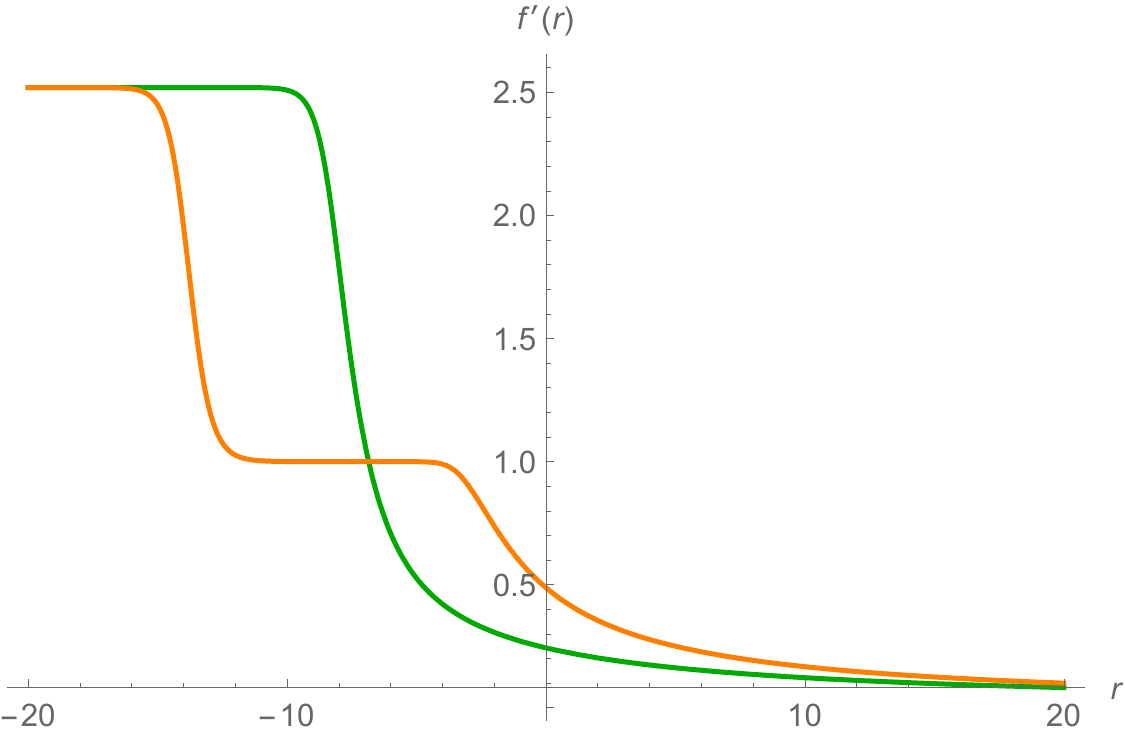}
                 \caption{Solution for $f'(r)$}
         \end{subfigure}
         \caption{Examples of RG flows from an $N=2$ non-conformal field theory to superconformal quantum mechanics dual to $AdS_2\times H^3$ geometry (green) and to four-dimensional $N=2$ SCFT and superconformal quantum mechanics (orange) with $g=2$.}\label{fig61}
 \end{figure}

In this section, we perform a similar analysis in the case of supersymmetric $AdS_5$ black hole solutions with a near horizon geometry given by $AdS_2\times \mc{M}_3$. We will consider the case of $\mc{M}_3$ being a constant curvature $3$-manifold in the form of $H^3$ or $S^3$. The metric ansatz is given by 
\begin{equation}
ds^2=-e^{2f(r)}dt^2+dr^2+e^{2h(r)}\left[d\psi^2+f_\kappa^2(\psi)(d\theta^2+\sin^2\theta d\phi^2)\right]
\end{equation}
with $f_\kappa(\psi)$ defined in \eqref{f_def}. For convenience, we note the vielbein and spin connection of this metric as follows
\begin{eqnarray}
& & e^{\hat{0}}=e^fdt,\qquad e^{\hat{r}}=dr,\qquad e^{\hat{\psi}}=e^hd\psi,\nonumber \\
& & e^{\hat{\theta}}=e^hf_\kappa(\theta)d\theta,\qquad e^{\hat{\phi}}=e^hf_\kappa(\theta)\sin\theta d\phi
\end{eqnarray}
and
\begin{eqnarray}
& &{\omega^{\hat{0}}}_{\hat{r}}=f'e^{\hat{0}},\qquad {\omega^{\hat{\psi}}}_{\hat{r}}=h'e^{\hat{\psi}},\qquad {\omega^{\hat{\theta}}}_{\hat{r}}=h'e^{\hat{\theta}},\qquad {\omega^{\hat{\phi}}}_{\hat{r}}=h'e^{\hat{\phi}},\nonumber \\
& &{\omega^{\hat{\phi}}}_{\hat{\psi}}=\frac{f_\kappa'(\psi)}{f_\kappa(\psi)}e^{-h}e^{\hat{\phi}},\qquad {\omega^{\hat{\phi}}}_{\hat{\theta}}=\frac{e^{-h}}{f_\kappa(\psi)}\cot\theta e^{\hat{\phi}},\qquad {\omega^{\hat{\theta}}}_{\hat{\psi}}=\frac{f_\kappa'(\psi)}{f_\kappa(\psi)}e^{-h}e^{\hat{\theta}}\, .\qquad \label{BH_spin_con}
\end{eqnarray}
To perform a topological twist, we turn on $SO(3)$ gauge fields corresponding to $A^3$, $A^4$ and $A^5$. We then consider $SO(3)$ invariant scalars. There is only one $SO(3)$singlet scalar from $SO(5,3)/SO(5)\times SO(3)$ corresponding to the non-compact generator
\begin{equation}
Y_s=Y_{31}+Y_{42}+Y_{53}\, .
\end{equation}
The coset representative is given by
\begin{equation}
\mc{V}=e^{\varphi Y_s}\, .
\end{equation}
\indent The relevant terms in the composite connection are given by
\begin{equation}
{Q_i}^j=-\frac{i}{2}gA^3{(\sigma_2\otimes \mathbf{I}_2)_i}^j+\frac{i}{2}gA^4{(\sigma_3\otimes \sigma_1)_i}^j-\frac{i}{2}gA^5{(\sigma_1\otimes \sigma_1)_i}^j\, .
\end{equation}
To cancel the components of the spin connection along $\mc{M}_3$, given by the second line of \eqref{BH_spin_con}, we take the ansatz for the gauge fields of the form
\begin{equation}
A^3=-a_3\cos\theta d\phi,\qquad A^4=-a_4f'_\kappa(\psi)\sin\theta d\phi,\qquad A^5=-a_5f'_\kappa(\psi)d\theta\, . 
\end{equation}
We perform the twist by imposing the following projectors
\begin{eqnarray}
& &\gamma_{\hat{\theta}\hat{\psi}}\epsilon_i=-i{(\sigma_1\otimes \sigma_1)_i}^j\epsilon_j,\nonumber \\ 
& &\gamma_{\hat{\phi}\hat{\psi}}\epsilon_i=-i{(\sigma_3\otimes \sigma_1)_i}^j\epsilon_j,\nonumber \\
& &\gamma_{\hat{\phi}\hat{\theta}}\epsilon_i=-i{(\sigma_2\otimes \mathbf{I}_2)_i}^j\epsilon_j\label{BH_pro}
\end{eqnarray}
and twist conditions
\begin{equation}
a_3g=1,\qquad a_4g=-1,\qquad a_5g=1\, .
\end{equation}
We also note that only two projectors in \eqref{BH_pro} are independent. Accordingly, the near horizon geometry $AdS_2\times \mc{M}_3$ preserves four supercharges. As in the case of black string solutions, all the two-form fields can be consistently set to zero. It is useful to note the field strength tensors for the gauge fields
\begin{equation}
F^3=\kappa ae^{-2h}e^{\hat{\theta}}\wedge e^{\hat{\phi}},\quad F^4=-\kappa ae^{-2h}e^{\hat{\psi}}\wedge e^{\hat{\phi}},\quad F^5=\kappa a e^{-2h}e^{\hat{\psi}}\wedge e^{\hat{\theta}}
\end{equation}
in which we have written $a_5=a_3=-a_4=a$ and used the relations $f''_\kappa(\psi)=-\kappa f_\kappa(\psi)$ and $1-{f'_\kappa(\psi)}^2=\kappa f^2_\kappa(\psi)$.
\\
\indent Using the projector \eqref{gamma_r_string}, we find the following BPS equations
\begin{eqnarray}
\varphi'&=&\frac{1}{2}\Sigma^{-1}e^{-2h-3\varphi}(e^{2\varphi}-1)(ge^{2h}-a \kappa \Sigma^2 e^{2\varphi}),\\
\Sigma'&=&\frac{1}{3}\left[ge^{-2\varphi}(\cosh\varphi+2\sinh\varphi)+\sqrt{2}g_1\Sigma^3-3\kappa a \Sigma^2e^{-2h}\cosh\phi_3\right],\\
h'&=&-\frac{1}{6}\Sigma^{-1}\left[ge^{-3\varphi}(1-3e^{2\varphi})+\sqrt{2}g_1\Sigma^3-6\kappa a \Sigma^2 e^{-2h}\cosh\varphi\right],\\
f'&=&-\frac{1}{6}\Sigma^{-1}\left[ge^{-3\varphi}(1-3e^{2\varphi})+\sqrt{2}g_1\Sigma^3+6\kappa a \Sigma^2 e^{-2h}\cosh\varphi\right].
\end{eqnarray}
These equations admit one $AdS_2\times \mc{M}_3$ fixed point solution given by
\begin{equation}
\varphi=0,\qquad \Sigma=-\sqrt{2}\left(\frac{g}{g_1}\right)^{\frac{1}{3}},\qquad h=\frac{1}{2}\ln \left[-\frac{2a\kappa}{(gg_1^2)^{\frac{1}{3}}}\right].
\end{equation}
The solution exists only for $\kappa=-1$ giving rise to $AdS_2\times H^3$ geometry. This solution is also the same as that of pure $N=4$ gauged supergravity \cite{AdS5_BH_Romans}. By setting $\varphi=0$, we find examples of numerical solutions interpolating between the $AdS_5$ vacuum and this $AdS_2\times H^3$ geometry as shown in figure \ref{fig6} with three different values of $g=2,4,6$. We also point out that all values of $g\neq 0$ lead to physically equivalent solutions. These solutions describe supersymmetric black holes in asymptotically $AdS_5$ space with $AdS_2\times H^3$ near horizon geometry. Holographically, the solutions correspond to RG flows across dimensions from $N=2$ SCFT in four dimensions to superconformal quantum mechanics via twisted compactifications on $H^3$. Upon uplifted to eleven dimensions, these solutions lead to $AdS_2\times H^3\times H^2\times S^4$ geometry in M-theory \cite{Gauntlett_pure_5DN4_from_11D}. Similar to the black string solutions, turning on the scalar $\varphi$ from vector multiplets leads to solutions describing RG flows from an $N=2$ non-conformal field theory, arising from M5-branes wrapped on $H^2$, and $N=2$ SCFT in four dimensions to superconformal quantum mechanics as shown by the green and orange lines in figure \ref{fig61}, respectively. These solutions describe black holes in asymptotically domain wall space-time.
\\
\indent We end this section by giving the entropy of the black hole using the formulae
\begin{equation}
S_{\textrm{BH}}=\frac{A}{4G_N^{(5)}}\, .
\end{equation}
Using $G_N^{(5)}$ given in \eqref{GN_5} and $A=\textrm{vol}(H^3)e^{3h_0}$, we find the entropy of the black hole
\begin{equation}
S_{\textrm{BH}}=\frac{N^2|\tilde{g}-1|\textrm{vol}(H^3)}{2^{\frac{7}{5}}\pi^2 g^3}\, .\label{S_BH}
\end{equation}        
    
\section{Conclusions and discussions}\label{conclusion}
In this paper, we have studied various holographic solutions from five-dimensional $N=4$ gauged supergravity with $SO(2)\times ISO(3)$ gauge group. The gauged supergravity admits a unique $N=4$ supersymmetric $AdS_5$ vacuum dual to an $N=2$ SCFT in four dimensions. The $N=2$ SCFT arises from M5-branes wrapped on a Riemann surface with genus higher than one, $H^2$. We have found solutions describing holographic RG flows preserving eight supercharges between this $N=2$ SCFT and non-conformal phases with $SO(2)\times SO(3)$ and $SO(2)\times SO(2)$ symmetries. The solution with only the dilaton non-vanishing describes an RG flow from the $N=2$ SCFT in the UV to an $N=2$ non-conformal phase in the IR. Both the five-dimensional solution and the uplifted eleven-dimensional one contain a physically acceptable IR singularity according to the criteria of \cite{Gubser_Sing} and \cite{MN_nogo}. This RG flow is driven by a vacuum expectation value of the dual operator of dimension $\Delta=2$. The other $N=2$ solutions correspond to RG flows from $N=2$ non-conformal phases with $SO(2)\times SO(3)$ and $SO(2)\times SO(2)$ symmetries to an $N=2$ conformal fixed point in the IR. These RG flows are due to the deformations involving an irrelevant operator of dimension $\Delta=6$ in the presence of non-vanishing vacuum expectation values of marginal and relevant operators of dimensions $\Delta=4$ and $\Delta=2$. Upon uplifted to eleven dimensions, the $N=2$ non-conformal field theories in the UV can be identified as decompactifcation limits of $N=(0,2)$ six-dimensional field theory on the world-volume of M5-branes wrapped on $H^2$. We have also found an RG flow solution preserving four supercharges and $SO(2)_{\textrm{diag}}$ symmetry from the $N=2$ non-conformal phase to $N=2$ SCFT in the IR. In this case, the RG flow is driven by the deformations due to irrelevant operators of dimensions $\Delta=5$ and $\Delta=6$. The UV field theory in this cas is also expected to arise from a decompactification limit of six-dimensional theories on the world-volume of M5-branes wrapped on $H^2$.
\\
\indent Another class of solutions are Janus interfaces described by $AdS_4$-sliced domain walls. We have studied these solutions within the $SO(2)_{\textrm{diag}}$ truncation. There do not exist regular Janus solutions interpolating between $AdS_5$ vacua on both sides of the interfaces at least within the truncation considered here. However, we have found solutions interpolating between non-conformal phases. The entire solutions preserve four supercharges while the non-conformal phases on both sides preserve eight supercharges. These solutions would describe conformal interfaces induced by position-dependent deformations and position-dependent vacuum expectation values of operators within $N=2$ non-conformal field theories arising from M5-branes wrapped on $H^2$. The solution provides the first example of non-conformal Janus solutions in five-dimensional gauged supergravities. We have also given examples of solutions interpolating between the non-conformal phase and a singularity as well as between singularities. We expect these solutions to describe conformal boundaries as pointed out in \cite{BCFT_Gutperle}.   
\\
\indent As a final class of solutions, we have considered supersymmetric $AdS_5$ black string and black hole solutions. By performing an $SO(2)$ twist on a Riemann surface, we have found an $AdS_5$ black string preserving four supercharges with $AdS_3\times H^2$ near horizon geometry. On the other hand, by turning on $SO(2)\times SO(2)$ gauge fields to implement the topological twist, we have found supersymmetric $AdS_5$ black strings preserving two supercharges with both $AdS_3\times H^2$ and $AdS_3\times S^2$ near horizon geometries. These solutions holographically describe RG flows across dimensions from $N=2$ SCFT in four dimensions to two-dimensional $N=(2,2)$ and $N=(0,2)$ SCFTs in the IR. We also note that the near horizon geometries enhance supersymmetry to eight and four supercharges, respectively.  
\\
\indent It turns out that the existence of $AdS_3\times \Sigma$ solutions requires vanishing of scalar fields from vector multiplets. Therefore, the near horizon geometries and the entire flow solutions with only the dilaton non-vanishing are the same as those of pure $N=4$ $SU(2)\times U(1)$ gauged supergravity. Upon uplifted to eleven dimensions, these geometries would lead to $AdS_3\times H^2\times H^2\times S^4$ and $AdS_3\times S^2\times H^2\times S^4$ solutions of eleven-dimensional supergravity \cite{Gauntlett_pure_5DN4_from_11D}. By turning on scalars from vector multiplets dual to irrelevant and marginal operators, we have found solutions interpolating between $AdS_3\times \Sigma$ and an $N=2$ non-conformal phase in four dimensions. We interpret these solutions as holographic RG flows from an $N=2$ non-conformal field theory arising from M5-branes wrapped on $H^2$ to $N=(2,2)$ and $N=(0,2)$ SCFTs in two dimensions. For particular values of the boundary condition, the solutions flow from this $N=2$ non-conformal field theory to $N=2$ SCFT and then to $N=(2,2)$ and $N=(0,2)$ SCFTs in two dimensions. Similarly, the only $AdS_2\times H^3$ near horizon geometry of black holes is that of pure $N=4$ gauged supergravity. As in the case of black strings, when scalars from vector multiplets participate in the flows, we have found solutions interpolating between the $AdS_2\times H^3$ near horizon geometry and an $N=2$ non-conformal phase with and without intermediate $N=4$ $AdS_5$ geometry. These solutions are expected to describe holographic RG flows from $N=2$ non-conformal field theory in four dimensions obtained from M5-branes wrapped on $H^2$ to $N=2$ SCFT and superconformal quantum mechanics in the IR.    
\\
\indent It would be interesting to extend the present study to other types of holographic solutions such as line defects within $N=2$ SCFT considered recently in \cite{line_5Ddefect_Petri} with non-vanishing two-form fields and solutions describing strings and black holes with the near horizon geometries involving spindles or topological disks as in \cite{Gauntlett_D3_spindle,Minwoo_D3_spindle}. It could also be interesting to identify the field theory duals of the gravity solutions given in this paper. In particular, the field theory interpretation of the Janus solution between non-conformal phases would be highly desirable. In addition, recovering the black hole entropy given in \eqref{S_BH} by using the twisted index of the $N=2$ SCFT on $H^3$ as in \cite{Twisted_index1,Twisted_index2,Twisted_index3} is worth considering as well. We hope to come back to some of these issues in future works.  
\begin{acknowledgments}
The author would like to thank Jerome P. Gauntlett for helpful correspondences and valuable comments. Extensive discussions with Carlos Nunez together with various useful comments and suggestions from an anonymous referee are gratefully acknowledged.
\end{acknowledgments}
\appendix
\section{Eleven-dimensional metric}
In this appendix, we give some useful relations used to derive the eleven-dimensional metric in the main text. The procedure is to uplift the five-dimensional metric to seven dimensions by using the results of \cite{ISO3_5D_N4_gauntlett} and then further uplift the resulting seven-dimensional metric to eleven dimensions using the $S^4$ truncation of eleven-dimensional supergravity given in \cite{M-theory_on_S4_2}.
\\
\indent In the notation of \cite{ISO3_5D_N4_gauntlett}, for vanishing gauge fields, the seven-dimensional metric is given by
\begin{equation}
ds^2_{11}=\Delta^{\frac{1}{3}}ds^2_7+\frac{1}{m^2}\Delta^{-\frac{2}{3}}T^{-1}_{\hat{a}\hat{b}}D\mu^{\hat{a}}D\mu^{\hat{b}}
\end{equation}
with
\begin{equation}
ds_7^2=e^{-4\phi}ds^2_5+e^{6\phi}ds^2(H_2)\, .
\end{equation}
$m$ is the seven-dimensional gauge coupling constant, and the warp factor $\Delta$ is defined by 
\begin{equation}
\Delta=T_{\hat{a}\hat{b}}\mu^{\hat{a}}\mu^{\hat{b}},\qquad \hat{a},\hat{b}=1,2,\ldots,5
\end{equation}
with $T_{\hat{a}\hat{b}}$ being a symmetric $SL(5)$ matrix and $\mu^{\hat{a}}$ are constrained coordinates on $S^4$ satisfying $\mu^{\hat{a}}\mu^{\hat{a}}=1$. We also note that $D\mu^{\hat{a}}=d\mu^{\hat{a}}+mA^{\hat{a}\hat{b}}\mu^{\hat{b}}$. Although all the solutions given here have vanishing five-dimensional vector fields, there is a non-vanishing gauge field in seven dimensions namely $A^{12}=\frac{1}{m}\omega_H$ with $\omega_H$ being the spin connection on $H^2$. This leads to $D\mu^1=d\mu^1+\omega_H$ and $D\mu^2=d\mu^2-\omega_H$ while $D\mu^{3,4,5}=d\mu^{3,4,5}$.
\\
\indent In the truncation considered in \cite{ISO3_5D_N4_gauntlett}, the matrix $T_{\hat{a}\hat{b}}$ decomposes as 
\begin{equation}
T_{\hat{a}\hat{b}}=\begin{pmatrix}
e^{-6\lambda} & & \\
 & e^{-6\lambda} & \\
  &   & e^{4\lambda}\mc{T}_{\alpha\beta}
\end{pmatrix}
\end{equation}
with $\mc{T}_{\alpha\beta}$, $\alpha,\beta=1,2,3$ is a symmetric matrix parametrizing $SL(3)/SO(3)\subset SL(5)/SO(5)$ submanifold. In terms of $SL(3)/SO(3)$ coset representative $V$, we have $\mc{T}_{\alpha\beta}=(VV^t)_{\alpha\beta}$. 
\\
\indent We also note that the $SO(5,3)$ invariant tensor used in \cite{ISO3_5D_N4_gauntlett} is off-diagonal of the form
\begin{equation}
\tilde{\eta}_{MN}=\begin{pmatrix} \mathbf{I}_2 & 0 & 0 \\
							0 & 0  &  \mathbf{I}_3\\
							0 &	\mathbf{I}_3 & 0 \end{pmatrix}
\end{equation}
in which we have rearranged some rows and columns to match the present convention. $\tilde{\eta}_{MN}$ is related to the diagonal $\eta_{MN}$ used in this paper via a transformation matrix of the form
\begin{equation}
\mc{U}=\begin{pmatrix}
\mathbf{I}_2 & 0 & 0 \\
0 &  -U  &  U \\
0 &  U  &  U
\end{pmatrix}
\end{equation}
with
\begin{equation}
U=\frac{1}{\sqrt{2}}\begin{pmatrix}
0 & 0 & 1\\
0 & 1 & 0\\
1 & 0  & 0
\end{pmatrix}.
\end{equation}
We also note some useful relations $\mc{U}=\mc{U}^{-1}=\mc{U}^t$.

\subsection{$SO(2)\times SO(3)$ invariant scalars}
In this case, we have $\mc{T}_{\alpha\beta}=\delta_{\alpha\beta}$, and the $SO(5,3)/SO(5)\times SO(3)$ coset representative given in \cite{ISO3_5D_N4_gauntlett} becomes
\begin{equation}
\tilde{\mc{V}}=\begin{pmatrix}
\mathbf{I}_2  &  & \\
  &   e^{-\varphi_3}\mathbf{I}_3 &  \\
     &   &   e^{\varphi_3}\mathbf{I}_3
\end{pmatrix}.
\end{equation}
By transforming to the basis with diagonal $\eta_{MN}$ using $\mc{U}$, we precisely recover the coset representative \eqref{SO2d_coset} with only $\phi_3$ and $\phi_5=\phi_3$ non-vanishing. We then identify $\varphi_3$ with $\phi_3$. Using the relations given in \cite{ISO3_5D_N4_gauntlett}
\begin{equation}
\varphi_3=3\phi-\lambda\qquad \textrm{and}\qquad \Sigma=e^{-\phi-3\lambda},
\end{equation}  
we can determine $\phi$ and $\lambda$ in terms of the scalars $\Sigma$ and $\phi_3$ in section \ref{RG_flows} as
\begin{equation}
\phi=\frac{1}{10}(3\phi_3-\ln\Sigma)\qquad \textrm{and}\qquad \lambda=-\frac{1}{10}(\phi_3+3\ln \Sigma).
\end{equation}

\subsection{$SO(2)\times SO(2)$ invariant scalars}
In this case, we have
\begin{equation}
\mc{T}_{\alpha\beta}=\textrm{diag}(e^w, e^w, e^{-2w}).
\end{equation}
After trasforming to the basis with diagonal $\eta_{MN}$, we find the following identification
\begin{equation}
\phi_3=\varphi_3+\frac{w}{2}\qquad \textrm{and}\qquad \phi_5=\varphi_3-w
\end{equation}
or
\begin{equation}
w=\frac{2}{3}(\phi_3-\phi_5)=\frac{2}{3}\varphi_2\qquad \textrm{and}\qquad \varphi_3=\frac{1}{3}(\phi_5+2\phi_3)=\frac{1}{6}(3\varphi_1+\varphi_2).
\end{equation}

\end{document}